\documentclass[aps,prd,nofootinbib,twocolumn,reprint,superscriptaddress,showpacs,showkeys,longbibliography]{revtex4-1}
\pdfoutput=1
\usepackage {amsmath}
\usepackage {amssymb}
\usepackage {amsfonts}
\usepackage {amsthm}
\usepackage {mathrsfs}
\usepackage {natbib}
\usepackage {latexsym}
\usepackage {graphicx}
\usepackage {dsfont}
\usepackage {txfonts}
\usepackage {rotating}
\usepackage {wasysym}
\usepackage {multirow}
\usepackage {hhline}
\usepackage {graphicx}
\usepackage {hyperref}
\usepackage[dvipsnames,usenames]{color}
\usepackage {bm}
\usepackage{appendix}
\usepackage{acronym}
\usepackage{dcolumn}   \usepackage{enumitem}
\usepackage{bigdelim}
\usepackage {url}
\usepackage{caption}
\usepackage {subcaption}
\usepackage{multirow}

\usepackage{framed}

\newcommand\arcmin{\mbox{$^\prime$}}\newcommand\farcs{\mbox{$.\!\!^{\prime\prime}$}}
\newcommand{\dcc}{LIGO-P1400217-v3}

\newcommand{\sref}[1]{Section~\ref{#1}}
\newcommand{\aref}[1]{Appendix~\ref{#1}}
\newcommand{\fref}[1]{Figure~\ref{#1}}
\newcommand{\tref}[1]{Table~\ref{#1}}
\newcommand{\Sref}[1]{Section~\ref{#1}}

\newcommand{\Fref}[1]{Figure~\ref{#1}}
\newcommand{\Tref}[1]{Table~\ref{#1}}
\newcommand{\Tasc}{T_{\text{asc}}}
\newcommand{\Tsft}{T_{\text{sft}}}
\newcommand{\Tcoh}{T_{\text{coh}}}
\newcommand{\Tmax}{T_{\text{max}}}
\newcommand{\Tobs}{T_{\text{obs}}}
\newcommand{\reftime}{t_{\text{ref}}}
\newcommand{\tbin}{t_{\text{bin}}}
\newcommand{\Ndet}{N_{\text{det}}}

\newcommand{\Npair}{N_{\text{pairs}}}

\newcommand{\asini}{a\sin i}
\newcommand{\Porb}{P}
\DeclareMathOperator{\Var}{Var}
\DeclareMathOperator{\erf}{erf}
\DeclareMathOperator{\erfc}{erfc}

\makeatletter{}\def\commitID{commitID: 32a7b7fa3c955d13df9b101e580ab6927764285f}
\def\commitDATE{ Wed Apr 22 17:54:12 2015 +0100}

\def\commitSTATUS{CLEAN}

\begin{document}

\title{Gravitational waves from Sco X-1: A comparison of search methods\\
  and prospects for detection with advanced detectors}

\author{C.~Messenger}
\email{chris.messenger@astro.cf.ac.uk}
\affiliation{School of Physics and Astronomy, Cardiff University, Queen's Buildings, The Parade, CF24 3AA}
\affiliation{SUPA, School of Physics and Astronomy, University of Glasgow, Glasgow G12 8QQ, United Kingdom}
\author{H.~J.~Bulten}
\affiliation{Nikhef National Institute for Subatomic Physics,
  Science~Park~105, Amsterdam, The~Netherlands}
\author{S.~G.~Crowder}
\affiliation{Department of Physics, University of Minnesota,
  Minneapolis, MN, USA}
\author{V.~Dergachev}
\affiliation{LIGO Laboratory, California Institute of Technology, MS 100-36, Pasadena, CA 91125, USA}
\author{D.~K.~Galloway}
\affiliation{Monash Centre for Astrophysics (MoCA) School of Physics and Astronomy, Monash University,
  VIC 3800, Australia}
\author{E.~Goetz}
\affiliation{Max-Planck-Institut f\"{u}r Gravitationphysik, Callinstr. 38,
  30167 Hannover, Germany}
\affiliation{University of Michigan, Ann Arbor, MI 48109, USA}
\author{R.~J.~G.~Jonker}
\affiliation{Nikhef National Institute for Subatomic Physics,
Science~Park~105, Amsterdam, The~Netherlands}
\author{P.~D.~Lasky}
\affiliation{Monash Centre for Astrophysics (MoCA) School of Physics and Astronomy, Monash University,
  VIC 3800, Australia}
\affiliation{School of Physics, University of Melbourne, Parkville, VIC 3010,
  Australia}
\author{G.~D.~Meadors}
\affiliation{Max-Planck-Institut f\"{u}r Gravitationphysik, Callinstr. 38,
  30167 Hannover, Germany}
\affiliation{University of Michigan, Ann Arbor, MI 48109, USA}
\author{A.~Melatos}
\affiliation{School of Physics, University of Melbourne, Parkville, VIC 3010,
Australia}
\author{S.~Premachandra}
\affiliation{Monash Centre for Astrophysics (MoCA) School of Physics and Astronomy, Monash University,
  VIC 3800, Australia}
\author{K.~Riles}
\affiliation{University of Michigan, Ann Arbor, MI 48109, USA}
\author{L.~Sammut}
\affiliation{School of Physics, University of Melbourne, Parkville, VIC 3010,
  Australia}
\author{E.~H.~Thrane}
\affiliation{LIGO Laboratory, California Institute of Technology, MS 100-36, Pasadena, CA 91125, USA}
\affiliation{Monash Centre for Astrophysics (MoCA) School of Physics and Astronomy, Monash University,
  VIC 3800, Australia}
\author{J.~T.~Whelan}
\affiliation{Max-Planck-Institut f\"{u}r Gravitationphysik, Callinstr. 38,
  30167 Hannover, Germany}
\affiliation{School of Mathematical Sciences
  and
  Center for Computational Relativity and Gravitation,
  Rochester Institute of Technology,
    Lomb Memorial Drive, Rochester NY 14623, USA}
\author{Y.~Zhang}
\affiliation{School of Physics and Astronomy
  and
  Center for Computational Relativity and Gravitation,
  Rochester Institute of Technology,
    Lomb Memorial Drive, Rochester NY 14623, USA}

\date{\today}
\date{\commitDATE\\\mbox{\small{\commitID} \commitSTATUS}\\\mbox{\dcc}}

\begin{abstract}
The low-mass X-ray binary Scorpius X-1 (Sco X-1) is potentially the most
luminous source of continuous gravitational-wave radiation for interferometers
such as LIGO and Virgo.  For low-mass X-ray binaries this radiation would be
sustained by active accretion of matter from its binary companion.
With the Advanced Detector Era fast approaching, work is underway to develop an
array of robust tools for maximizing the science and detection potential of Sco
X-1.  We describe the plans and progress of a project designed to compare the
numerous independent search algorithms currently available.  We employ a
mock-data challenge in which the search pipelines are tested for their
relative proficiencies in parameter estimation,
computational efficiency, robustness, and most importantly, search sensitivity. 
The mock-data challenge data contains an ensemble of 50 \ac{ScoX1} type
signals, simulated within a frequency band of 50--1500 Hz.
Simulated detector noise was generated assuming the expected best strain
sensitivity of Advanced LIGO\cite{aligo} and Advanced VIRGO\cite{adv} ($4\times
10^{-24}$ Hz$^{-1/2}$). A distribution of signal amplitudes was then chosen so
as to allow a useful comparison of search methodologies. A factor of 2 in
strain separates the quietest detected signal, at $6.8\times 10^{-26}$ strain,
from the torque-balance limit at a spin frequency of 300 Hz, although this
limit could range from $1.2\times 10^{-25}$ (25 Hz) to $2.2 \times 10^{-26}$
(750 Hz) depending on the unknown frequency of \ac{ScoX1}.
With future improvements to the search algorithms and using advanced detector
data, our expectations for probing below the theoretical torque-balance strain
limit are optimistic.
\end{abstract} \maketitle

\acrodef{NS}[NS]{Neutron Star}
\acrodef{GW}[GW]{gravitational-wave}
\acrodef{LMXB}[LMXB]{low-mass X-ray binary}
\acrodefplural{LMXB}[LMXBs]{low-mass X-ray binaries}
\acrodef{GPU}[GPU]{graphical processor unit}
\acrodefplural{GPU}[GPU]{graphical processor units}
\acrodef{AMXP}[AMXP]{accreting millisecond X-ray pulsar}
\acrodef{EM}[EM]{Electromagnetic}
\acrodef{CW}[CW]{continuous wave}
\acrodef{MDC}[MDC]{mock-data challenge}
\acrodef{PSD}[PSD]{power spectral density}
\acrodef{SNR}[SNR]{signal-to-noise ratio}
\acrodef{CPU}[CPU]{central processing unit}
\acrodef{SFT}[SFT]{short Fourier transform}
\acrodef{IFO}[IFO]{interferometer}
\acrodef{GPS}[GPS]{Global Positioning System}
\acrodef{ScoX1}[Sco~X-1]{Scorpius~X-1}

\section{Introduction}\label{sec:intro}

Low-mass X-ray binaries (LMXBs)\acused{LMXB} are one of the most promising
sources of continuous \ac{GW} emission for ground-based \ac{GW} detectors.
This precedence is motivated by the availability of an accretion-driven power source in
these systems potentially capable of generating and supporting non-axisymmetric
distortions in the \ac{NS}
component~\cite{1978MNRAS.184..501P,1984ApJ...278..345W,1998ApJ...501L..89B,2000MNRAS.319..902U,2002PhRvD..66h4025C}.
\acp{LMXB}, and specifically sources such as \ac{ScoX1} and Cygnus X-2~\cite{PremachandraThesis2015} are prime targets for \ac{GW} searches. 
Since the second LIGO Science Run, numerous
searches have been performed for \ac{ScoX1} using varied data analysis
strategies~\cite{2007PhRvD..76h2003A,2007PhRvD..76h2001A,2011PhRvL.107A1102A,PhysRevD.90.062010,S5sideband},
resulting in non-detections, but with increasing sensitivity.  \ac{ScoX1} is
identified as the most likely, strongest \ac{GW} emitter of the currently-known
\acp{LMXB} due to its relative proximity to Earth and its high accretion rate.
The accretion rate is used to infer the possible amplitude of \acp{GW} emitted
according to the torque-balance model proposed in~\cite{1984ApJ...278..345W}.
With the forthcoming and unprecedented sensitivity from the advanced \ac{GW}
detectors~\cite{2015CQGra..32g4001T,2013arXiv1304.0670L,Harry:2010zz,TheVirgo:2014hva}, our goal is detecting this source or performing
more astrophysically constraining non-detections.  In the latter case,
the analyses would
eventually be probing signal amplitudes that are below the current
torque-balance limit and hence constraining \ac{LMXB} accretion models.

The parameters governing the expected phase evolution of a continuous \ac{GW}
signal from \ac{ScoX1} are only partially constrained.  The \ac{ScoX1}
system is believed to contain a \ac{NS}, but unlike a subset of other
\acp{LMXB}~\cite{2015ApJ...800L..12R,2014arXiv1412.5155D,2014ApJ...791...77T,2014arXiv1412.5145B}, the \ac{NS} 
exhibits neither persistent nor intermittent pulsations in any
electromagnetic band, and hence the spin frequency of the \ac{NS} is unknown.  This
non-pulsating property has consequences for the estimation of the orbital
parameters of the system, which are currently constrained through optical
observations of the lower-mass companion
object~\cite{2002ApJ...568..273S,2014ApJ...781...14G}.  Additionally, there are
relatively large uncertainties in the intrinsic spin evolution of the \ac{NS}
since it is constantly under the influence of a high rate of accretion from its
companion.  Consequently, the volume of the search parameter space is vast and
computationally prohibitive for the most sensitive type of approach---the
fully-coherent, matched-filter search over a bank of filters.

Other approaches to the detection problem attempt to maximize detection
probability with a limited computational cost and are the best strategy for this
problem. Numerous such methods have been developed within the \ac{GW}
community over the past decade.  Most have been designed with other types of
continuous \ac{GW} sources as targets, but many are also suitable, with
appropriate tuning, to the \ac{ScoX1} problem.  For this reason, we
performed the study presented in this article.  The principal objective is to compare and contrast the
detection capabilities and parameter estimation properties of the numerous
search methods presently available for \ac{ScoX1}.  A \ac{MDC} is the best
approach to identify commonalities and differences between analysis
methods. The \ac{MDC} includes many \ac{ScoX1}-type signals (with parameter values
unknown to the partcipants)
that are simulated in noise and analyzed by the various search pipelines in
parallel.  Since this is the first \ac{MDC} of its kind for \ac{ScoX1},
the focus here is on a comparison between pipelines rather than including
astrophysically realistic signal amplitudes. We anticipate a future \ac{MDC}
that employs more realistic signal parameters in order to more fully
approximate a true search for continuous \acp{GW} from \ac{ScoX1}.

This article is organized as follows.  \Sref{sec:source} is a
description of the \ac{ScoX1} system, with focus on the possible emission
mechanisms and on the state of knowledge of those parameters that influence the
form of a continuous \ac{GW} signal.  In \sref{sec:methods} brief
descriptions and relevant references to the search pipelines that have
participated in the \ac{MDC} are given.  \Sref{sec:comparison} contains
a qualitative comparison of the search pipelines and the design and
implementation of the \ac{MDC} itself is presented in \sref{sec:mdc}.
The results from each search pipeline are reported in \sref{sec:results}
and the manuscript concludes with \sref{sec:discussion} containing a
discussion of our findings and plans for future pipelines, pipeline
improvements and a more realistic future \acp{MDC}.

\section{Scorpius X-1}\label{sec:source}

\ac{ScoX1} is a binary system with an orbital period of approximately 18.9~h,
likely consisting of a $\sim$$1.4M_\odot$ \ac{NS} that accretes mass from a
$0.42M_\odot$ companion~\cite{2002ApJ...568..273S}. With a long-term average
X-ray flux of $3.9 \times 10^{-10}\,\mathrm{W} \,
{\mathrm{m}}^{-2}$~\cite{2008MNRAS.389..839W}, it is the brightest continuous
extrasolar X-ray source on the sky, indicating a comparatively high accretion
rate.

It has been proposed~\cite{1998ApJ...501L..89B} that in a stable, X-ray
luminous \ac{NS}
binary system like \ac{ScoX1}, the angular momentum transferred from the low-mass
companion to the \ac{NS} and the energy loss due to gravitational radiation are
in equilibrium. Since the former can be deduced from the X-ray flux,
torque-balance leads to a \ac{GW} strain amplitude as a function of the spin
frequency $\nu_{\text{s}}$ for \ac{ScoX1}
of~\cite{1998ApJ...501L..89B,2014ApJ...781...14G}
\begin{equation} 
  h_{0} \approx 3.5 \times 10^{-26} \sqrt{
\frac{300\,\mathrm{Hz}}{\nu_{\text{s}}} }\,.\label{eq:torquebalance}  
\end{equation}
It is possible that the system could temporarily be in a state where accretion
torque exceeds the \ac{GW} torque while maintaining the long term torque
balance on average. This could result in a temporary increase in the strength of
\ac{GW} emission~\cite{2015arXiv150106039H}.  Considering the long term
average, if the spin frequency is between $25\,\mathrm{Hz}$ and
$750\,\mathrm{Hz}$, the torque-balance strain is between $2.2 \times 10^{-26}$
and $1.2 \times 10^{-25}$. 

There is significant astrophysical uncertainty in the torque-balance limit.
Its derivation assumes accretion of mass at the radius of the neutron star, but
the effective accretion radius for angular momentum transfer may be closer to
the Alfv\'en radius, leading to a higher strain limit. On the other hand, its
derivation also assumes negligible angular momentum loss from the star other
than from \ac{GW} emission and hence may be too high.

In a \ac{GW} interferometer, this strain would be recorded (circular-orbit
approximation) as $h(t)$:
\begin{subequations}
\begin{align}
\begin{split}
h(t) =&\  h_0 F_+ (t, \alpha, \delta, \psi) \frac{1+\cos^2 (\iota)}{2} \cos [\Phi(t)] \\
  &+ h_0 F_\times (t, \alpha, \delta, \psi) \cos (\iota) \sin [\Phi(t)]\,,
\end{split}
\\
\Phi(t) =&\ \Phi_0 + 2\pi f_0 (\tbin - \reftime)
+\delta\Phi_\textup{spin-wander}
\\
\tbin =&\ t - d(t) - (\asini) \sin [2\pi(t - \Tasc)/\Porb] \,.
\end{align}
\end{subequations}
\noindent where $h_0$ is strain in the solar system barycenter, $F_+$
and $F_\times$ are detector antenna patterns for plus- and
cross-polarizations, $t$ is time the signal is received at the
detector, $\alpha$ and $\delta$ are respectively right ascension and
declination, $\psi$ is polarization angle, $\iota$ is the inclination
angle of the neutron star with respect to the line of sight, $f_0$ is
the intrinsic signal frequency, $\Phi_0$ is the \ac{GW} phase at
reference time $t_\textup{ref}$, $d(t)$ and $\asini$ are the
projections respectively of the detector's separation relative to the
solar system barycentre and the orbital semimajor axis onto the line of sight
(where $i$ is the inclination angle of the \ac{LMXB} orbit
with respect to the line of sight), both
measured in units of time, $\Porb$ is the orbital period,
$\Tasc$ is the time of the orbital ascending node,
and
$\delta\Phi_\textup{spin-wander}$ is a unknown quantity accounting for 
spin-wandering induced by the short-term variation in accreted mass from the
companion star.

\subsection{The parameter space}\label{sec:params}

\ac{ScoX1} has been studied widely due to its prominence in the \ac{LMXB}
population. It is relatively
nearby, at a distance (estimated from radio parallax measurements) of
$2.8\pm0.3$~kpc~\cite{1538-4357-512-2-L121}. 
Thanks in part to the relatively low extinction, the optical counterpart,
V818~Sco, is also unusually bright for a LMXB
($V\approx12.5$; \cite{lmxb07}).

The parameters that completely describe the binary system
(for the purposes of
the gravitational wave searches)
are the
orbital period $\Porb$; reference phase $\Tasc$ (the ascending node,
i.e. the time at which
the compact object crosses the plane tangent to the sky,
moving away from the observer);
and the projected semi-major axis $\asini$, where $i$ is the angle of the
orbit's axis relative to our line of sight
(\Tref{tab:scoparams}).
In addition, it may be necessary
to consider the limits on the system eccentricity $e$,
(e.g. \cite{2014ApJ...781...14G, 2015arXiv150200914L}), may require more than one
template to span the parameter uncertainty interval.

The most precise orbital parameter measurements have been made from
analysis of the Bowen blend emission lines around 4640~\AA\ in the
optical spectrum, arising from N~{\sc iii} and C~{\sc iii}
\cite{2002ApJ...568..273S}. These emission lines are known to arise
from the heated side of the companion facing the neutron star, and so
by repeat measurements of their radial velocity, the orbital period
and phase can be measured. The most recent effort combined two epochs
of radial velocity measurements over a 12-yr baseline, to obtain an
orbital period of $\Porb=0.7873114\pm0.0000005$~d and a time of
inferior conjunction of the companion of
$T_0=2454635.3683\pm0.0012$~HJD \cite{2014ApJ...781...14G}.

Because these measurements track the companion (rather than the
neutron star that is the source of the \ac{GW} emission) the reference
epoch must be shifted by for the purposes of \ac{GW} searches.  To
convert from $T_0$ (when the companion is closest to the observer) to
$\Tasc$ (when the compact object crosses the plane of the sky moving
away from the observer, one must take
$\Tasc = T_0 - \Porb/4 = 2454635.1715\pm0.0012$~HJD.
Furthermore, because the reference phase
is defined at a particular epoch (depending upon the span of data used in the
radial velocity fits), the effective uncertainty in $\Tasc$ increases
towards earlier and later times, and this increase must be taken into account
for future \ac{GW} searches. This effect was quantified by
\cite{2014ApJ...781...14G}, including the effects of additional observational
efforts.

The projected semi-major axis of the neutron star orbit $\asini$ is the
most challenging to measure. It can be obtained in principle from the 
velocity amplitude of the Bowen emission region on the face of the companion,
but this requires a correction first to the companion's center of mass, and then
to the neutron star, which requires constraints on the companion radius as well
as the mass ratio of the binary components. This parameter is
estimated instead from the symmetric component of the Doppler tomogram
of the broad emission lines in the system as $1.44 \pm 0.18$~lt-s
(derived from a velocity amplitude of $K_1=40\pm5\ {\rm km\,s^{-1}}$)
\cite{2002ApJ...568..273S}. However,
the Doppler tomogram derived from the subsequent epoch of optical spectroscopy
analyzed by \cite{2014ApJ...781...14G}, exhibited significantly different
morphology, such that it was not possible to (for example) combine the two
datasets to improve the precision of the $\asini$ estimate.

While further incremental improvements on $\Porb$ and $\Tasc$ can be achieved
relatively easily with additional optical spectroscopic measurements, improving
the estimate of $\asini$ will likely require a deeper understanding of how
the emission line morphology in the system evolves in response to secular
variations.

In contrast to the binary system parameters, the spin frequency 
of the neutron star is unknown. No persistent or intermittent X-ray pulsations
have been detected from \ac{ScoX1}. 
While the accreting source is thought to be a neutron star, no thermonuclear
(``type-I'') bursts have ever been detected from the source, and hence no
``burst oscillations'' have been observed.
Non-detections for X-ray
pulsations have been reported for searches up to frequencies of 256~Hz, using
data obtained with the {\it European X-ray Observatory Satellite}\/ ({\it
EXOSAT}; 
\cite{1986ApJ...306..230M}), and up to 512~Hz using
observations by {\it Ginga}\/ 
\cite{1991ApJ...379..295W,1992ApJ...396..201H,1994ApJ...435..362V}.
A much larger set (approximately 1.3~Ms) of high-time resolution (down to $1\
\mu s$) data is available from the {\it Rossi X-ray Timing Explorer}\/
({\it RXTE}; \cite{xte96}) mission (1996--2012). While unsuccessful searches
of these data have almost certainly taken place 
(due to both the prominence of \ac{ScoX1} among the \ac{LMXB} population, and the
high priority for pulsation searches for this mission)
no limits have been reported.
Analysis of these data are hampered by the high count-rate of the source, which
necessitates 
non-standard
data modes, as well as introducing substantial
effects from instrumental ``dead-time''. 

The likely frequency range for the spin period has been estimated based on the
separation of a pair of high-frequency quasi-periodic oscillations (QPOs),
measured in the range 240--310~Hz
\cite[]{1996ApJ...469L...1V,1997ApJ...481L..97V,2000MNRAS.318..938M}.
In sources that exhibit pulsations or burst oscillations in addition to pairs
of kHz QPOs, the QPO frequency separation is roughly equal to the spin
frequency (or half that value). 

\begin{table}
\caption{Scorpius X-1: system parameters\label{tab:scoparams}}
\begin{tabular}{lccr}
\hline
\ac{ScoX1} parameter & Value & Uncertainty & Ref\\
\hline
Period & 68023.70 sec & 0.04 sec & \cite{2014ApJ...781...14G} \\
Orbital semi-major axis & 1.44 sec & 0.18 sec & \cite{2007PhRvD..76h2001A,2002ApJ...568..273S}\\
Time of ascension & 897753994 & 100 sec & \cite{2014ApJ...781...14G} \\
Orbital eccentricity & $<0.068$ & $3\sigma$ & \cite{2014ApJ...781...14G,
2015arXiv150200914L}\\
Right Ascension & $16^\mathrm{h}19^\mathrm{m}55^\mathrm{s}.067$ & $0\farcs06$ & \cite{2mass06} \\
Declination & $-15^\circ38\arcmin25\farcs02$ & $0\farcs06$ & \cite{2mass06} \\
System inclination & $44^\circ$ & $6^\circ$ & \cite{2001ApJ...558..283F} \\
Companion mass & 0.42 M$_\textup{Sol}$ & & \cite{2002ApJ...568..273S} \\
X-ray flux & $3.9\times 10^{-10} \mathrm{Wm}^{-2}$ & & \cite{2008MNRAS.389..839W} \\
\hline
\end{tabular}
\caption*{Note that the time of ascension ($T_{\rm asc}$) refers to the
neutron star, and is calculated as
$T_0-P/4$, where $T_0$ is the epoch of inferior conjunction of the
companion from \cite{2014ApJ...781...14G}.
The radial velocity data from this paper were also the source of the 
eccentricity limit, which was calculated by the authors.
}
\end{table}

Accreting neutron stars exhibit ``spin wandering'' (gradual changes
in spin frequency; 
e.g. \cite{1997ApJS..113..367B,1993AandA...267..119B}), attributed primarily to
variations in the accretion rate.
The accretion rate in turn varies on timescales of minutes to decades,
with most notably, transient sources exhibiting outbursts during which the
accretion rate increases by several orders of magnitude compared to the
quiescent level \cite{2006csxs.book..507K}.
As a result, \ac{GW}
searches for \ac{LMXB} systems are necessarily limited to
a coherence time equal to the maximum timescale over
which the spin evolution can be well modelled.

Although observations of the radio jets from \ac{ScoX1} can be used under
model-dependent assumptions to constrain the orientation of the neutron star
spin axis \cite{2001ApJ...558..283F}, here we assume no {\it a priori} knowledge of the axis
direction.
\section{Current and future methods}\label{sec:methods}

In this section we give an overview of the current search algorithms (or
pipelines) available for searches targeting \ac{ScoX1}. For additional technical
details we either refer the reader to the corresponding methodological papers
for each algorithm, where possible, or to a corresponding Appendix.
In the following sections, we describe
six algorithms: four which were used in our original comparison study, one
for which the analysis infrastructure was completed after the initial deadline
and run in self-blinded mode on the same data set in the subsequent months
as described in \tref{tab:submissionTimeline},
and one that has been proposed for future analyses.

\subsection{Polynomial Search}\label{sec:polynomial}

The Polynomial Search~\cite{2010JPhCS.228a2005V} is a generic all-sky
method for finding \acp{GW} from continuously emitting sources, such
as \acp{NS} in binary systems, in \ac{GW} interferometric data. It is
based on the assumption that the phase of an expected signal due to
these sources in a ground-based \ac{GW} detector can be approximated
by a third-order polynomial in time during short stretches of time.
If the binary orbit is the dominant source of Doppler modulation in
the signal, this holds for periods up to one quarter of the binary
period.

For each input \ac{SFT}, the algorithm generates a set of templates of
signals with a phase $\Phi_t (t)$ that evolves as a polynomial in time.
\begin{equation}\label{PolynomialSearchTemplate}
        \Phi_t (t) = 2 \pi \left[ f_0 t + \frac{c_1}{2} t^2 + \frac{c_2}{6} t^3 \right]
\end{equation}
The range for the polynomial coefficients $f_0$, $c_1$ and $c_2$
are chosen prior to analysis based on the properties of expected
signals. The initial phase is matched implicitly by allowing for an
offset in time between data and template.

The correlation of each template with the data segment is calculated
as a function of offset time by multiplication in the frequency domain.
The offset time that yields the largest correlation value is then
recorded.

The probability that one or more templates yield a correlation exceeding
a thresold value $C_{\text{t}}$ due to noise is
\begin{equation}\label{PolynomialSearchPerSFTPvalue}
  p_{\mathrm{SFT}} (C_{\text{t}}) = 1 - { \left [ \frac{1}{2} + \frac{1}{2}
      \erf\left( \frac{C_t}{\sqrt{2} \sigma} \right)
    \right] }^{N_e}
\end{equation}
where $\sigma$ is the square root of the average power contained in an
\ac{SFT} frequency bin and $N_e$ is the effective number of templates.
While the number of templates $N$ is known exactly, there is some degree
of overlap between successive templates and $N_e$ corrects for this
overlap. It can be determined by fitting the measured false alarm rate
versus correlation threshold $C_{\text{t}}$ on the analysis results of
a data set that contains only noise.

When analyzing $N$ \acp{SFT} with pure noise, the probability $p(n)$ that
$n$ or more \acp{SFT} have one or more templates with a correlation
exceeding the threshold is governed by a cumulative binomial
distribution with a per-trial probability given by
equation~\eqref{PolynomialSearchPerSFTPvalue}. This is the single-trial
test statistic for detection. In order to test against a chosen false
alarm probability threshold, the threshold is divided by the number of
frequency bins to correct for the multiple comparisons problem
\cite{Bonferroni35,Bonferroni36}.

The Polynomial Search is an all-sky search and it does not benefit from
detailed knowledge of the source that only influences the phase or the
amplitude of the signal, however orbital parameters put a constraint on
the time derivatives of the frequency in the data and the orbital period
of \ac{ScoX1} can be exploited by using longer \acp{SFT} than would be
feasible for an all-sky search, increasing the coherence time and
therefore also the
sensitivity.

\subsection{Radiometer}\label{sec:radiometer}

The Radiometer
analysis~\cite{Ballmer2006CQG,2007PhRvD..76h2003A,2011PhRvL.107A1102A}
cross-correlates data from pairs of detectors to detect \ac{GW} point sources
with minimal assumptions about the signal, and uses an estimator given by
\begin{equation}
\hat{Y}=\int_{-\infty}^{\infty}df\int_{-\infty}^{\infty}df'\delta_T(f-f'){\tilde
s}^{*}_{1}(f){\tilde s}_{2}(f'){\tilde Q}(f')\label{eq:Ystat} 
\end{equation}
with variance
\begin{equation}
\sigma^2_Y \approx \frac{T}{2}\int_{0}^{\infty}dfP_1(f)P_2(f)|{{\tilde
Q}(f)}|^2\,.\label{eq:Yerr}
\end{equation}
Here, $\delta_T(f-f')$ is the finite$\mbox{-}$time approximation to the Dirac
delta function, ${\tilde s}_{1}$ and ${\tilde s}_{2}$ are Fourier transforms of
time$\mbox{-}$series strain data for each detector in the pair, $T$ is the
detector pair live$\mbox{-}$time, and $P_1$ and $P_2$ are one$\mbox{-}$sided
strain power spectral densities for each detector.  The
cross-correlation is performed with an optimal filter function, ${\tilde
Q}(f)$, which weights time and frequency bins based on their sensitivity.  The
filter depends upon the modeled strain power spectrum and is normalized by the
strain noise power spectra of the detector pairs.
Also included in the filter is a phase factor which takes into account the time
delay between the two detector sites.  The detection statistic, the Radiometer
\ac{SNR} known as the `$Y$-statistic', is calculated using a weighted average
of data from many segments and multiple detector pairs:
\begin{equation}
{\hat{Y}_{\rm tot}=\frac{\sum_l\hat{Y}_l\sigma_l^{-2}}{\sum_l\sigma_l^{-2}}}
\end{equation}
with total variance
\begin{equation} 
{\sigma_{\rm tot}^{-2}=\sum_l\sigma_l^{-2}\,}
\end{equation}
where $l$ sums over time segments and/or detector pairs for the variables defined in \eqref{eq:Ystat} and \eqref{eq:Yerr}. It is expected to be normally distributed from the central limit
theorem, and indeed this is born out
empirically~\cite{2007PhRvD..76h2003A,2011PhRvL.107A1102A}.  The Radiometer
$Y$-statistic is the optimal maximum likelihood estimator for a
cross-correlation search~\cite{Ballmer2006CQG}.

In practice, the Radiometer method has also been shown to yield robust results
in the presence of realistic (non-Gaussian)
noise~\cite{2007PhRvD..76h2003A,2011PhRvL.107A1102A}.  The Radiometer search
does not use a matched filter; so there are no assumptions about time
evolution,
except that the signal frequency remains within the 0.25~Hz frequency
bin searched (0.25~Hz frequency bins were chosen based on the convention of other applications of the Radiometer algorithm and are not optimized for the Sco X-1 analysis). By not employing a matched filter, the Radiometer loses the sensitivity
possible from using prior knowledge about the waveform.  For the same reason,
however, it is sensitive to arbitrary signal models (within the observing
band), and is therefore robust.

\subsection{Sideband search}\label{sec:sideband}

The ``Sideband'' search~\cite{Messenger2007CQG,2014PhRvD..89d3001S} is based on an
approach used in the detection of radio pulsars and low-mass X-ray
binaries emitting \ac{EM} radiation~\cite{2003ApJ...589..911R}.  It
uses the fact that the power in a continuously emitted signal from a
source in a binary system, when observed over many orbits, becomes
distributed among a finite number of frequency-modulated sidebands.  These
sidebands have the property that there is always a fixed number of
sidebands for a given source.  This number is dependent only upon the
intrinsic emission frequency and the orbital radius.  The frequency
separation of the sidebands is the inverse of the orbital period, and
the phasing relation between sidebands is a function of the orbital
phase.  The power spectrum of a timeseries containing such a signal
will therefore be independent of the orbital phase.  For a source of
known orbital period and reasonably well constrained orbital radius
(e.g. \ac{ScoX1}) the characteristic ``comb''-like structure is well-defined and the data analysis task becomes one of locating this
frequency domain structure.

In this case it is computationally efficient to construct a template
in the frequency domain that closely matches the main features of this
structure i.e. the width of the comb and the separation of the teeth.
This template is then convolved with the
$\mathcal{F}$-statistic~\cite{1998PhRvD..58f3001J} yielding a
frequency series that contains the summed power from all sidebands as
a function of central intrinsic emission frequency.
The $\mathcal{F}$-statistic is used instead of the power spectrum due to the quadrupole 
emission of \acp{GW} coupled with the time-varying detector response.  This is computed
as a function of frequency and for the known fixed sky position allowing the effects of the
motion of the detector relative to the source to be removed from the data.
The convolution of the template with the $\mathcal{F}$-statistic, known as the $\mathcal{C}$-statistic, 
is given by
\begin{align}
  \mathcal{C}(f)&=\sum\limits_{j}2\mathcal{F}(f_{j})\mathcal{T}(f_{j}-f)\nonumber\\
  &=\left(2\mathcal{F}\ast\mathcal{T}\right)(f)
\end{align}
where $2\mathcal{F}$ is the $\mathcal{F}$-statistic and $\mathcal{T}$
is the comb template. Although this statistic is the incoherent sum of
power from many sidebands, and is hence less sensitive than a fully
coherent search, it does have the following qualities.  It is very
efficient to compute since its only search dimension is frequency and
therefore only requires the computation of a fixed set of Fourier transforms.
Also, unlike other semi-coherent search algorithms, its sensitivity
does not scale with the fourth-root of the observation time.  Its
incoherent summation occurs in the frequency domain and its
sensitivity is therefore proportional to the fourth-root of the number
of sidebands.  It maintains a square-root sensitivity relation to the
observation time.

\subsection{TwoSpect}\label{sec:twospect}

The TwoSpect method~\cite{2011CQGra..28u5006G} relies on computing a
sequence of power spectra  whose coherence length is short enough so
that a putative signal would remain in a single frequency bin during a
single spectrum. 
Thus the coherence length is typically no longer than 1800 seconds;
for the Scorpius X-1 search, it is either 360 or 840 seconds,
depending on search frequency.
For comparison, the sampling frequency of raw data is typically 
16384 Hz, and observation runs are millions of seconds long.
After the sequence of power spectra are corrected for
the Doppler shift caused by Earth's motion, a periodogram is
created. 
For each frequency bin in the periodogram, 
we compute a second Fourier Transform, for which the
integrated variable is the time of each power spectrum.
When a continuous \ac{GW} signal is present in
the data, the second power spectra will contain excess power at
frequencies corresponding to the \ac{GW} signal frequency in the first
spectra computed, and also the inverse of the binary orbital period in
the second power spectrum. The name TwoSpect is given to this
algorithm because two successive Fourier transforms are computed.

Gravitational wave detector data are analyzed by creating templates
that mimic the putative signal pattern. A detection statistic, $R$, is
computed by a weighted sum of $M$ pixel powers in the second Fourier
transform $Z_i$, subtracting
estimated noise $\lambda_i$, where the weights $w_i$ are
determined by the template values for $M$ pixels:
\begin{equation}
  R = \frac{\Sigma_{i=0}^{M-1} w_i \left[ Z_i - \lambda_i
\right]}{\Sigma_{i=0}^{M-1} \left[ w_i \right]^2}\,.\label{eq:Rstat}
\end{equation}
To create a template for a circular orbit binary system, the
putative \ac{GW} signal frequency, binary orbital period, and
amplitude of the frequency modulation are given. Orbital phase is
an unimportant parameter due to the nature of the analysis: computing successive
power spectra from the \acp{SFT} and then, importantly, the second second Fourier transforms remove dependence on orbital
phase.

Although the original design of TwoSpect was an all-sky search
method~\cite{2011CQGra..28u5006G,PhysRevD.90.062010}, it can also be used as a directed
search algorithm.  By design, it is robust against phase jumps of the
signal between successive power spectra, and TwoSpect is unaffected by
spin wandering of sources because of the semi-coherent nature of the
method and because of the short coherence time of the first Fourier
transform. The choice of the coherence length of the first series of
power spectra is given by the putative amplitude and period of the
frequency modulation caused by the motion of the source.  With these
features, the TwoSpect method is a computationally efficient and
robust algorithm capable of analyzing long stretches of gravitational
wave data and detecting continuous \ac{GW} signals from \acp{NS} in
binary systems.

Running TwoSpect as a directed search algorithm involves calculating
the $R$-detection statistic in the parameter space that might contain
the \ac{GW} signal (the incoherent harmonic sum stage of TwoSpect,
used for the all-sky search, was bypassed entirely). The orbital
period of \ac{ScoX1} is sufficiently well-known to restrict the search to
the two dimensions of putative signal frequency and frequency
modulation. The grid spacing, inversely proportional to spectrum
coherence time, was chosen to allow a mismatch of no more than $0.2$ in
the detection statistic.
Mismatch, in this context, means the relative loss in the the detection
statistic, $R$, where the spacing was informed by studies in the TwoSpect methods paper~\cite{2011CQGra..28u5006G}.

\subsection{Cross-Correlation}\label{sec:crosscorr}

The cross-correlation
method~\cite{Dhurandhar:2007vb,Chung:2011da,LMXBCrossCorr}, henceforth
referred to as the CrossCorr method, is a modification of
the directed stochastic-background search described in
\sref{sec:radiometer}.  By using the signal model, it is able to
coherently combine not just data taken by different detectors at the same time,
but also data taken at different times, by the same or different detectors.
Since this signal model depends on parameters such as frequency and binary
orbital parameters, the search must be repeated over a grid of points in
parameter space.  In order to control computational costs associated with
parameter space resolution, the method limits the time offset between pairs of
data segments included in the construction of the statistic, allowing a
tradeoff between computation time and sensitivity.

The data from each detector are divided into segments of length
$\Tsft$, which are then Fourier transformed.  The index $K$ is used to label
an \ac{SFT} (so that it encodes both the time of the data segment and the
detector from which it's taken).  We construct a statistic
\begin{equation}
  \label{e:CCrhodef}
  \rho = \sum_{KL\in\mathcal{P}} (W_{KL}z_K^*z_L + W_{KL}^*z_Kz_L^*)
\end{equation}
using the product of data from \acp{SFT} $K$ and $L$, where $KL$ (or $LK$)
is in a list of allowed pairs $\mathcal{P}$, defined by $K\ne L$ and
$\left\lvert T_K-T_L\right\rvert \le \Tmax$, i.e., the timestamps of
the two different data segments should differ by no more than some
specified lag time.  The weighting factor $W_{KL}$ is determined by
the expected signal and noise contributions to the cross-correlation,
and the frequency bin or bins used to create the normalized data value
$z_K$ out of \ac{SFT} $K$ is determined by the Doppler-shifted signal
frequency associated with the modelled signal parameters.  The linear
combination of cross-correlation terms is normalized so that
$\Var(\rho)=1$, and weighted to maximize $E[\rho]$ in the presence of
the modelled signal.  With this choice of weighting,  the
expected statistic value for a given $h_0$ scales like
\begin{equation}
  E[\rho]
  \sim (h_0^{\text{eff}})^2
  \sqrt{\Tobs\Tmax
    \left\langle
      \frac{(\Gamma^{\text{ave}}_{KL})^2}{S_KS_L}
    \right\rangle_{KL\in\mathcal{P}}
  }
\end{equation}
where $h_0^{\text{eff}}$ is the combination of $h_0$ and $\cos\iota$
defined in \eqref{eq:h0eff}, $S_K$ is constructed from the noise power
spectrum and $\Gamma^{\text{ave}}_{KL}$ from the antenna patterns for
detectors $K$ and $L$ at the appropriate times.  The search can be
made more sensitive by increasing $\Tmax$, but at the cost of additional
computing cost, as detailed in \sref{sec:compcost}.

\subsection{Future method: Stacked $\mathcal{F}$-statistic}\label{sec:stackedF}

We now describe a method which has not yet been implemented or run on
the \ac{MDC} data, but holds promise for the future.
The distributed computing project
Einstein@Home~\cite{eath,2013PhRvD..87d2001A} was originally designed
to undertake the highly computational task of searching for unknown
isolated continuous \ac{GW} sources.  This involves a wide frequency
band, all-sky search using computing power volunteered by participants
across the globe.  In recent years this power has been shared between
algorithmically similar searches of radio data from the
Arecibo and Parkes radio
telescopes~\cite{Knispel:2011ss,Knispel:2013da,Allen:2013sua}
and gamma-ray data from the the Fermi telescope.\cite{Pletsch:2013iva}

The \ac{GW} search uses an algorithm that coherently computes a
maximum likelihood statistic over a finite length of data on a bank of
signal templates.  This statistic is known as the
$\mathcal{F}$-statistic~\cite{1998PhRvD..58f3001J,2007PhRvD..75b3004P}
which is then incoherently summed (or stacked) over segments in such a
way as to track a potential signal between
segments~\cite{2000PhRvD..61h2001B,Wette:2011jr,2012PhRvD..85h4010P}.  The computational cost of this
search is primarily controlled via the ratio of the coherence length to
total data length.  This is tuned to return approximate optimal
sensitivity for the fixed and large computational
power available from the Einstein@Home project.

Currently under development is a fixed-sky-location, binary-source
version of this search algorithm.  In this case the binary parameter
space replaces that of the sky.  A recent feasibility study
\cite{2015arXiv150200914L} investigated the potential sensitivity to
\ac{ScoX1}.

\section{Comparison of methods}\label{sec:comparison}

In this section, we discuss the general properties of the five
algorithms taking part in the \ac{MDC} in terms of i) their dependence
on the parameter space, ii) their intrinsic parameter estimation
ability, and iii) their computational cost.

\subsection{Parameter space dependence}

Each algorithm's performance in relation to computational cost, search
sensitivity and how the search is setup, depends on the \ac{ScoX1} parameter
space. Both the Polynomial and the Radiometer searches have the least parameter
space dependence while the Sideband search has the most.  The TwoSpect and
CrossCorr searches fall between these extremes.

Since the Polynomial search does not explicitly model any of the source orbital
parameters, changes in the parameter space boundaries only have indirect effects.
For effective detection, the template parameters need to approximate the time
derivatives of the phase of the signal as received by the detector. The
contributions due to the binary orbit to these derivatives are proportional to the
$a \sin i$ and inversely proportional to $P^2$ and $P^3$ for the first and
second derivative of frequency with respect to time, respectively. Additionally,
both derivatives scale with $f_0$. Therefore, the boundaries of the $c_1$ and
$c_2$ template parameters need to be set to reflect the range of values of $P$
and $a \sin i$ compatible with measurements.

As long as the data contains at least one full binary period, the time of
ascension should not affect the Polynomial search's sensitivity.

Like the Polynomial search, the Radiometer search does not explicitly model and
is largely insensitive to the orbital parameters.  It operates under the
assumption that the instantaneous received frequency of the signal resides
within a single 0.25~Hz bin for the duration of the observation.  The total
expected variation of the instantaneous frequency is proportional to the
product of the intrinsic frequency, the orbital semi-major axis and the orbital
angular frequency.  However, the intrinsic frequency is uncertain over a large
range and hence at values in excess of $\sim$1~kHz it is increasingly likely
that the assumption that the signal is restricted to a single bin is
invalidated.  The corresponding effects on sensitivity (and the related conversion factors for $h_0$ estimates and their uncertainties) are discussed in
\aref{sec:radiometertech}. The current version of the search also assumes
that the signal is circularly polarized.  This assumption does not make the
search insensitive to other polarizations, however it does affect resulting
estimates of the signal amplitude $h_0$.  To account for the assumption on
polarization an average conversion factor can be applied to $h_0$ estimates and
the associated uncertainties (see \aref{sec:radiometertech}). If information
on the polarization were available, this would change the conversion factor and
reduce the associated uncertainty.  

The Sideband search in its current form is heavily restricted to the analysis
of signals with well-known orbital periods and sky positions. The orbital
period defines the spacing between the frequency-domain template ``teeth'' and
knowledge of the sky location allows the coherent demodulation of the detector
motion with respect to the source binary barycenter.  The search is as
sensitive to the source sky location as a fully coherent search. For a
1-year-long observation of a source with frequency $1$~kHz, the sky position must be
known to a precision of $\sim$0.1~arcsec.  In reality, spin wandering limits observation 
times for Sideband searches of \ac{ScoX1} to $\sim$10~days \cite{S5sideband}, which significantly
relaxes the restriction on the sky position.  The fractional orbital
period uncertainty must be $<(4\pi f\asini T/\Porb)^{-1}$ which is $\leq 4\times
10^{-6}$ for a 10 day observation of \ac{ScoX1}. The orbital semi-major axis
determines the width of the frequency domain template which needs to be known to
$\sim$10\% precision. Post-processing techniques allow using any level of
prior knowledge of the \ac{NS} orientation parameters to be folded into our
parameter estimation. The search is completely insensitive to knowledge of the
orbital phase of the source, but is extremely sensitive to spin-wandering since
the intrinsic frequency resolution is ${\approx}1/T$ where $T$ is the total
observation time. 

The TwoSpect search is sensitive to projected semi-major axis and orbital period, but is
insensitive to initial orbital phase.
The two Fourier transforms in TwoSpect preserve only power information at present, ignoring orbital phase.
Orbital period can be explored, with a template
spacing~\cite{2011CQGra..28u5006G} of $\Delta P \simeq P_0^2 / (\alpha_\textup{TS}
T_{obs})$ for an allowed detection statistic mismatch of 0.2 in the
templates; the empirical value $\alpha_\textup{TS} = 2.7 (\Tsft / 1800) + 1.8$ is
derived from simulations.
Taking \ac{ScoX1}'s estimated period as $P_0$ and a 1-year-long data set of
360- or 840-s \acp{SFT} yields template spacing of 50 to 65 s, much greater than the uncertainty in \ac{ScoX1}'s orbital period.
For this reason, TwoSpect does not attempt to infer orbital period in this \ac{MDC}.
Similar to the Radiometer search, the TwoSpect search is optimized
for a circularly polarized signal.  The search is nonetheless sensitive to arbitary
polarizations and the details of the corresponding $h_{0}$ sensitivity
dependence are detailed in \sref{sec:twospect}. As is the case for the
Sideband search, TwoSpect post-processing of search results can be optimized by
the inclusion of prior information on \ac{NS} orientation
parameters.
In this \ac{MDC}, however, we assume no prior information on orientation or polarization.

The CrossCorr search is a template-based method, in that the
weights and particularly the phases with which cross-correlation terms
are combined depend on the assumed signal parameters.  The search is
sensitive to frequency, projected semi-major axis and time of
ascension, and requires a search over a grid of points in this
three-dimensional parameter space, laid out according to the metric
constructed in~\cite{LMXBCrossCorr}.  The same is in principle true
for orbital period, but the prior constraints on this parameter for
the \ac{MDC} were tight enough that the search could be performed with
the a priori most likely value.  The response of the search is
insensitive to initial \ac{GW} phase and only weakly sensitive to
polarization angle.  It is sensitive to both the intrinsic amplitude
$h_0$ and the inclination angle $\iota$ between the neutron star spin
and the line of sight; the amplitude weighting $\Gamma^{\text{ave}}_{KL}$
selects the part of the wave which is robust in $\iota$ and therefore
the quantity to which the search is sensitive is $h_0^{\text{eff}}$
defined in~\eqref{eq:h0eff}.  This choice of weighting produces
an unknown systematic offset in the other parameters, and was the
limiting error on frequency estimates in the MDC.

\subsection{Parameter estimation}

Each pipeline can reveal information about the physical parameters of \ac{ScoX1}
in the event of a detection or a null result. In the latter case, in principle,
constraints can be placed on the amplitude, source orientation and polarisation
parameters, however in practice this is limited to upper limits on the
amplitude only. All of the searches in this
comparison are insensitive to initial \ac{GW} phase. Other parameters can
nevertheless, in principle, be estimated, including \ac{GW} strain amplitude
$h_0$, neutron star inclination angle $\iota$ and projected orientation angle
$\psi$, \ac{GW} radiation frequency $f$, projected orbital semi-major axis $a
\sin i$, time of ascension $\Tasc$, and orbital period $P$.
This MDC has assumed the orbit of \ac{ScoX1} to be circular, but a non-zero
eccentricity would also add two dimensions to the parameter space: the
eccentricity itself and the argument of periapse.

The Polynomial search models templates with a frequency and frequency
time-derivatives over short data segments.  The intrinsic \ac{GW} frequency of a
source in a binary system can be estimated from the average frequency of templates
that correlate relatively strongly with data. For a template to contribute towards
the estimate it must satisfy two conditions. First, the frequency of the
template must be in the bin in which the signal was detected. Second, the
correlation must exceed the threshold value that corresponds to a $10\%$
per-\ac{SFT} false alarm rate.  The standard deviation of the template frequencies
is representative of the uncertainty in the intrinsic frequency estimation. The
orbital period can potentially be extracted similarly from the times of sequential
zero points in the second derivatives of the frequency with respect to time, but
this is currently not implemented in the search pipeline.

Currently, the Radiometer search can be sensitive either to sky location or tuned for a
narrowband search, as for \ac{ScoX1} (though work is in progress on an all-sky narrowband search).It is not, at present, sensitive to orbital semi-major axis,
orbital period or time of ascension and hence these parameters are not
estimated. The $Y$-statistic (defined in \eqref{eq:Ystat}) in each frequency
bin (0.25~Hz in width) can be converted to a strain $h_0$. This is
done via a normalization from root-mean-squared strain and the application of a
correction for the assumption of circular polarization. Strain is reported for
the loudest frequency bin and hence, in the event of detection, the intrinsic
\ac{GW} frequency is estimated with an uncertainty of 0.25~Hz and the
amplitude $h_{0}$ is returned. For non-detection, upper-limits on $h_{0}$ are
reported based on the loudest event in the total search band.

The Sideband search estimates a detection statistic at each frequency bin of
width $5\times 10^{-7}$Hz (the inverse of twice the observation span).
However, signals trigger multiple non-sequential but equally spaced frequency
bins. Consequently, signal frequency estimation ability is conservatively
reduced by ${\sim}4$ orders of magnitude. At present, $\asini$ is not estimated from the search, but estimates
can be derived by follow-up analyses that vary the width of the comb template.
Such a procedure could also be enhanced by exchanging the flat comb template
for a more accurate version. The orbital period is assumed to be known and the
time of ascension is analytically maximized over in the construction of the
Sideband statistic and hence neither are estimated. Future algorithm
developments may allow time of ascension to be determined. In the event of a
detection the corresponding statistic is processed to yield an estimate of the
signal amplitude $h_{0}$.  In the event of a null result the loudest statistic
is used to compute an upper-limit on the amplitude. 

The TwoSpect search, in its directed search mode with fixed sky location, tests
templates with a model of $f$, $\asini$, and $P$.  The orbital period is
fixed for the \ac{ScoX1} search since its uncertainty is small.  TwoSpect is
insensitive to the time of ascension.  Signal parameters are estimated from a
detection based on the most extreme single-template $p$-value from any one interferometer.
Here, single-template $p$-value is the probability of the TwoSpect detection
statistic, $R$, being as large or greater if the given template is applied to Gaussian noise.
This $p$-value is not corrected for correlations or trials factors, so it does not directly correspond to an overall false alarm probability of detection, but it is locally useful for ascertaining the best-matching template.
The amplitude $h_0$ is proportional to the fourth-root of the $R$ statistic (see
\eqref{eq:Rstat}) and estimates and upper-limits of $h_{0}$ are determined
as described 
a forthcoming methods paper~\cite{TwoSpectMDCMethods2015}.
Uncertainty on the estimate of $h_0$
is largely due to the unknown value of the \ac{NS} inclination angle $\iota$.
Uncertainties in estimates of $f$ and $\asini$ are empirically derived from 
signal injections and are on the scale of the template grid except for marginally-detected pulsars. 
More precisely, since the estimates are the $f$ and $\asini$ values of the highest-statistic template, there true $f$ and $\asini$ are somewhere between that template and its neighbors, approaching a uniform distribution for fine grid-spacing. 
If a signal is an extremely marginal detection, it is possible for noise to change which template has the highest statistic, adding further uncertainty. 
For most detected pulsars, however, the uncertainty is dominated by the spacing between neighboring templates, a grid scale of $1/(2\Tsft)$ in $f$ and $1/(4\Tsft)$ in $\asini$.
This scale is set by prior simulations~\cite{2011CQGra..28u5006G}.

The CrossCorr search is performed over a grid of templates in $f$,
$\asini$ and $\Tasc$, whose spacing is determined by the metric
given in \cite{LMXBCrossCorr}, and in particular becomes finer in each
direction if the maximum allowed time separation $\Tmax$ between pairs
of \acp{SFT} is increased.  As described in
\sref{sec:CrossCorrimpl}, parameter estimates can be obtained
that are more accurate than the spacing of the final parameter grid
by fitting a quadratic function to the highest statistic values and
reporting the peak of that function.  The errors in estimating these
parameters come from three sources: a systematic offset depending on
the unknown value of the inclination angle $\iota$, a standard
statistical uncertainty due to the noise realization, and a residual
error associated with the interpolation procedure.

\subsection{Computational Cost}
\label{sec:compcost}

The volume of the \ac{ScoX1} signal parameter space makes a fully coherent search
intractable and has motivated the development of the algorithms described in
this paper. In designing these algorithms compromises between computation time
and sensitivity have been made in order to maximize detection probability with
realistic computational resources. For all searches that are part of this
study, with the exception of the Sideband search, computation cost scales
linearly with the length (in time) of the data analyzed.  The
Polynomial Search and the present version of TwoSpect analyze data from different detectors independently and hence
the computation time required scales with the number of interferometers.
Radiometer, Sideband, and CrossCorr instead analyze combined datasets and
therefore scale
with the number of combinations. The main component of the Sideband search
involves the convolution of the data with a template in the frequency domain
and consequently scales as ${\sim}T\log{T}$ where $T$ is the observation time.

As the spin frequency of \ac{ScoX1} is currently unknown, the frequency bandwidth
is a substantial factor in the search cost for most methods.  The cost of the
Sideband and Polynomial searches scales linearly with the size of the \ac{GW}
frequency search band.
For TwoSpect the number of templates grows in proportion
to the search frequency $f$ and hence the total number of templates $N_\textup{template}$, and also
therefore computational cost, scales with the maximum search frequency $f_\textup{max}$ squared, 
for wide band searches starting at $f_\textup{min}$.
To be precise, let one the duration on a short Fourier transform containing the data be $T_\textup{sft}$ (sometimes denoted $T_\textup{coh}$, because for TwoSpect this is the coherence length.)
Let also the analysis be split into subsections, each analyzing a frequency band of $f_\textup{bw}$, typically much less than $(f_\textup{max} - f_\textup{min})$.
The astrophysical period is $P$ and the uncertainty in the projected semimajor axis is $a \sin i$.
Then the number of templates is precisely~\cite{TwoSpectMDCMethods2015},
\begin{equation}
  \begin{split}
    N_{\textup{template}} & = 2 \left(T_\textup{sft} + \frac{1}{f_{\text{bw}}}\right) \\
    & \times \left[ 1+\frac{4 \pi T_\textup{sft}}{P} (6\sigma_{a \sin i}) (f_{\text{max}} + f_{\text{min}} + f_{\text{bw}})\right] \\
    & \times (f_{\text{max}} - f_{\text{min}}),
  \end{split}
\end{equation}
\noindent for a template grid spacing of $1/(2
T_\textup{sft})$ in $f$ and $1/(4 T_\textup{sft})$ in $a \sin{i}$ along with a search to
$\pm3\sigma$ in $\asini$. An empirical estimate of 3 \ac{CPU}-seconds per
template holds on modern \ac{CPU} cores at the time of the \ac{MDC}.

For CrossCorr the situation is more extreme, as the density of
templates in each orbital direction ($\asini$ and $\Tasc$) grows
proportional to the frequency, so the number of templates scales with
the cube of the maximum search frequency.  However, this can be
mitigated somewhat by reducing the coherence time $\Tmax$ as a
function of frequency, since the density of templates in each of the
three parameter space directions also scales approximately as $\Tmax$.
Overall, the computing cost of the CrossCorr method scales
approximately as the number of templates times the number of SFT
pairs.  For a search of $\Ndet$ detectors each with observing time
$\Tobs$, carried out using \acp{SFT} of duration $\Tsft$ and maximum lag
time $\Tmax$, the number of \ac{SFT} pairs is
\begin{equation}
  \Npair \approx \Ndet^2\frac{\Tobs}{\Tsft}\frac{\Tmax}{\Tsft}
\end{equation}
The \ac{SFT} duration $\Tsft$ is limited by the potential loss of SNR due
to unmodelled phase acceleration during the \ac{SFT}, and must also be
reduced with increasing frequency.  (Note that the coherence time of
the search is $\Tcoh=\Tmax$ and \emph{not} $\Tsft$, so the question of
\ac{SFT} length is one of computational cost and not of sensitivity.)

The Radiometer
search is limited primarily by data throughput, which renders the frequency
bandwidth irrelevant to computational performance. Reductions in the
uncertainties on orbital parameters will not impact the Radiometer search. For
the Sideband search, refined measurements of the semi-major axis or time of
ascension could motivate algorithmic changes but would not affect computational
cost. The Polynomial and TwoSpect search costs would decrease in proportion to
improvements in semi-major axis estimates.

The Sideband method is limited in observation length by the possibility of spin
wandering within the \ac{ScoX1} and other \ac{LMXB} systems. For \ac{ScoX1} the
current observation limit is 10 days resulting in an analysis time of
${\sim}1500$ \ac{CPU} hours on a modern processor\footnote{Comparable in
performance to an Intel Xeon~3220 processor} for a full search. It is possible
that the Sideband search could play a role as a fast and relatively low-latency
first-look algorithm used to scan the data as it is generated. The other
search methods are not thought to be limited by possible spin wandering in
\ac{LMXB} systems due to their higher tolerance to small frequency
variations.  Hence, observation times of $O(\text{yrs})$ are feasible. For the
TwoSpect search the corresponding computational cost for a complete analysis is
estimated as between $5\times 10^4$--$5\times 10^{5}$ \ac{CPU} hours.
The computational cost of a CrossCorr search depends on the coherence
times used at different frequencies, but scaling up the cost of the
analysis described in this paper
to a 1500~Hz bandwidth gives
an estimated computing cost of $\sim 3\times 10^6$ \ac{CPU}
hours.Analysis
of a full year of data for the Polynomial search would require ${\sim}10^8$
\ac{CPU} hours, rendering analysis of part of the data the most viable option.
The Radiometer pipeline is by far the computationally cheapest method that is
able to use all available data. It would require ${\sim}600$ \ac{CPU}-hours to
search over all combinations of detectors in a 3--detector
network~\footnote{All computational cost estimates are based on extrapolations
of smaller-scale test analyses.}.

\section{Mock data challenge\label{sec:mdc}}

We have chosen an \ac{MDC} as our primary tool for evaluating the qualities of
the different search methodologies. The aims of the \ac{MDC} are to simulate
multiple realizations of \ac{ScoX1}-type signals under psuedo-realistic
conditions such that pipelines can be compared empirically using both
individual signals and signal populations.  The properties of the
detector noise, signal parameter distributions, and scope of the \ac{MDC}
(described below) are chosen based on a balance between the current development
level of the search and simulation algorithms, the computational cost of this
analysis, and the expected sensitivities of the search algorithms.

The \ac{MDC} is characterized by the observational parameters
and data output of the simulated detectors, the injection parameters
of the simulated signals, and the information provided to the
participating pipelines of the \ac{MDC}. The \ac{MDC} data and simulated
signals are created using the program {\it lalapps\_Makefakedata\_v5} of the
LIGO Analysis Library software package for \ac{GW} data
analysis\cite{lalsuite}. The properties of the data are described in
\tref{tab:mdcdata}.
\begin{table*}
\caption{Simulated data.\label{tab:mdcdata}}
\begin{tabular}{l c}
\hline
Parameter & Value \\
\hline
Detectors & LIGO Hanford (H1), LIGO Livingston (L1), and Virgo (V1) \\
Observing run duration & 00:00:00 1 January 2019 -- 00:00:00 1 January 2020\\
Duty factor\footnote{The \ac{MDC} contains gaps in the time-series
  consistent with the duty factor observed in the initial LIGO S5
  science runs.
  The actual timestamps files from these
  analyses are time shifted and used in the generation of the \ac{MDC}
  data.} $\left\{\begin{tabular}{@{\ }l@{}}
    H1 \\ L1 \\ V1
  \end{tabular}\right.$ & $\begin{tabular}{@{\ }l@{}}
    73.6\% \\ 61.8\% \\ 75.2\%
  \end{tabular}$\\
Data sampling rate & 4096 Hz \\
Detector strain noise\footnote{This is equivalent to the design sensitivity
  of the proposed advanced detectors in the frequency range $\sim
  100$--$500$ Hz.} & White, Gaussian noise, with noise spectral
density $\sqrt{S_{\text{h}}}=4\times 10^{-24}$ Hz$^{-1/2}$ \\
Data storage format & Time-series data in \ac{GW}
frame files~\cite{FrameLib} \\
Orbital parameters & Selected from Gaussian distributions using values given in \tref{tab:scoparams} \\
Frequency parameters & Distributed psuedo-randomly in the
range $50$--$1500$ Hz \\
\hline
\end{tabular}
\end{table*}

\begin{figure}
\centering
\includegraphics[width=\columnwidth]{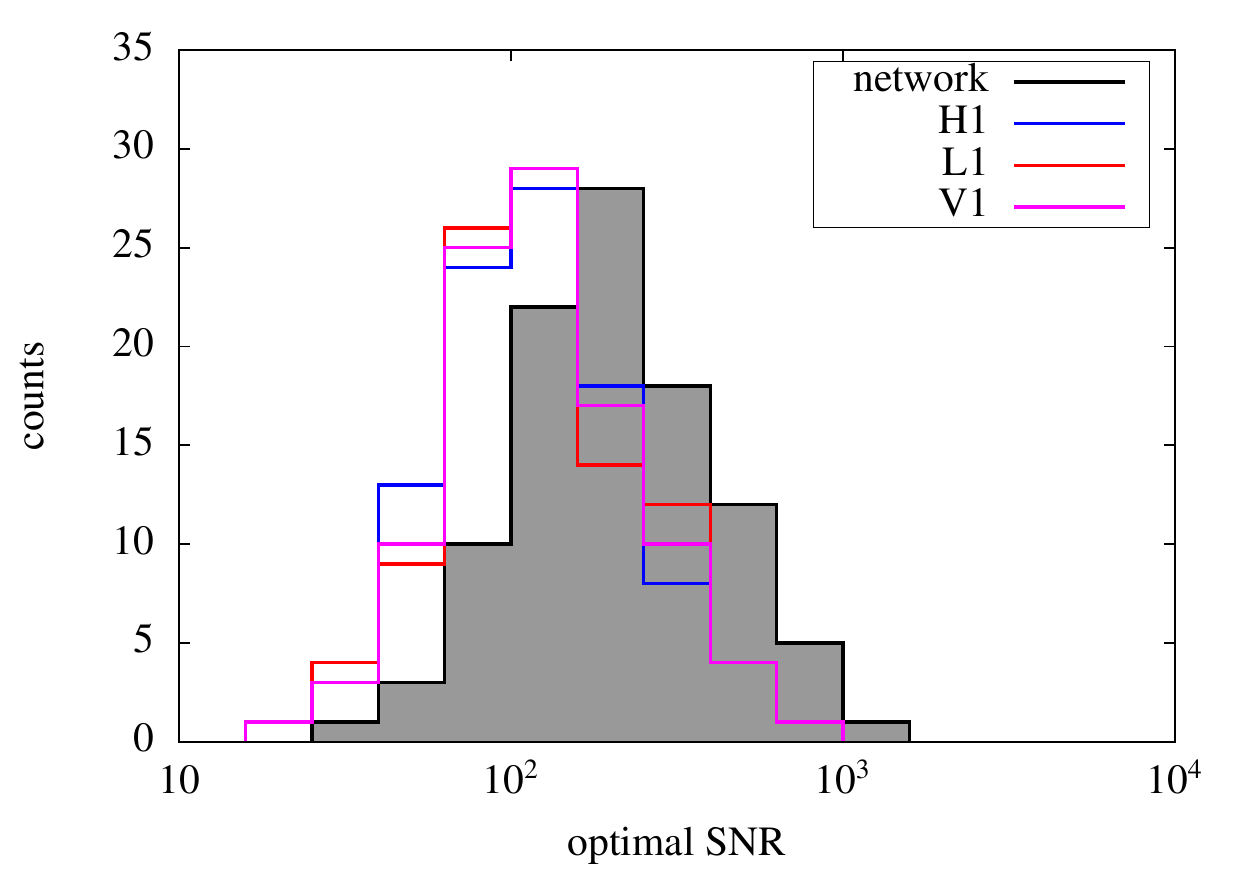}
\caption{The distribution of \acp{SNR} for the open and closed simulated signals.
  \label{fig:snrdist}}
\end{figure}

For this \ac{MDC}, 100 simulated \ac{ScoX1}-type signals were added to the
data, 50 of which were considered as ``open'' signals and 50 as
``closed''. The simulated detector noise was chosen to be Gaussian with no frequency
dependence and characterised by a noise spectral density broadly equivelent to
the advanced detector design
sensitivities~\cite{2015CQGra..32g4001T,2013arXiv1304.0670L,Harry:2010zz,TheVirgo:2014hva}.  
The parameters of the open signals were made available to
the challenge participants making these signals ideal for pipeline
tuning and validation. Detection and parameter estimation of the closed signals
constitute the goals of the \ac{MDC}. A list of the closed-signal parameters
are listed in \tref{tab:mdcparams}. All signals had the following
properties:
\begin{itemize}[itemsep=-1mm]
\item Sky location: Fixed equal to the best-known value for
  \ac{ScoX1} (see \sref{sec:params}).
\item Intrinsic frequency $f_{0}$: Each signal has an intrinsic frequency value
that is contained within a unique
  5--Hz band, selected pseudo-randomly in the frequency range of
  50--1500~Hz. There is a small bias towards lower frequencies in
  order to reduce the computational cost of the challenge\footnote{In
    general the computational cost of continuous wave search pipelines
    scales $f^{\alpha}$ where $\alpha$ is usually a positive
    integer.}. There is a minimum 5--Hz spacing between the boundaries of each 5--Hz band
containing a simulated signal. The
  intrinsic frequency is monochromatic and randomly chosen from a
  uniform distribution. There are no accretion induced spin-wandering effects of the
  \ac{GW} frequency.
\item \ac{NS} orientation $\cos{\iota}$, \ac{GW} polarization angle
  $\psi$, and initial \ac{NS} rotation phase $\phi_0$: randomly chosen
  from uniform distributions with $\cos{\iota}{\in}(-1,1)$ and
  $\psi,\phi_0{\in}(0,2\pi)$~rad.
\item Orbital parameters $P$, $\Tasc$, $\asini$: The values
  are randomly chosen from known Gaussian distributions with
  means and variances equal to the values given in
  \tref{tab:scoparams}\footnote{The version of the
orbital period measurement used at the time of generating the \ac{MDC} was from
an early draft of~\cite{2014ApJ...781...14G} in which the value was
$68023.7136\pm0.0432$ sec. We also acknowledge an inconsequential error in the
shifting of $T_{\text{asc}}$ to the midpoint of the simulated observation
resulting in an offset of half an orbital period in relation to the real Sco
X--1 system.}.
The time of orbital ascension was shifted to an epoch close to the mid-point of
the simulated observation and hence the \ac{MDC} value was $1245967384\pm250$ GPS
seconds.  The larger uncertainty on $\Tasc$ is consistent with additional
components due to the orbital period uncertainty and the time span between the
most recent Sco X--1 orbital measurements and the proposed \ac{MDC} observing
epoch. The orbit is assumed to be circular
  (eccentricity $e=0$).
\item \ac{GW} strain amplitude $h_0$: For a given signal with pre-selected
$\cos{\iota}$ and $\psi$, a value of $h_0$ is chosen to be consistent with the
3-detector multi-IFO optimal \ac{SNR} having been drawn from a log-normal distribution
with parameters $\mu=\log{200},\sigma=0.7$. The optimal \ac{SNR}$=(h|h)$, where $h$ is the signal
(multi-IFO) timeseries and $(x|y)$ is the usual scalar product 
(see~\cite{2007PhRvD..75b3004P} for a derivation).  These parameters define the mean
and standard deviation of the \ac{SNR} natural logarithm.  The distribution of
\acp{SNR} is shown in \fref{fig:snrdist}.
The \ac{SNR} distribution
parameters were originally selected in order to satisfy the requirement that
the weakest searches would detect $\mathcal{O}(5)$ signals and that the strongest
would fail to detect approximately the same fraction. Tuning was performed prior to the \ac{MDC}
to establish the distribution parameters based on the original 4 pipelines only
(excluding CrossCorr).  
\end{itemize}

\begin{table*}
\caption{Signal parameters for the closed part of the MDC.\label{tab:mdcparams}}
\begin{tabular}{cccccccccc}
\hline
index & band (Hz) & $f_{0}$ (Hz) & $\asini$ (sec) & $P$ (sec) &
$\Tasc$ (GPS sec) & $h_{0}$ ($10^{-25}$) & $\cos{\iota}$ & $\psi$ (rads) &
$\phi_{0}$ (rads)\\ \hline
\makeatletter{}1 & 50--55 & 54.498391348174 & 1.379519 & 68023.673692 & 1245967666.024 & 4.160101 & -0.611763 & 0.656117 & 4.184335\\
2 & 60--65 & 64.411966012332 & 1.764606 & 68023.697209 & 1245967592.982 & 4.044048 & -0.573940 & 4.237726 & 5.263431\\
3 & 70--75 & 73.795580913582 & 1.534599 & 68023.738942 & 1245967461.346 & 3.565197 & 0.971016 & 1.474289 & 4.558232\\
5 & 90--95 & 93.909518008164 & 1.520181 & 68023.681326 & 1245966927.931 & 1.250212 & -0.921724 & 0.459888 & 5.442296\\
11 & 150--155 & 154.916883586097 & 1.392286 & 68023.744190 & 1245967559.974 & 3.089380 & 0.323669 & 1.627885 & 3.402987\\
14 & 180--185 & 183.974917468730 & 1.509696 & 68023.755607 & 1245967551.047 & 2.044140 & 0.584370 & 3.099251 & 5.420183\\
15 & 190--195 & 191.580343388804 & 1.518142 & 68023.722885 & 1245967298.451 & 11.763777 & 0.028717 & 5.776490 & 1.844049\\
17 & 210--215 & 213.232194220000 & 1.310212 & 68023.713119 & 1245967522.541 & 3.473418 & 0.082755 & 5.348830 & 2.848229\\
19 & 230--235 & 233.432565653291 & 1.231232 & 68023.686054 & 1245967331.136 & 6.030529 & 0.224890 & 1.467310 & 0.046980\\
20 & 240--245 & 244.534697522529 & 1.284423 & 68023.742615 & 1245967110.972 & 9.709634 & -0.009855 & 3.008558 & 1.414107\\
21 & 250--255 & 254.415047846878 & 1.072190 & 68023.753262 & 1245967346.405 & 1.815111 & 0.292830 & 0.302833 & 0.449571\\
23 & 270--275 & 271.739907539784 & 1.442867 & 68023.685008 & 1245967302.288 & 2.968392 & -0.498809 & 1.367339 & 3.578383\\
26 & 300--305 & 300.590450155009 & 1.258695 & 68023.687437 & 1245967177.469 & 1.419173 & 0.817770 & 6.028239 & 0.748872\\
29 & 330--335 & 330.590357652653 & 1.330696 & 68023.774609 & 1245967520.825 & 4.274554 & 0.711395 & 4.832193 & 3.584838\\
32 & 360--365 & 362.990820993568 & 1.611093 & 68023.714448 & 1245967585.560 & 10.037770 & 0.295336 & 2.372268 & 1.281230\\
35 & 390--395 & 394.685589797695 & 1.313759 & 68023.671480 & 1245967198.049 & 16.401523 & 0.491537 & 4.023472 & 4.076188\\
36 & 400--405 & 402.721233789014 & 1.254840 & 68023.628720 & 1245967251.346 & 3.864262 & 0.210925 & 2.195660 & 1.662426\\
41 & 450--455 & 454.865249156175 & 1.465778 & 68023.695320 & 1245967225.750 & 1.562041 & -0.366942 & 2.712863 & 4.785230\\
44 & 480--485 & 483.519617972096 & 1.552208 & 68023.724831 & 1245967397.861 & 2.237079 & -0.889314 & 3.754288 & 5.584973\\
47 & 510--515 & 514.568399601819 & 1.140205 & 68023.714935 & 1245967686.805 & 4.883365 & -0.233705 & 3.645842 & 5.773243\\
48 & 520--525 & 520.177348201609 & 1.336686 & 68023.634260 & 1245967675.302 & 1.813016 & -0.241020 & 0.816681 & 2.908419\\
50 & 540--545 & 542.952477491471 & 1.119149 & 68023.750909 & 1245967927.484 & 1.092771 & 0.939190 & 4.031313 & 1.527390\\
51 & 550--555 & 552.120598886904 & 1.327828 & 68023.741431 & 1245967589.535 & 9.146386 & 0.120515 & 3.280902 & 0.382047\\
52 & 560--565 & 560.755048768919 & 1.792140 & 68023.831850 & 1245967377.203 & 2.785731 & 0.486566 & 4.530901 & 4.726265\\
54 & 590--595 & 593.663030872532 & 1.612757 & 68023.722670 & 1245967624.534 & 1.517530 & -0.819247 & 5.029020 & 0.539005\\
57 & 620--625 & 622.605388362863 & 1.513291 & 68023.736515 & 1245967203.215 & 1.576918 & 0.402573 & 3.365393 & 5.634876\\
58 & 640--645 & 641.491604906276 & 1.584428 & 68023.683124 & 1245967257.744 & 3.416297 & 0.149811 & 0.273787 & 5.120474\\
59 & 650--655 & 650.344230698489 & 1.677112 & 68023.696004 & 1245967829.905 & 8.834794 & 0.497028 & 3.148233 & 3.305762\\
60 & 660--665 & 664.611446618250 & 1.582620 & 68023.623412 & 1245967612.309 & 2.960648 & 0.825769 & 5.828391 & 6.093132\\
61 & 670--675 & 674.711567789201 & 1.499368 & 68023.712738 & 1245967003.318 & 6.064238 & 0.047423 & 3.616627 & 6.236046\\
62 & 680--685 & 683.436210983289 & 1.269511 & 68023.734889 & 1245967453.966 & 10.737497 & -0.070857 & 6.155982 & 3.343461\\
63 & 690--695 & 690.534687981171 & 1.518244 & 68023.681037 & 1245967419.389 & 1.119028 & -0.630799 & 2.583073 & 4.573909\\
64 & 700--705 & 700.866836291234 & 1.399926 & 68023.663565 & 1245967596.121 & 1.599528 & 0.052755 & 0.493210 & 0.457488\\
65 & 710--715 & 713.378001688688 & 1.145769 & 68023.749146 & 1245967094.570 & 8.473643 & 0.420557 & 1.782869 & 5.600087\\
66 & 730--735 & 731.006818153273 & 1.321791 & 68023.713215 & 1245967576.493 & 9.312048 & 0.596321 & 4.560452 & 5.114716\\
67 & 740--745 & 744.255707971300 & 1.677736 & 68023.702943 & 1245967084.297 & 4.579697 & 0.028568 & 3.060388 & 2.536793\\
68 & 750--755 & 754.435956775916 & 1.413891 & 68023.738717 & 1245967538.698 & 3.695848 & -0.401291 & 4.343783 & 0.034602\\
69 & 760--765 & 761.538797037770 & 1.626130 & 68023.662519 & 1245966821.545 & 2.889282 & 0.102754 & 3.302613 & 3.405741\\
71 & 800--805 & 804.231717847467 & 1.652034 & 68023.792724 & 1245967156.547 & 2.922576 & -0.263274 & 2.526713 & 5.884348\\
72 & 810--815 & 812.280741438401 & 1.196485 & 68023.718158 & 1245967159.077 & 1.248093 & 0.591815 & 2.341322 & 4.708392\\
73 & 820--825 & 824.988633484129 & 1.417154 & 68023.683539 & 1245967876.831 & 2.443983 & -0.169611 & 0.114125 & 1.081173\\
75 & 860--865 & 862.398935287248 & 1.567026 & 68023.746169 & 1245967346.324 & 7.678400 & 0.432360 & 0.574140 & 0.813485\\
76 & 880--885 & 882.747979842807 & 1.462487 & 68023.621227 & 1245966753.240 & 3.260143 & 0.447011 & 5.242454 & 0.560221\\
79 & 930--935 & 931.006000308958 & 1.491706 & 68023.642700 & 1245967290.057 & 4.680848 & 0.015637 & 5.686775 & 0.729836\\
83 & 1080--1085 & 1081.398956458276 & 1.198541 & 68023.740103 & 1245967313.935 & 5.924668 & 0.121699 & 3.760452 & 6.032308\\
84 & 1100--1105 & 1100.906018344283 & 1.589716 & 68023.763681 & 1245967204.150 & 11.608892 & -0.571199 & 2.310229 & 2.956547\\
85 & 1110--1115 & 1111.576831848269 & 1.344790 & 68023.748155 & 1245967049.350 & 4.552730 & 0.069526 & 0.365444 & 2.048360\\
90 & 1190--1195 & 1193.191890630547 & 1.575127 & 68023.773099 & 1245966914.268 & 0.684002 & -0.900467 & 0.195847 & 0.873581\\
95 & 1320--1325 & 1324.567365220908 & 1.591685 & 68023.703242 & 1245967424.756 & 4.293322 & 0.687636 & 4.543767 & 4.301401\\
98 & 1370--1375 & 1372.042154535880 & 1.315096 & 68023.760793 & 1245966869.917 & 5.404060 & -0.080942 & 4.895973 & 3.760856\\
 
\hline
\end{tabular}
\end{table*}

The participants of the challenge were given the following additional
information to guide them in the analysis of the data:
\begin{itemize}[itemsep=-1mm]
\item A list of the 5--Hz frequency bands that contain open signals or
  closed signals. The exact signal parameters for the open signals were
  also known.
\item Participants were required to assume that signals \emph{do} contain phase
  contributions due to spin wandering (although they do
  not). They were to assume that this wandering would have the
  characteristics of a time-varying spin frequency derivative of
  maximum amplitude $10^{-12}$~Hz$\cdot$s$^{-1}$ with variation timescale
  $10^{6}$ seconds.
\end{itemize}

The participants were requested to provide the following data products
from their analysis, in order to perform like-for-like comparisons
between pipelines:
\begin{itemize}[itemsep=-1mm]
\item Detectability: for the 50 closed signals, identify each as a
  detection or non-detection. Signal detection is defined as
  candidates recovered at a confidence equivalent to a
  $p$-value~$=10^{-2}$ accounting for multiple-trials over each
  5--Hz band. The $p$-value is generically defined as the probability
  of obtaining a given detection statistic from data containing only
  the non-astrophysical background noise.
\item Parameter estimation: If a signal is claimed as detected in a given 5--Hz
  frequency band, then the analysis pipeline must report on the
  measured signal parameters and associated uncertainties. Note that
  each individual pipeline has different abilities to measure signal
  parameters. In particular, no participating pipeline currently
  provided estimates of $\cos{\iota}$, $\psi$, or $\phi_0$.
\item Upper limits: for those 5--Hz frequency bands where a signal is
  not detected, then the pipeline must report the $95$\% confidence
  level upper limit on the \ac{GW} amplitude $h_0$ (also accounting
  for the multiple trials).
\end{itemize}

Additional, less-strict instructions were also suggested to participants
and included the sensible use of costly computational resources.  This
was stated so as to be able to compare pipelines under the assumption
of broadly similar computational costs.  Limiting each pipeline to
identical total computational resources is currently an unfeasible
restriction to enforce.

\subsection{Search Implementations}

In this section, a description is given of the different choices made by each
search pipeline specifically for this \ac{MDC}. 

\subsubsection{Polynomial}

For this \ac{MDC}, the Polynomial search analyzed $1.81\times
10^{6}$~s of
simulated data from the LIGO H1 (Hanford) interferometer, spread over a period
of $4.855\times 10^{6}$~s, starting at the \ac{GPS} time $1251698492$. The
length of the period was a compromise between sensitivity and use of
computational resources.  Only $10^{4}$~s long segments of data (without gaps)
were analyzed. The data was taken from the interval that had the largest duty
cycle in terms of uninterrupted \ac{SFT}-size segments.

For an all-sky search, Polynomial Search uses 1200~s
\ac{SFT}s, in order to be sensitive to a wide range of binary orbital periods.
Since the binary period of \ac{ScoX1} is known, the \ac{SFT} length can be
increased to up to one fourth of its period. If beam patterns were constant in
time, the sensitivity would scale with the square root of the \ac{SFT} length.
However, for longer \ac{SFT}s, evolution of the beam patterns negatively affect
sensitivity.  $10^{4}$~s was chosen as a compromise.

In total, 50 regions of 5~Hz each were searched, with template parameters in the
range $\pm 4.0 \times 10^{-5}$~Hz~s$^{-1}$ for the first derivative of signal
frequency with respect to time and $\pm 1.0 \times 10^{-8}$~Hz~s$^{-2}$ for the
second derivative with respect to time. The largest expected values of these
derivatives assuming 1--$\sigma$ uncertainties on the simulated signal
parameters are $\pm 2.1 \times 10^{-5}$~Hz~s$^{-1}$ and
$\pm 1.9 \times 10^{-9}$~Hz~s$^{-2}$, respectively.

Detection statistics were determined for each 0.5~Hz frequency bin based on
the number of \acp{SFT} in which one or more templates exceed the correlation 
threshold. The threshold required to attain a 1\% false alarm probability was
determined from the analysis results in a 5~Hz reference band of the \ac{MDC}
data known not to contain a signal (720--725 Hz).

\subsubsection{Radiometer}

The Radiometer search used all data from H1, L1, and V1 that was coincident
between pairs of detectors.  This was $\sim$185~days for H1--L1, $\sim$244~days
for H1--V1, and $\sim$185~days for L1--V1.

For each 0.25 Hz band the $p$-value was calculated under the assumption that
the corresponding $Y$-estimate (see
\eqref{eq:Ystat}) is Gaussian distributed, as expected from the central limit
theorem for the many independent segments.  This assumption has been shown to
be robust in studies with realistic
data~\cite{2007PhRvD..76h2003A,2011PhRvL.107A1102A}.  The single trial
$p$-value is given by 
\begin{equation} 
  p=\frac{1}{2}\left(1-\erf\left(\frac{{\rm SNR}}{\sqrt{2}}\right)\right)
\end{equation} 
from which the multi-trial $p$-value is computed via
\begin{equation} 
  p_{\rm multi}=1-(1-p)^N 
\end{equation} 
where ${\rm SNR}=\hat{Y}_{\rm tot}/\sigma_{\rm tot}$ and $N$ is number of
independent trials which, for a 5~Hz band with 0.25~Hz bins, is 21 (due to choice of bin start frequency, the Radiometer search here searched slightly beyond the 5~Hz band which resulted in 21 rather than 20 trials).
The Radiometer search results were converted to match the format presented in this paper.  The conversion process is described in \aref{sec:radiometertech}.

\subsubsection{Sideband}

The Sideband search analyzed a 10-day stretch of \ac{MDC} data (864000 sec)
using all 3 interferometers and with an initial \ac{GPS} time of 1245000000.
This was not an optimally selected 10-day stretch of data (as was done in
the~\cite{S5sideband}), with the duty factors for the three interferometers
being $70\%$, $58\%$ and $80\%$ for H1, L1 and V1 respectively.
Since the noise floor is constant in time, optimality in this case is dependent
upon the duty factors of the data combined with the diurnal time variation of
the antenna patterns in relation to the \ac{ScoX1} sky position.   
The ``optimal" 10--day data-stretch has subsequently been identified as starting
at \ac{GPS} time 1246053142 and
having duty factors $86\%$, $83\%$ and $94\%$. 

For Gaussian noise, each value of the $\mathcal{C}$-statistic is drawn from a
central $\chi^{2}_{4M}$ distribution, where $M=2{\rm ceil}(2\pi f_{0}\asini)
+1$ is the number of sidebands.  For $N$ independent trials, $p$-values are therefore
calculated as 
\begin{align}
  p=1-\left[F\left(\mathcal{C},\,4M\right)\right]^{N},\label{eqn:sidebandp}
\end{align}
where $F(\mathcal{C},4M)$ is the cumulative distribution function of a
$\chi_{4M}^2$ distribution evaluated at $\mathcal{C}$.  If one assumes each
trial is statistically independent and defines a target false alarm
probability, \eqref{eqn:sidebandp} allows us to determine a threshold value
of the maximum recovered $\mathcal{C}$-statistic, denoted $\mathcal{C}^{*}_{N}$
(for details see ~\cite{S5sideband}).  In practice, there is strong correlation
between  $\mathcal{C}$-statistic values due to the nature of the comb template.
In addition there are small deviations from the expected statistical behavior
of the $\mathcal{C}$-statistic due to approximations and noise normalization
procedures within the search algorithm.  Therefore, Monte Carlo simulations have
been used to identify a correction to $\mathcal{C}^{*}_{N}$ that corresponds to the
desired false alarm probability.  The corrected threshold statistic is given by
\begin{align} C^{\star}_{\kappa}=C^{\star}_{N}\left(1+\kappa\right)-4M\kappa,
\end{align}
where $\kappa=0.3$~\cite{S5sideband}. A detection is therefore claimed if the
maximum recovered value of the statistic satisfies
$\mathcal{C}>\mathcal{C}^{\star}_{\kappa}$.

\subsubsection{TwoSpect}

The TwoSpect pipeline analyzed all \ac{MDC} data from each interferometer
separately. For each interferometer, detection statistics and corresponding
single-template $p$-values were computed for each template. A set of most
significant $p$-value outliers in 5~Hz bands were produced for each
interferometer, subject to a $p$-value threshold inferred from Monte-Carlo
simulations in Gaussian noise (see 
a forthcoming methods paper~\cite{TwoSpectMDCMethods2015}
for
details). These sets were compared in pairwise coincidence (H1-L1, H1-V1, or L1-V1),
where coincidence required proximity within a few grid points in the parameter
space. Surviving outliers were classified as a detection at the predefined 1\%
false alarm threshold.

In the case of detection, the highest $p$-value from a single interferometer in
a given band was used to produce estimated signal parameters. Uncertainties in
these parameters were determined from the open signals within the \ac{MDC}.
For the intrinsic signal frequency and modulation
depth, we estimated the mean and standard deviation of parameter estimation
error in the open signals.
This error varied little for different injected signal strength $h_0$, so function was or could be estimated to yield more precise uncertainty measurements other than the mean error.
Since the parameter distribution for the closed signals was known 
to be the same as the open signals, we reported the mean error as our estimate of uncertainty.
Since some higher-frequency bands appeared to have greater error, a seperate mean error was estimated for those bands.
Further details to be reported in a forthcoming methods paper.
Confidence intervals calculated more rigorously for the signal amplitude.
Upper-limits on signal
amplitude were determined from an estimate of the 95\% confidence level of
non-detected open \ac{MDC} signals. The largest uncertainty in upper limits
and signal amplitude estimation derives from the ambiguity between true $h_{0}$
signal and $\cos\iota$ inclination. This ambiguity cannot be resolved with the
present algorithm and depends partially on the assumed prior distribution of
signal ampltitudes; the uncertainty was estimated by simulation. Complete
details of the parameter estimation and upper-limit setting procedure are detailed in the methods paper~\cite{TwoSpectMDCMethods2015}.

\subsubsection{CrossCorr}

\label{sec:CrossCorrimpl}

The CrossCorr pipeline analyzed all \ac{MDC} data from all three
interferometers together, calculating cross-correlation contributions
from each pair of \acp{SFT} for which the timestamps differed by less than
a coherence time $\Tmax$.  In order to control computational costs,
different values of $\Tmax$ were used for bands in different frequency
ranges, and also for different parts of orbital parameter space within
each frequency band, as detailed in \aref{app:CrossCorrtech}.
Each 5 Hz frequency band was divided into 100 frequency slices and
eight regions of orbital parameter space, described in more detail in
\sref{app:CrossCorrTmax}.  The resulting 800 parameter space
regions were then searched using a cubic lattice with a metric
mismatch of $0.25$ (as defined in \cite{LMXBCrossCorr}), and the
highest resulting statistic values combined into a ``toplist'' for the
entire band.  Local maxima over parameter space were in principle
considered as candidate signals, although in practice each band
contained high statistic values clustered around a single global
maximum.

A ``refinement'' was performed around each such maximum, decreasing
the grid spacing by a factor of 3 and limiting attention to a cube 13
grid spacings on a side.  The resulting maximum statistic value was
high enough to declare a confident signal detection for each of the 50
bands, but for some of the weaker detected signals, a followup was
performed with an even finer parameter space resolution and a longer
coherence time, which approximately doubled the statistic value.

Since the CrossCorr statistic is a sum of contributions from
many \ac{SFT} pairs, and is normalized to have unit variance and zero mean
in the absence of a signal, the nominal significance of a detection
can be estimated using the cumulative distribution function of a
standard Gaussian distribution.  A false alarm probability for the
loudest statistic value in a 5 Hz band can be estimated by assuming
that each of the templates in the original grid was an independent
trial and multiplying the single-template $p$-value by the associated
trials factor.  The $p$-values generated by this procedure are not
reliable false alarm probabilities, however, since with typical trials
factors of $10^8$, the relevant single-template $p$-values are
$10^{-10}$ or smaller, for which the Gaussian distribution is no
longer a good approximation.  Therefore, the nominal multi-template
$p$-value corresponding to an actual false alarm probability of 1\%
was estimated by running the first stage of the pipeline on forty-nine
5~Hz bands containing no signal.  Comparing this value to those
associated with the detected closed signals showed the latter all
to be detections.  For more details, see \sref{app:CrossCorrPval}
of \aref{app:CrossCorrtech}.

For each detected signal, the best-fit values of $f_0$, $\asini$ and
$\Tasc$ were determined by interpolation, fitting a multivariate
quadratic to the 27 statistic values in a cube centered on the highest
value in the final grid, and reporting the peak of this function.
Parameter uncertainties were a combination of: residual errors from
the interpolation procedure, statistical errors associated with the
noise contribution to the detection statistic, and a systematic error
associated with parameter offset associated with the unknown value of
$\cos\iota$.  Additionally, analysis of the open signals showed a
small unexplained frequency-dependent bias in the $\asini$
estimates. To produce conservative errorbars, the size of the
empirical correction for this bias was added in quadrature with the
other errors.  The procedure is described in further detail in
\aref{app:CrossCorrtech} and \cite{CrossCorrMDC}.

\section{Results\label{sec:results}}

Participants in the \ac{MDC} were asked to submit their results on the 50
closed signals no later than 30 April 2014 in the form described in
\tref{tab:submission}. Four pipelines (Polynomial, Radiometer, Sideband and
TwoSpect) completed their analysis of the closed signals on or near the
original deadline of the \ac{MDC}, at which point the previously secret
parameters were made available. Some of the final post-processing analyses took
place after the initial submissions in order to provide the full final
submission. A fifth analysis method, the CrossCorr pipeline, was not in place
soon enough to participate in the original challenge, but carried out a
subsequent opportunistic analysis.  This ``self-blinded'' analysis was
conducted and a submission table prepared without looking at the parameters of
the closed signals.  \Tref{tab:submissionTimeline} summarizes these submission
dates.

From the submission tables of each pipeline, we have generated a
number of comparison figures and tables.  The description of results
are divided
into the topics of detection, upper-limits and parameter estimation. 
\begin{table*}
\caption{The MDC submission parameters.\label{tab:submission}}
\begin{tabular}{lccll}
\hline
parameter & symbol & units & description &\\ \hline
PULSAR INDEX &  &   & the index of the closed pulsar
&\rdelim\}{5}{3mm}\multirow{5}{*}{for all signals}\\
PULSAR FSTART&  & Hz & the lower bound on the search frequency band &\\
PULSAR FEND &  & Hz & the upper bound on the search frequency band &\\
DETECTION &  &   & please state either yes or no &\\
P VALUE & $\log{p}$ &   & natural-log of the multi-trial statistical significance of the loudest event found &\\
\hline
\multirow{2}{*}{H0 UL} & \multirow{2}{*}{$h_{0}^{95\%}$} &   & \multirow{2}{*}{95\% confidence upper limit on $h_{0}$, the
dimensionless strain tensor amplitude} &
\rdelim\}{2}{3mm}\multirow{2}{*}{\begin{tabular}{@{\ }l@{}}for non-detected
\\signals only\end{tabular}}\\
& & & &\\
\hline
H0 EST & $h_{0}$ &   & best estimate for $h_{0}$, the dimensionless strain
tensor amplitude & \rdelim\}{10}{3mm}\multirow{10}{*}{\begin{tabular}{@{\
}l@{}}for detected \\signals only\end{tabular}}\\
H0 ERR & $\Delta h_{0}$ &   & uncertainty on the best estimate of $h_{0}$&\\
F0 ESTIMATE & $f_{0}$ & Hz & best estimate for $f_{0}$, the intrinsic \ac{GW} frequency &\\
F0 ERROR & $\Delta f_{0}$ & Hz & uncertainty on the best estimate of $f_{0}$&\\
ASINI EST & $\asini$ & sec & best estimate for
the product of the orbital radius and the sin of the inclination &\\
ASIN ERR & $\Delta (\asini)$ & sec & uncertainty on the best estimate of $\asini$&\\
PERIOD EST & $P$ & sec & best estimate for the orbital period &\\
PERIOD ERR & $\Delta P$ & sec & uncertainty on the best estimate of $P$&\\
TASC EST & $\Tasc$ & GPS sec & best estimate for the time of ascension & \\
TASC ERR & $\Delta \Tasc$ & GPS sec & uncertainty on the best estimate
of the time of ascension& \\
\hline
\end{tabular}
\end{table*}

\begin{table*}
\setlength{\tabcolsep}{12pt}
\caption{Dates of submitted results for the \ac{MDC}.\label{tab:submissionTimeline}}
\begin{tabular}{l c c c c c}
\hline
Submission deadline 30 April 2014 & TwoSpect & Polynomial & Radiometer & Sideband & CrossCorr \\
\hline
Initial submission & 30 April 2014 & 1 May 2014 & 1 May 2014 & 19 May 2014 & 19 Dec. 2014 \\
Final submission & 22 Aug. 2014 & 1 Oct. 2014 & 29 Mar. 2015 & 27 June 2014 & 16 Jan. 2015 \\
\hline
\end{tabular}
\end{table*}

\subsection{Detection}

\begin{figure*}
\centering
\includegraphics[width=\textwidth]{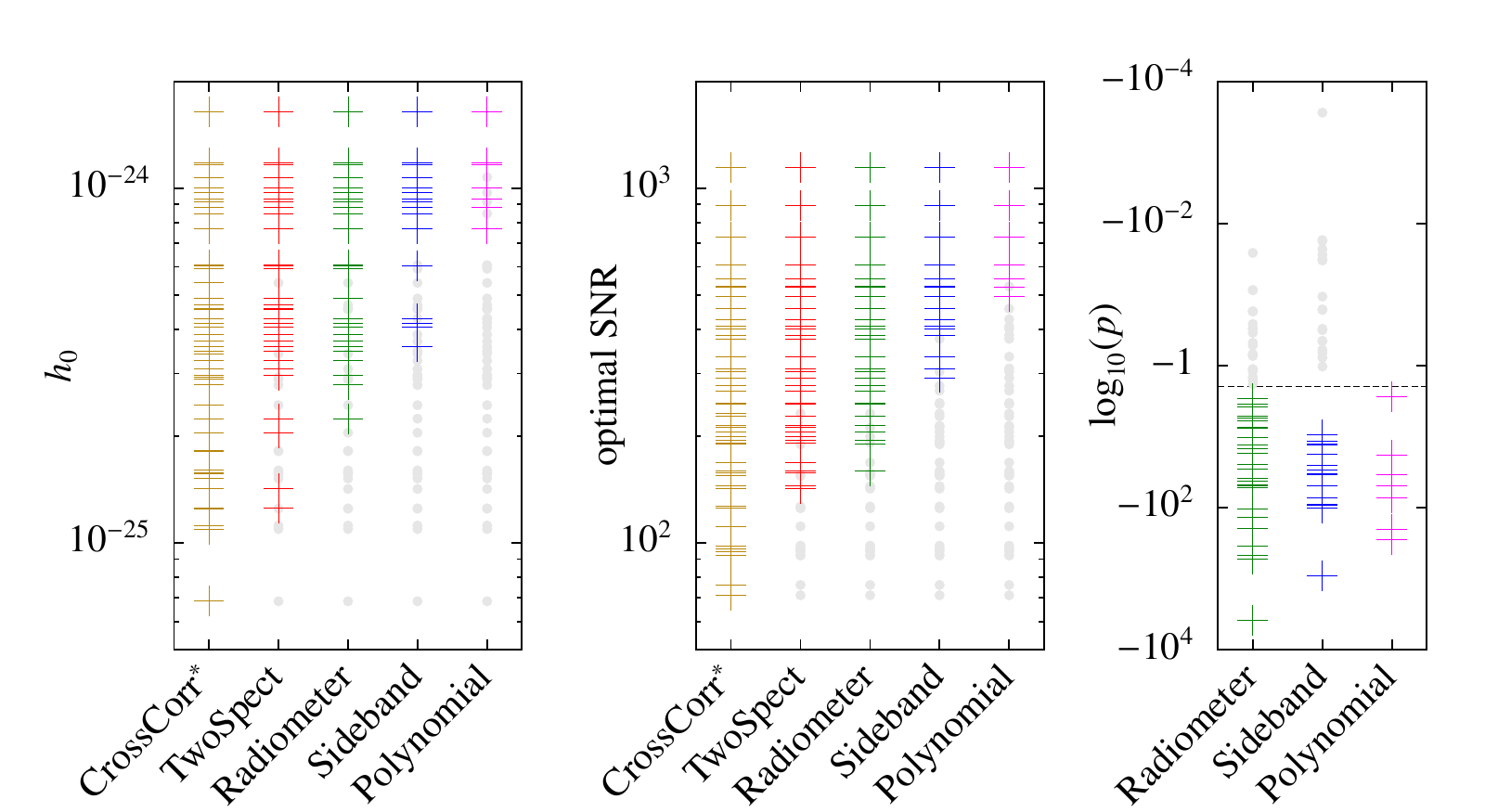}
\caption{A comparison of detected signal properties for each search
  pipeline.  We plot the values of $h_{0}$, optimal \ac{SNR}, and
  estimated $\log_{10}(p)$ value for the detected (color) and
  non-detected (grey) signals from each pipeline.  The 3rd panel shows
  the $\log_{10}(p)$ values from the search pipelines which were able
  to estimate reliable values, with the black horizontal dashed line
  representing the detection threshold of $\log_{10}(p)=-2$.  Note
  that the TwoSpect and CrossCorr pipelines generated nominal
  $p$-values, but as they were known not to be quantitatively
  accurate, they are not shown here.  Note also that the
  CrossCorr pipeline obtained results seven months beyond the original
  deadline for the \ac{MDC}, as detailed in \tref{tab:submissionTimeline}.
  The full list of detected and non-detected signals is given in
  \aref{sec:fullresults}
  \label{fig:detoverview}}
\end{figure*}

An overview of the detectability of the \ac{MDC} signals is shown in
\fref{fig:detoverview}.  The list of specific signals detected by each
pipeline are given in \aref{sec:fullresults}.  Three different
figures of merit are plotted: the detection success as a function of $h_{0}$, as a
function of optimal \ac{SNR}, and as a function of reported
$\log_{10}(p)$.  Of the original four pipelines that ran in the \ac{MDC} (see \tref{tab:submissionTimeline}), TwoSpect was able to
detect the most signals and detect signals of lower intrinsic strain
and \ac{SNR} than the other three pipelines. The CrossCorr
pipeline, which ran an opportunistic, ``self-blinded'' analysis
in the months following the
original \ac{MDC}, was able to detect all 50 signals.

Specifically,
the CrossCorr, TwoSpect, Radiometer, Sideband, and Polynomial
pipelines detect 50, 34, 28, 16, and 5 respectively with ratios of
1, 1.83, 3.27, 5.21, and 11.2 between the weakest
detected $h_0$ values from each pipeline and the weakest signal present.
Equivalent ratios in detectable optimal
\ac{SNR} are
1, 2.0, 2.2, 4.1, and 7.0.  We also plot the estimated value of
the $\log_{10}(p)$, the (base 10) logarithm of the $p$-value as defined
in \sref{sec:mdc}, for all signals (detected and non-detected) in
the third panel of \fref{fig:detoverview}.

Among the four original pipelines,
we note that all detected signals from the Polynomial pipeline are a subset of
those detected by the Sideband pipeline which in turn are a subset of those
detected by the Radiometer pipeline which, with the exception of pulsar 52, are
a subset of those detected by TwoSpect.(TwoSpect saw an above-threshold statistic in the pulsar 52 band for V1, 
but not in coincidence with H1 or L1, so no detection was
declared). While the CrossCorr pipeline was the most successful, detecting all 50
closed signals, it is also the least mature.  In particular CrossCorr has not
yet been used for an astrophysical analysis of \ac{GW} detector data as TwoSpect, Sideband and
Radiometer have, and its behavior in the face of non-Gaussianity and
other instrumental noise features has not been probed by this
idealized \ac{MDC}.  

Due to the relatively low number of simulated signals in
the \ac{MDC} we are aware that we do not deeply probe the interesting
boundaries in sensitivity between pipelines.  In particular, the closed
signal detections give no indication of the lower limit of
detectability for the CrossCorr pipeline.  Some insight can be taken
from the open signal data, in which CrossCorr was able to find 49 of
the 50 open signals.  The one ``missed'' signal had $h_0=3.81\times
10^{-26}$ and an optimal \ac{SNR} of $33$.  The ``quietest'' of the 49 open
signals which CrossCorr detected had $h_0=4.96\times 10^{-26}$ and an
optimal \ac{SNR} of $48$.  For comparison, the weakest of the closed
signals were pulsar 90 with $h_0=6.84\times 10^{-26}$ and pulsar 64
with an optimal \ac{SNR} of $71$.

\begin{figure} \centering
\includegraphics[width=\columnwidth]{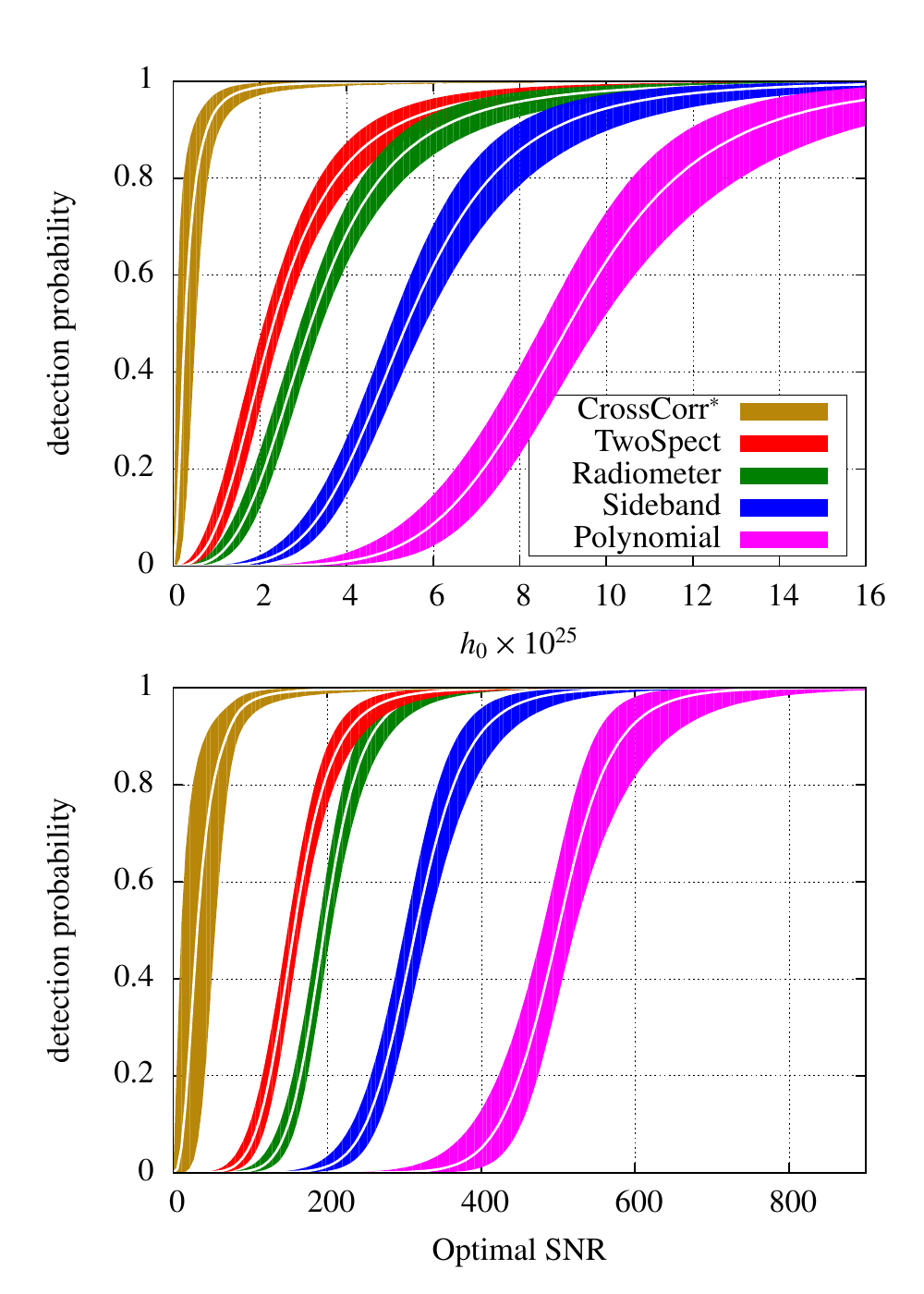} 
\caption{The upper and lower plots show the detection efficiencies for
  each pipeline as a function of $h_0$ and multi-detector optimal
  \ac{SNR} respectively. The shaded regions represent the 50\%
  uncertainty (inter-quartile range) after
  marginalizing over the parameters of a basic sigmoid
  function $f(x)=(1+e^{-\alpha (\log(x)-\beta)})^{-1}$ using
  the posterior distribution generated from the 50
  detection/non-detection results from the closed signal bands.
  The posterior was constructed using a Jeffreys prior on $\alpha$
  and $\beta$ so that the inferred efficiencies are all less than
  unity, even for the CrossCorr pipeline (whose results were
  obtained in self-blinded mode several months after the others and
  half a year beyond the nominal end date of the \ac{MDC}, as detailed in
  \tref{tab:submissionTimeline}), which detected all 50 closed signals.
  However, the exact turnover point of the CrossCorr efficiency curve
  is more uncertain and less robust against changes of the fitting
  procedure than the others.
  \label{fig:efficiency}}
\end{figure}

Further comparison between pipelines is shown in
\fref{fig:efficiency} where detection efficiency versus the \ac{GW}
strain $h_{0}$, and the optimal \ac{SNR} respectively are plotted.
Detection efficiency is defined as the fraction of signals claimed as
detected at the chosen confidence ($p<10^{-2}$) as a function of the
value indicated on the $x$-axis.  For example, the Sideband search
achieves a detection efficiency of $\approx 0.8$ at $h_{0}=7\times
10^{-25}$. The efficiency curves and their uncertainties are obtained
by marginalizing over the parameters of a basic sigmoid function using
the posterior distribution generated from the 50
detection/non-detection results from the closed signal bands.  (Note
that although the CrossCorr pipeline detected all 50 signals, the
Jeffreys prior used for the sigmoid parameters prevents the posterior
from implying 100\% efficiency at all signal strengths, as it would
with a maximum likelihood method.  The inferred sigmoid parameters for
CrossCorr are still somewhat arbitrary and dependent on this choice of
prior, however.)
At 50\% detection efficiency, the scaling in $h_{0}$ sensitivity
relative to the weakest signal in the closed data set
is $\approx$0.40, 3.4, 4.6, 7.9, and 13
for CrossCorr, TwoSpect, radiometer, sideband and polynomial,
respectively.  In terms of \ac{SNR} at 50\% efficiency, these numbers
are $\approx$0.44, 2.2, 2.8, 4.4, and 7.0.

\subsection{Parameter Estimation}
\label{sec:param}

The parameter estimation abilities of each pipeline are varied and range from
the minimum state of inference: only estimating the signal frequency, up to the
maximum state: estimation of frequency, orbital semi-major axis and time of
ascension, and the strain amplitude.  None of the pipelines performed
additional parameter inference on $\cos\iota, \psi$ or $\phi_{0}$.  None
treated the orbital period as a search parameter and hence they do not refine
this estimate beyond the initial known prior distribution.  There was only very
limited candidate follow-up analyses to potentially enhance parameter
estimation via, {\it e.g.}, analysis of additional data, or deeper analysis
over localized parameter space regions around candidates.  (CrossCorr employed
a limited narrow-parameter-band analysis with a longer coherence time on three
of the quietest detections.  This method was developed to confirm marginal
detections and was used for that purpose on one of the open signals.  While it
was not necessary for the closed-signal detections, it did provide more
accurate parameter estimates as well as more confident detections, and could in
principle have been used more widely.) The details of the estimated parameter
values from all pipelines can be found in \aref{sec:fullresults}.
\begin{figure*}
\centering
\includegraphics[width=\textwidth]{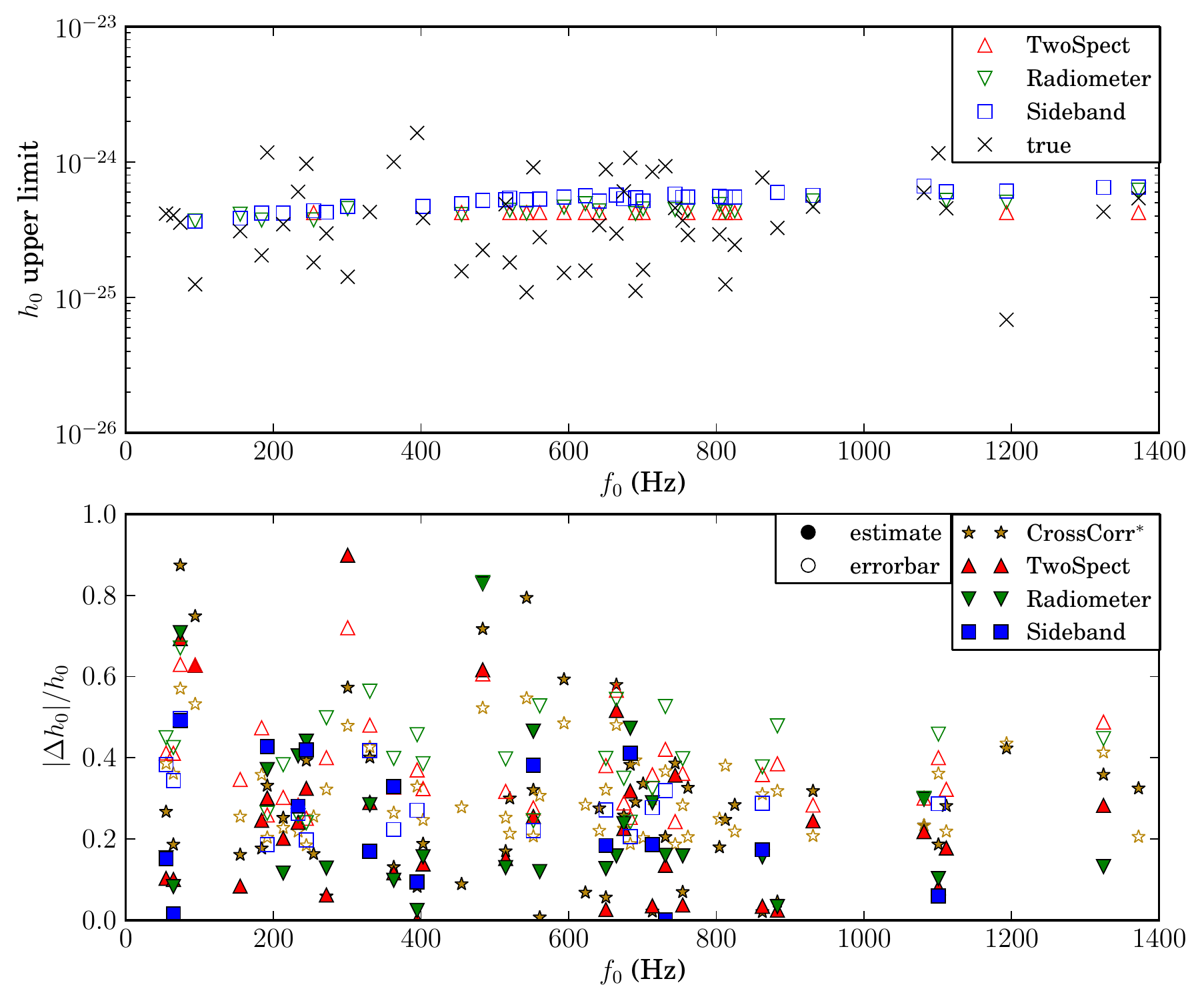}
\caption{Comparison of signal amplitude estimates (from detected
  signals) and upper limits (from non-detected signals) from each
  pipeline as a function of intrinsic signal frequency.
  The top panel
  shows the $95\%$ confidence upper-limits on $h_{0}$ for non-detected
  signal from each pipeline that provided such results.  The black
  crosses indicate the true value of $h_{0}$.
  In the bottom panel, the solid
  symbols in the top panel show the fractional errors in $h_{0}$
  estimates, and open symbols show the quoted one-sigma errorbars,
  again divided by the true $h_{0}$ value.  The uncertainty in $h_{0}$
  is comparable for all pipelines which provided estimates, since all
  were dominated by the unknown value of $\cos\iota$.  The complete details of
  the amplitude estimates and upper limits can be found in
  \aref{sec:fullresults}.\label{fig:ampcomp}}
\end{figure*}
\begin{figure*}
\centering
\includegraphics[width=\textwidth]{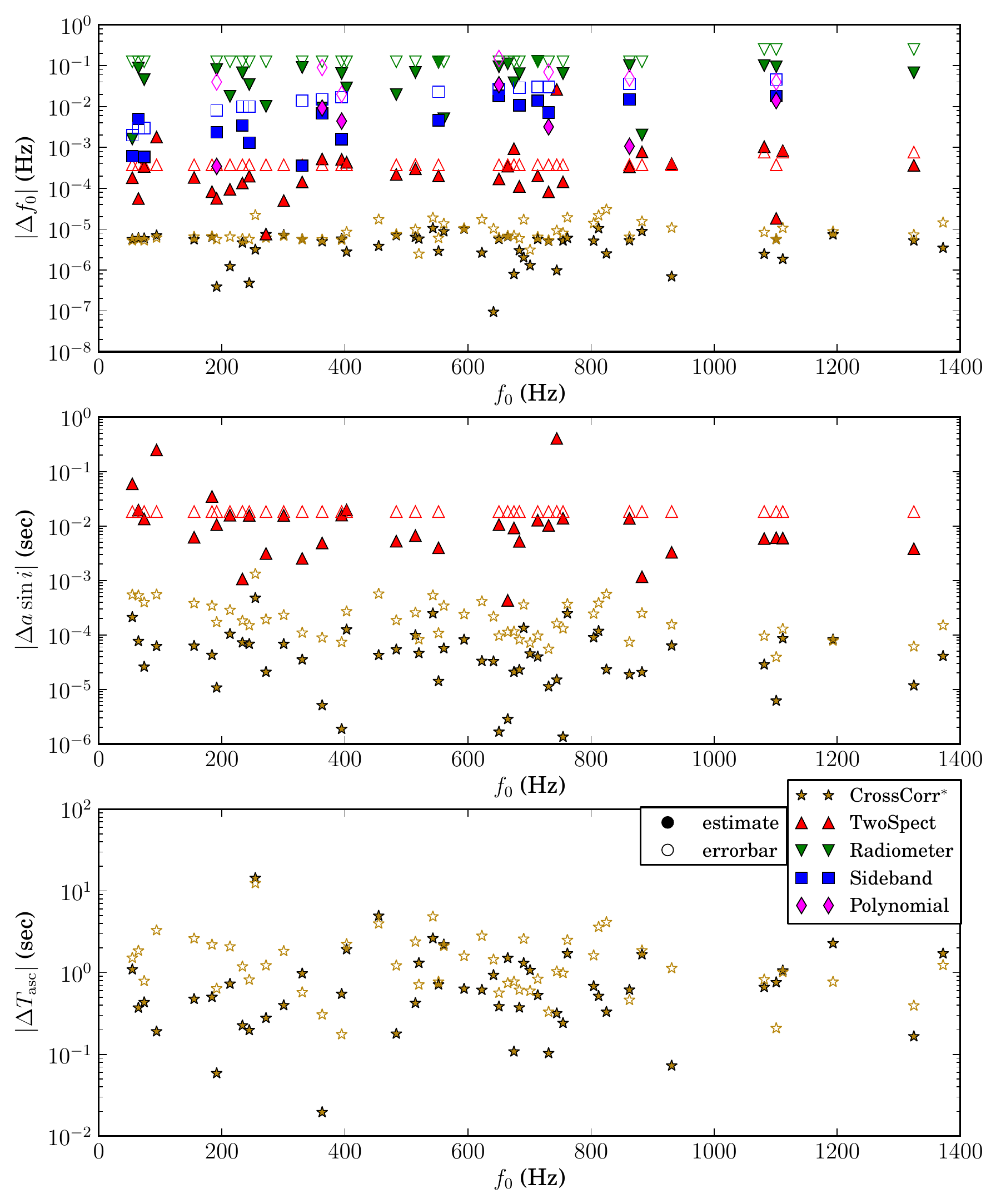}
\caption{Comparisons of parameter estimation for detected signal from
  each pipeline as a function of intrinsic signal frequency.  The
  solid symbols in each plot show the difference between the true and
  estimated values of intrinsic signal frequency, projected semimajor
  axis, and time of passage of the ascending node for the detected
  signals for pipelines which report those quantities.  The open
  symbols show the quoted one-sigma errorbars corresponding to each
  estimate.  The complete details of the parameter estimates can be
  found in \aref{sec:fullresults}.  The causes of
  notable outliers seen in the estimation of the orbital semi-major
  axis by TwoSpect are discussed in
  \sref{sec:param}. \label{fig:paramcomp}}
\end{figure*}

\Fref{fig:ampcomp} shows the fractional error in the estimates
of $h_{0}$, along with the quoted errorbars, for the detected signals
and the upper-limits set for those signals not
detected.  Note that the TwoSpect pipeline sets a fixed upper-limit value for
each non-detected signal at the level of $4.23\times 10^{-25}$, and the
Polynomial pipeline does not produce estimates or upper-limits for $h_{0}$.
(The CrossCorr
pipeline detected all 50 signals and therefore had no upper limits to report.)
It is clear that for non-detections, there is very little spread between pipelines
in the resulting upper-limits.  Typically, these values vary between pipelines
by of order of tens of percent with TwoSpect and the Radiometer searches consistently setting
the most stringent upper-limits. As can be seen from the second panel in
\fref{fig:ampcomp}, in all cases for detected signals, the estimated $h_0$
values are consistent with the true values given each pipeline's reported
uncertainties.
Additionally, the $h_0$ uncertainties are comparable for all
pipelines.  This is because these searches are all sensitive not to
$h_0$ but to a combination of $h_0$ and $\cos\iota$ known as
$h_0^{\text{eff}}$ and given by
\begin{equation}
  \label{eq:h0eff}
  (h_0^{\text{eff}})^2
  = h_0^2\,\frac{[(1+\cos^2\iota)/2]^2 + [\cos\iota]^2}{2}
\end{equation}
This is equal to $h_0^2$ for circular polarization ($\iota=0^\circ$ or
$180^\circ$) and $\frac{h_0^2}{8}$ for linear polarization
($\iota=90^\circ$), and has an average value of $\frac{2}{5}h_0^2$
when averaged isotropically over the inclination angle $\iota$.
The uncertainty in the value of $\cos\iota$
dominates the other measurement errors for $h_0$ in each of the
pipelines.

The first panel of \fref{fig:paramcomp} gives the difference between the true
and estimated intrinsic \ac{GW} frequency.  Note that for search frequencies
below $\sim$1~kHz, the Radiometer search returns a fixed estimate for frequency uncertainty
of $\pm 0.125$~Hz based on the size of the frequency bins used in the Radiometer
analysis. Beyond this frequency, where the signal is likely to span two frequency bins, the uncertainty is increased to $\pm 0.25$~Hz.  These uncertainties are conservative.  
In
all but 1 case, the Radiometer analysis correctly identifies the signal frequency bin.
For pulsar 65, two adjacent bins yielded $p$-values below the detection
threshold and the lower of the two was selected as the candidate signal.
The true signal location, however, was within the rejected bin. The TwoSpect
search has the best frequency accuracy of the original four
pipelines of the \ac{MDC}, and in the majority of cases
is consistent with the true values to within their quoted error bars.
There is one notable outlier, however, for
a low-\ac{SNR} detection.  The Sideband search's claimed uncertainties are
conservative and appear to have overestimated uncertainties since all 18
detected signals lie within the 1-$\sigma$ error bars. The Polynomial search
frequency estimates are consistent within its estimated uncertainties with the
true values for all 7 of its corresponding detected signals.  The CrossCorr
search produces considerably smaller error bars than all of the
original 4 \ac{MDC} pipelines, and the errors in its frequency estimates are consistent with
those uncertainties.  CrossCorr's parameter space precision is
due in part to its method of finding the best fit parameters by interpolation
rather than reporting the grid point with the highest statistic value.

The final two panels of \fref{fig:paramcomp} represent the orbital
parameter estimation ability of the TwoSpect and CrossCorr pipelines.
These two searches
reported the projected orbital semi-major axis, while CrossCorr was the
only pipeline to estimate the time of ascension $\Tasc$.
The interpolation performed by CrossCorr allows it to obtain parameter
estimates with smaller error bars, with a resolution finer than its final grid
spacing.
For TwoSpect, of the
majority of signals that are detected, the $\asini$ estimate is
consistent with the true values and uncertainties are
$\mathcal{O}(0.02)$~s (representing a $\approx$1.5\% error). This indicates a
potential improvement over the known prior observational uncertainty by a factor of
${\sim}$10. There are however, 2 notable outliers in which $\asini$ is
significantly underestimated.
This occurs when a strong signal in one detector matches a weak signal in another;
reading parameters off the highest $p$-value template does not always yield
an accurate estimate in this marginal case. 
Further refinement of coincident parameter estimation, or parameter estimation
in a future coherent mode, may correct this problem.

\begin{figure*}
  \centering
  \includegraphics[width=\textwidth]{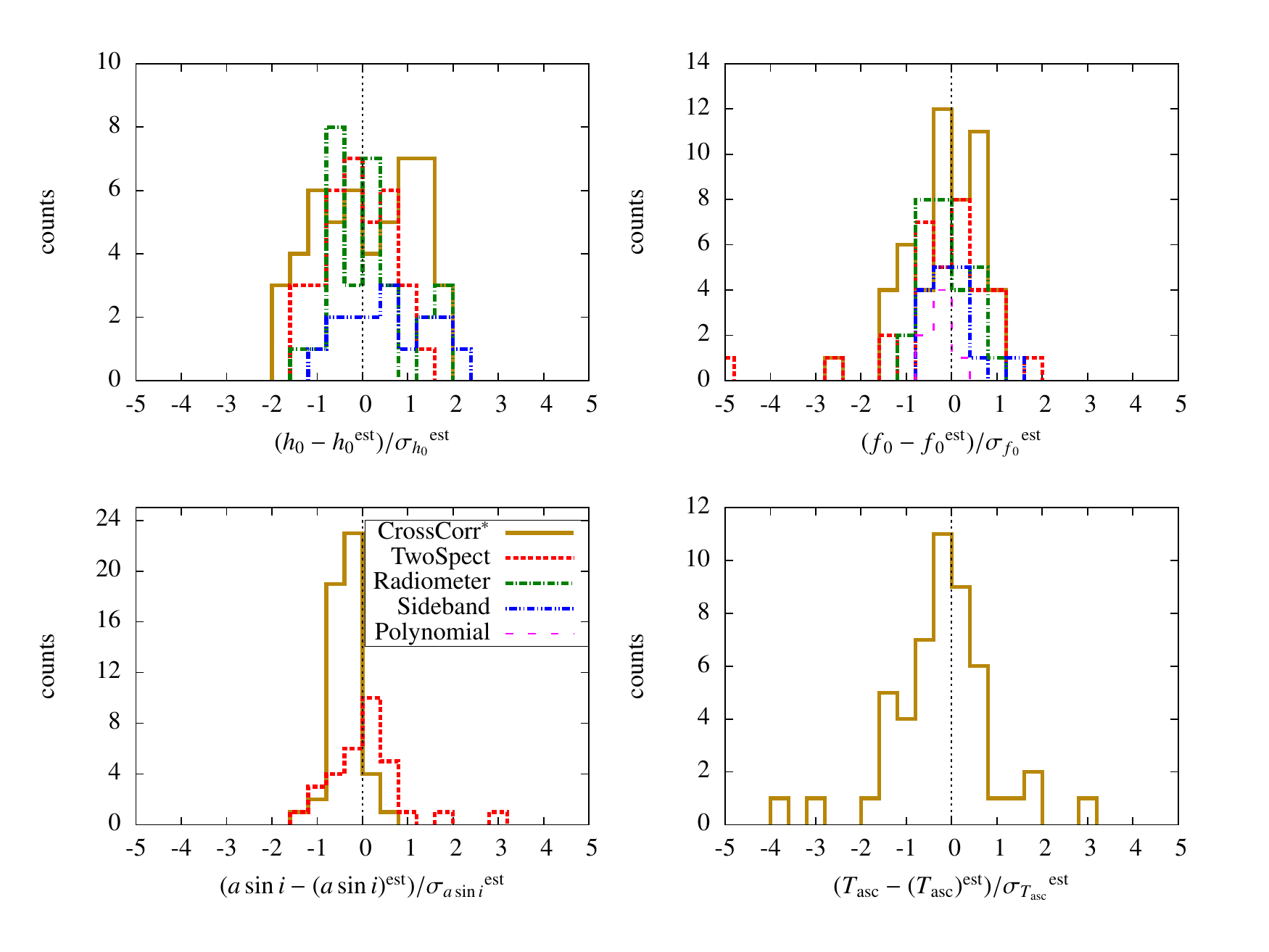}
  \caption{Distributions of estimated parameter offsets relative to
    their estimated measurement uncertainties for the detected
    \ac{MDC} signals.  Only TwoSpect and CrossCorr return an estimate
    of the projected semimajor axis, and only CrossCorr returns an
    estimate of time at the ascending node.  None of the pipelines
    return period estimates.}\label{fig:allerrdist}
\end{figure*}

In \fref{fig:allerrdist}, distributions of parameter estimation
offsets from the true values and rescaled by the estimated measurement
uncertainty are shown.  It is expected to observe
distributions that are proportional to zero mean, unit
variance, Gaussian distributions.  For the estimates of the signal strain
amplitude $h_{0}$ for which we have results from four algorithms, the
distributions are generally consistent with the expected Gaussian.
This is also the behavior for the estimates of $\asini$ from the
TwoSpect algorithm, although it should be noted that 2 outliers
(visible in the middle panel of \fref{fig:paramcomp}) have been
omitted from this plot.  The $\asini$ errorbars reported by
CrossCorr appear to be somewhat larger than the typical actual
offsets, which is to be expected from the inclusion of the bias
correction as a conservatively-estimated source of error.
For the intrinsic
\ac{GW} frequency estimation, deviations are observed from the expected behavior for
the Sideband, Polynomial, and Radiometer approaches whereas the TwoSpect
and CrossCorr algorithms
are broadly consistent with expectations.  In all cases, matching the
expectations means that one can infer that the 1--$\sigma$ uncertainty estimates are valid.  For the
Polynomial, Sideband, and Radiometer frequency results, there is an apparent bias towards
overestimation of the 1--$\sigma$ errors (this is expected in the case of the Radiometer).  It should be noted that in the
Polynomial case, there are only seven detected signals from which the
distribution can be constructed and hence these results are subject to large
statistical uncertainties. 

\begin{figure}
  \centering
  \includegraphics[width=\columnwidth]{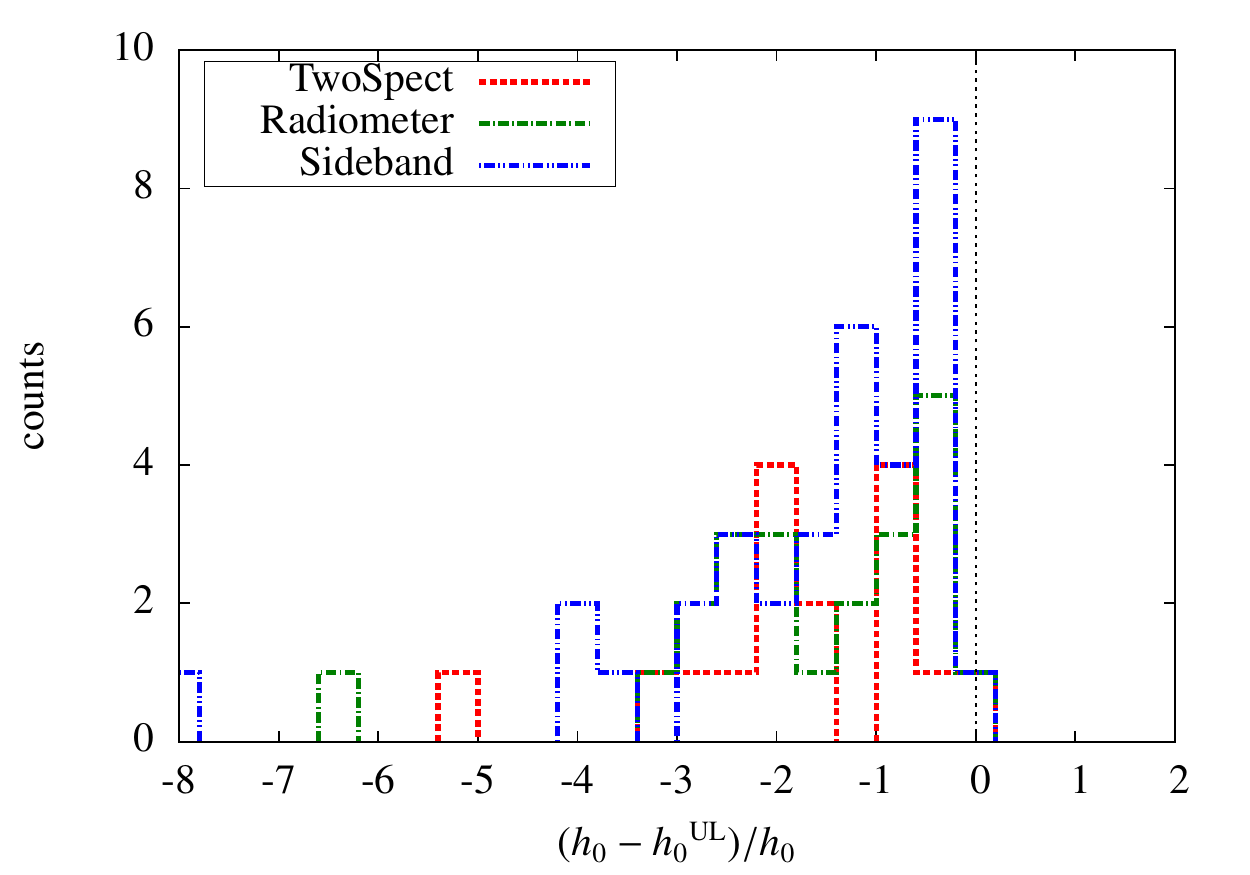}
  \caption{Distribution of offsets between $h_0$ upper limits and
    actual values non-detected \ac{MDC} signal. The polynomial search
    returns only frequency estimation for detection and does not
    return an $h_0$ upper-limit, while CrossCorr had no non-detections
    among the 50 closed signals.}\label{fig:ULerrdist}
\end{figure}

For the Sideband frequency estimates, a narrow
distribution is obtained implying that the algorithm produces overly conservative
uncertainties. This is expected due to the difficulty of
converting the intrinsically multi-modal Sideband detection statistic
into an equivalent single-mode uncertainty. In the limit of high
\ac{SNR}, the maximum Sideband detection statistic is expected to
belong to the frequency bin coincident with the true intrinsic \ac{GW}
frequency.  This frequency bin has a relatively narrow width in
comparison to other algorithms and is given by $\approx 1/T$ where $T$
is the length of the observation (so in this case $\sim 10^{-6}$~Hz).
For lower \ac{SNR} (still detectable signals), the maximum could originate
from Sideband statistics from integer multiples of $\pm 1/P$~Hz away
from the true value.  The total segmented space will contain
$\mathcal{O}(10^{-6})$~Hz but for \ac{ScoX1} will span
$\mathcal{O}(10^{-4})$~Hz. Based on the range of offsets between the
true frequency and those associated with the maximum statistic
observed in the open data set a conservative value of
$0.042(f/1\,\text{kHz})$~Hz was chosen for the error on
frequency.

In \fref{fig:ULerrdist}, the distribution
of $h_{0}$ upper-limit offsets relative to the true $h_{0}$ value are shown.
The expected form of such a distribution is an unknown function of the
original \ac{SNR} distribution of the \ac{MDC} signals and the search
algorithm in question. An expected property of this
distribution, however, is that, given a 95\% confidence on
the upper-limit value, 5\% of the quoted values should be greater
than the true $h_{0}$ value.  Given that each of the 3 algorithms that
reported $h_{0}$ upper-limits only did so on a limited number of
undetected signals it is expected that $\mathcal{O}(1)$ of the upper limits would
lie below the true strain value, which is consistent with observations.

\subsection{Computational cost of the \ac{MDC} analysis}

In one sense it would be desirable to compare each algorithm at fixed
computational cost, therefore separating the efficacy of the algorithm
from the computing power used.  The implicit restriction on
computational expense applied to the \ac{MDC} analysis was limited by
the availability of computational resources (equally available to all
participants) and the length of the challenge itself.  

The current implementations of both the Radiometer and the Sideband
searches are not computationally limited. For the Radiometer search
this means that choices made in the algorithm design result in the
ability to analyze \emph{all} \ac{MDC} data over the entire parameter
space on a single machine in $O(\text{hours})$.  This constitutes a
tiny fraction of the available computational power. The
post-processing of the results is equally cheap due to the relative
insensitivity of the Radiometer in parameter estimation.  For this
particular \ac{MDC} much effort through additional, relatively cheap
but time consuming injection studies (restricted to the Radiometer
analysis) was required to calibrate the Radiometer, designed for
stochastic signals, for continuous waves. Future Radiometer
implementations will benefit from source-specific tunings which will
likely increase the computational cost but will maintain the algorithm
status as not computationally limited.

The Sideband search is also not computationally limited.  However,
this is due to its high sensitivity to spin-wandering and hence there
is a self-imposed limit of 10 days observation.  The two main search
stages can then be run in a $O(10)$~hours using $\sim$100
machines (within a cluster).  Post-processing of the results for
parameter estimation adds an additional comparable cost.  Whilst this
is considerably more costly than the Radiometer algorithm, it
similarly costs far less than the available computational power.

The Polynomial search is highly computationally limited, primarily due
to its design as an all-sky search; the number of templates cannot be
reduced substantially using the known parameters available for \ac{ScoX1}.
Furthermore, the \ac{SFT} length was increased compared to the all-sky
search strategy, which increases sensitivity at the cost of further
increasing computation cost. The relative immaturity as an established
algorithm may have contributed further to the time required. For this
\ac{MDC}, only 56~days of simulated data from just one of the three
available interferometers was analyzed (H1). This allows
all potential \ac{MDC} signals to be analyzed, but at a reduced
sensitivity compared to analysis of the full data set. With this,
restriction, the computational cost was limited to approximately $10^6$~core-hours.

For TwoSpect the \ac{MDC} search over the 100 signals (open and
closed) required $\sim$few weeks of wall-clock processing time using $O(10^3)$
\acp{CPU}. Directed search post-processing was developed as part of
the \ac{MDC}, and the additional cost required was $\sim$few days in
total on one \ac{CPU}.  The main tunable algorithmic parameters of TwoSpect in a 
directed search for \ac{ScoX1} are 
template spacings for the tested frequency and projected semi-major axis, 
along with the \ac{SFT} coherence time. The template spacings were chosen
to give a mismatch no greater than 0.20.
The \ac{SFT} length was chosen to be 360~s for higher frequency bands, above 360~Hz, and 840~s for lower frequency bands.
These \acp{SFT} kept most of the modulated signal frequency contained within one bin per \ac{SFT}.
Ideally, \acp{SFT} are as long as possible, just short of where spectral leakage would occur; we chose to restrict ourselves to two sets of \acp{SFT} since the cost of generating more would be high for relatively low gains in sensitivity.

As described in \sref{sec:compcost}, the computational cost for
CrossCorr grows with search frequency, but can be tuned by reducing
the coherence time $\Tmax$.  To
maintain approximately the same computing cost for each signal band, a
range of $\Tmax$ values were used, as listed in
\tref{tab:sftparams}.  The total computing cost for the fifty
closed signal bands in this setup was approximately $20,\!000$~\ac{CPU}-days.

\section{Discussion}\label{sec:discussion}

We have considered five search pipelines presently available to the \ac{GW}
community that are capable of searching for the continuous \ac{GW} emission
from \ac{ScoX1}. A general overview of each pipeline has been presented with
regards to sensitivity, computational efficiency, parameter space dependencies,
and parameter estimation capabilities. To compare these methods, an \ac{MDC}
was performed that included 50 unknown simulated signals consistent with the
known \ac{ScoX1} parameter space. Each algorithm has presented its results of the
50 signals in terms of detection status and signal parameter estimation in the
event of a detected signal exceeding a predefined false-alarm threshold. These
results were then used to compare the algorithms and to elucidate unforeseen
strengths and weaknesses in each approach. We expect each team will employ
improvements to the pipelines in future versions.

Perhaps the most critical figure of merit for each pipeline is the detection
efficiency. Among the four original pipelines that ran in the \ac{MDC}, the clear leader in this category was the
TwoSpect algorithm, which detected 34 of the 50 closed signals. The next most sensitive algorithm was
the Radiometer search with 28 detections, followed by the Sideband search with
16, and finally the Polynomial approach with 7. The CrossCorr pipeline, which completed its
preliminary, self-blinded analysis seven months after the original MDC deadline,
as detailed in \tref{tab:submissionTimeline}
detected all 50 closed signals. The definition for detection in
the \ac{MDC} is less stringent than typically used in continuous \ac{GW}
searches, and, in the presence of non-Gaussian noise (as is usually the case
with real \ac{GW} detectors), a higher threshold would be used. Since the focus
of the \ac{MDC} was on algorithm comparison, this choice acted as a
discriminator between pipelines but should not necessarily be used as a true
indicator of detectable signal strengths in real data.

Whilst it is difficult in general to rank the algorithms with respect to some
measure of sensitivity at fixed computational cost, it is very clear that the
Radiometer algorithm does the best with limited computational power.
It uses less computational power than
the Sideband search and yet claims detection on more signals.  In comparison to
the TwoSpect and CrossCorr algorithms, which consumed the most computational
resources, the Radiometer search detected $82\%$ of the TwoSpect
signals and $56\%$ of the CrossCorr signals using $<1\%$ of the
computational power of either search.

The weakest signal in this relatively small \ac{MDC} sample had
strain $h_{0}\approx 6.8\times 10^{-26}$ at $1190$\,Hz, and the weakest
detected by the four pipelines which ran during the original \ac{MDC} time
frame had $h_{0}\approx1.4\times 10^{-25}$ at $300$ Hz which is slightly greater
than the torque-balance limit for \ac{ScoX1} at a \ac{GW} signal frequency of 50
Hz and almost an order of magnitude higher than the corresponding limit at a
\ac{GW} signal frequency of 1500 Hz.  From this perspective it is clear that
improvements must be made to these algorithms in order to make detection a
possibility and to start to set astrophysically interesting constraints on
\ac{GW} emission from \ac{ScoX1}.  This is also true of the CrossCorr pipeline,
which detected all of the closed signals, but missed one open signal with
$h_0=3.8\times 10^{-26}$, indicating that its detectability threshold was
likely slightly above the torque balance level as well, at least for the
coherence times considered in this analysis.

The Sideband search suffers the most from the potential effects of
spin-wandering in the \ac{ScoX1} system.  This issue has not been considered in
this analysis other than to acknowledge its possible effects and to assume,
rather than model, its presence.  Multiple 10-day data segments can, in principle, 
be added semi-coherently to improve the sensitivity of the Sideband search \cite{S5sideband}.
Other algorithms will most likely, at some
point in their development, have to account for this feature.  To do so, we must
be able to accurately quantify its realistic behavior and to model it in our
simulations.  The point at which an algorithm attains a frequency resolution at
the level at which spin-wandering is expected to vary is the point at which new
data analysis techniques are required. For the Radiometer and Polynomial
searches this is unlikely to ever be the case which makes them attractive in
their robustness.  For CrossCorr, the level of spin wandering
described in the \ac{MDC} is likely to limit the coherence time $\Tmax$
to about 12\,hours\cite{LMXBCrossCorr}, which could pose limitations on
future searches with even longer coherence times.  For TwoSpect  the
effect requires further study and for the Sideband search, the 10~day
observation limit is already a constraint imposed by spin-wandering.

In a realistic search, outliers would typically be further analyzed
using different follow-up methods. This can be as simple as analyzing
the same set of data using other analysis pipelines to verify the
presence of a putative signal, or the follow-up can involve multiple,
hierarchical steps to refine and improve the signal-to-noise ratio of
a candidate signal.

All of the algorithms presented here have planned improvements for the advanced
detector era.  It is currently unclear how much the detection sensitivity could
improve from each of the analyses. Investigations into the following specific
enhancements are already underway.
\begin{itemize}[itemsep=-1mm]
\item \textit{TwoSpect}: coherently combining \acp{SFT} from H1, L1,
and V1 prior to the second Fourier transform step should enable
improvements in the detectable $h_0$. Indeed, initial tests indicate an
additional 7 could be detected of the 16 closed signals that were
originally missed by TwoSpect. A coherent analysis of the second Fourier
transform should also yield non-negligible improvement in detection efficiency
as well. It is also possible that a finer grid spacing and/or interpolation could be used
to improve the uncertainties in parameter estimates, although
the improvement in $f_{0}$ and $\asini$ estimation may be limited by noise
fluctuations.
\item \textit{Radiometer}: applying a variable-size frequency window rather
than fixed $0.25$ Hz bins will improve sensitivity.  The width of the window
will scale proportionally with frequency and be tuned to the expected
modulation depth of the \ac{ScoX1} signal.  It will also be overlapped to reduce
the chance of mismatch with a possible signal.
\item \textit{Sideband}: Combining the results from multiple short observations
(rather than relying on a single short observation) will improve sensitivity.
There is also the possibility to perform partial orbital phase demodulation in
the coherent $\mathcal{F}$-statistic stage of the analysis.  This would take
advantage of prior orbital phase knowledge which is currently ignored by all
algorithms taking part in this \ac{MDC}.
\item \textit{Polynomial}: With improved algorithm and implementation efficiency,
as well as the use of \acp{GPU} it will be possible to analyze a larger subset
of the data, therefore improving sensitivity. For parameter estimation a
secondary search can be launched with smaller template grid spacing to refine
the frequency of signal candidates. It may also be possible to use multiple
detectors to triangulate a signal and estimate the sky location. This would also
enable an estimate of the $h_0$ of the \ac{GW}. Work to extend templates to span
multiple coherent intervals is also planned.
\item \textit{CrossCorr}: since the search is computationally limited,
  the primary planned approach to improve the sensitivity is to
  increase the speed of the code.  Any such increase in speed would
  allow longer coherence times, and therefore improved sensitivity, at
  the same computing power.  Possible avenues range from
  reorganization of the computation of weightings related to spectral
  leakage, to leveraging vectorized hardware instructions such as AVX
  or SSE (currently employed by the TwoSpect pipeline), to a more fundamental reorganization of the
  calculation to be performed in the time domain with the
  ``resampling'' method defined in
  \cite{1998PhRvD..58f3001J,T0900149-v5}.  Enhancements not related to
  speed may include filtering with multiple templates to make the
  search sensitive to other combinations of the amplitude parameters
  besides $h_0^{\text{eff}}$.
\end{itemize}

In addition to the algorithms that took part in this \ac{MDC}, the
stacked $\mathcal{F}$-statistic approach discussed
in \sref{sec:stackedF} has great potential to
exceed the performance of the five main algorithms presented in this paper.
Recent sensitivity estimates published for a hypothetical search with
days-long coherence time\cite{2015arXiv150200914L} are especially encouraging.
It should be stressed that this search in development
should be compared to the future sensitivities of the existing algorithms.
Likewise, the CrossCorr algorithm, which produced the best results
on the \ac{MDC} data included in this paper, had the advantage of running
later, so some care should be taken when comparing its results to the
four methods run during the \ac{MDC} timeframe, which have been
undergoing enhancements since then.  Also, the CrossCorr pipeline is
still relatively immature, and its performance on actual
interferometer data is yet to be tested.

The work presented here has been a valuable first step towards validating our
algorithms, understanding their uncertainties, quantifying our detection
criteria (albeit in Gaussian noise), and gauging our best sensitivity to Sco
X-1.  We intend to build on this work, and at the time of writing this
manuscript, are deciding on the format and features of the next \ac{MDC}.
Among other improvements, the three main advancements we plan to make are the
addition of spin-wandering to our simulated signals, the inclusion of
non-Gaussian noise (most likely, rescaled real 1st generation detector noise),
and the use of signal amplitudes at, or below, the torque balance limit.  This
will constitute a far greater challenge to the participants, but will allow us
to transition from primarily comparing our pipelines to being able to make
predictions about astrophysically realistic scenarios in the advanced detector
era.

\begin{acknowledgments}
  The authors are grateful to Badri Krishnan, Paola Leaci, and
  Reinhard Prix for useful
  discussions and comments.  YZ acknowledges the hospitality of the
  Max Planck Institute for Gravitational Physics (Albert Einstein
  Institute) in Hannover, Germany.  SGC was supported by NSF grant PHY-1204944.  JTW was supported by NSF grants
  PHY-0855494 and PHY-1207010.  YZ was supported by NSF grant
  PHY-1207010. GDM was supported by NSF grants PHY-0855422 and
  PHY-1205173.  PDL and AM were supported by Australian Research
  Council (ARC) Discovery Project DP110103347.  PDL is also supported
  by ARC DP140102578. HJB and RJ are supported by the research programme of
  the Foundation for Fundamental Research on Matter (FOM), which is part of
  the Netherlands Organisation for Scientific Research (NWO).
    The analysis for several of the searches in this project were
  performed on the Atlas computing cluster at AEI Hannover, which was
  funded by the Max Planck Society and the State of Niedersachsen,
  Germany.
This paper has been assigned LIGO Document Number
  \dcc.
\end{acknowledgments}

\appendix

\section{Complete MDC results\label{sec:fullresults}}

This section contains tables given the complete results of the
searches, which are summarized in \sref{sec:results}.  The specific
signals detected by each pipeline, summarized in
\fref{fig:detoverview} and \fref{fig:efficiency},
\fref{fig:allerrdist}, and \fref{fig:ULerrdist} are listed in
Table~\ref{tab:detres}.  (Note that the details of the signals
themselves are in \tref{tab:mdcparams}.)  The upper limits (for
non-detected signals) and estimates (for detected signals) of $h_0$,
summarized in \fref{fig:ampcomp}, are detailed in \tref{tab:h0res} and
\tref{tab:ULres}, respectively.  The estimates on the parameters
$f_0$, $\asini$ and $\Tasc$, summarized in \fref{fig:ampcomp} and
\fref{fig:allerrdist}, are shown in \tref{tab:f0res},
\tref{tab:asinires}, and \tref{tab:Tascres}, respectively.
\begin{table}
\caption{Comparison of signal detections.\label{tab:detres}}
\makeatletter{}\begin{tabular}{cc|ccccc}
\hline
index & opt SNR & CrossC.$^*$ & TwoSp. & Radiom. & Sideb. & Polyn. \\
\hline
1 & 335 & yes & yes & yes & yes & no \\
2 & 310 & yes & yes & yes & yes & no \\
3 & 427 & yes & yes & yes & yes & no \\
5 & 142 & yes & yes & no & no & no \\
11 & 168 & yes & yes & no & no & no \\
14 & 157 & yes & yes & no & no & no \\
15 & 526 & yes & yes & yes & yes & yes \\
17 & 159 & yes & yes & yes & no & no \\
19 & 292 & yes & yes & yes & yes & no \\
20 & 407 & yes & yes & yes & yes & no \\
21 & 96 & yes & no & no & no & no \\
23 & 205 & yes & yes & yes & no & no \\
26 & 144 & yes & yes & no & no & no \\
29 & 385 & yes & yes & yes & yes & no \\
32 & 554 & yes & yes & yes & yes & yes \\
35 & 1142 & yes & yes & yes & yes & yes \\
36 & 194 & yes & yes & yes & no & no \\
41 & 92 & yes & no & no & no & no \\
44 & 246 & yes & yes & yes & no & no \\
47 & 248 & yes & yes & yes & no & no \\
48 & 94 & yes & no & no & no & no \\
50 & 127 & yes & no & no & no & no \\
51 & 400 & yes & yes & yes & yes & no \\
52 & 190 & yes & no & yes & no & no \\
54 & 155 & yes & no & no & no & no \\
57 & 96 & yes & no & no & no & no \\
58 & 155 & yes & no & no & no & no \\
59 & 607 & yes & yes & yes & yes & yes \\
60 & 304 & yes & yes & yes & no & no \\
61 & 269 & yes & yes & yes & no & no \\
62 & 457 & yes & yes & yes & yes & no \\
63 & 92 & yes & no & no & no & no \\
64 & 71 & yes & no & no & no & no \\
65 & 528 & yes & yes & yes & yes & no \\
66 & 729 & yes & yes & yes & yes & yes \\
67 & 192 & yes & yes & no & no & no \\
68 & 227 & yes & yes & yes & no & no \\
69 & 125 & yes & no & no & no & no \\
71 & 155 & yes & no & no & no & no \\
72 & 98 & yes & no & no & no & no \\
73 & 111 & yes & no & no & no & no \\
75 & 495 & yes & yes & yes & yes & yes \\
76 & 214 & yes & yes & yes & no & no \\
79 & 211 & yes & yes & no & no & no \\
83 & 277 & yes & yes & yes & no & no \\
84 & 891 & yes & yes & yes & yes & yes \\
85 & 199 & yes & yes & no & no & no \\
90 & 76 & yes & no & no & no & no \\
95 & 376 & yes & yes & yes & no & no \\
98 & 232 & yes & no & no & no & no \\
\hline
\end{tabular}
 
\caption*{The detection status (at 1\% false alarm probability) for
  each search algorithm as a function of signal index and optimal
  \ac{SNR}.  Note that the CrossCorr pipeline results were
  obtained in self-blinded mode several months after the others and
  half a year beyond the nominal end date of the \ac{MDC}, as detailed in
  Table~\ref{tab:submissionTimeline}.
}
\end{table}
\begin{table}
\caption{Comparison of $h_{0}$ upper-limits.\label{tab:ULres}}
\makeatletter{}\begin{tabular}{cc|ccc}
\hline
index & $h_{0}\times 10^{25}$ & TwoSpect & Radiometer & Sideband \\
\hline
1 & 4.160 & -- & -- & -- \\
2 & 4.044 & -- & -- & -- \\
3 & 3.565 & -- & -- & -- \\
5 & 1.250 & -- & $3.67$ & $3.646$ \\
11 & 3.089 & -- & $4.12$ & $3.853$ \\
14 & 2.044 & -- & $3.73$ & $4.206$ \\
15 & 11.764 & -- & -- & -- \\
17 & 3.473 & -- & -- & $4.188$ \\
19 & 6.031 & -- & -- & -- \\
20 & 9.710 & -- & -- & -- \\
21 & 1.815 & $4.23$ & $3.74$ & $4.369$ \\
23 & 2.968 & -- & -- & $4.262$ \\
26 & 1.419 & -- & $4.53$ & $4.714$ \\
29 & 4.275 & -- & -- & -- \\
32 & 10.038 & -- & -- & -- \\
35 & 16.402 & -- & -- & -- \\
36 & 3.864 & -- & -- & $4.713$ \\
41 & 1.562 & $4.23$ & $4.10$ & $4.944$ \\
44 & 2.237 & -- & -- & $5.196$ \\
47 & 4.883 & -- & -- & $5.270$ \\
48 & 1.813 & $4.23$ & $4.41$ & $5.420$ \\
50 & 1.093 & $4.23$ & $4.19$ & $5.282$ \\
51 & 9.146 & -- & -- & -- \\
52 & 2.786 & $4.23$ & -- & $5.312$ \\
54 & 1.518 & $4.23$ & $4.64$ & $5.523$ \\
57 & 1.577 & $4.23$ & $4.92$ & $5.644$ \\
58 & 3.416 & $4.23$ & $4.36$ & $5.155$ \\
59 & 8.835 & -- & -- & -- \\
60 & 2.961 & -- & -- & $5.691$ \\
61 & 6.064 & -- & -- & $5.356$ \\
62 & 10.737 & -- & -- & -- \\
63 & 1.119 & $4.23$ & $4.17$ & $5.449$ \\
64 & 1.600 & $4.23$ & $4.51$ & $5.168$ \\
65 & 8.474 & -- & -- & -- \\
66 & 9.312 & -- & -- & -- \\
67 & 4.580 & -- & $4.46$ & $5.796$ \\
68 & 3.696 & -- & -- & $5.488$ \\
69 & 2.889 & $4.23$ & $4.38$ & $5.507$ \\
71 & 2.923 & $4.23$ & $4.86$ & $5.557$ \\
72 & 1.248 & $4.23$ & $4.27$ & $5.516$ \\
73 & 2.444 & $4.23$ & $4.36$ & $5.516$ \\
75 & 7.678 & -- & -- & -- \\
76 & 3.260 & -- & -- & $5.974$ \\
79 & 4.681 & -- & $5.17$ & $5.671$ \\
83 & 5.925 & -- & -- & $6.641$ \\
84 & 11.609 & -- & -- & -- \\
85 & 4.553 & -- & $5.21$ & $6.035$ \\
90 & 0.684 & $4.23$ & $5.05$ & $6.112$ \\
95 & 4.293 & -- & -- & $6.494$ \\
98 & 5.404 & $4.23$ & $6.20$ & $6.523$ \\
\hline
\end{tabular}
 
\caption*{The simulated signal strain amplitude $h_{0}$ compared with 95\%
confidence upper-limits provided by all pipelines with the exception of the
Polynomial search, which did not provide $h_0$ estimates, and the CrossCorr
search, which detected all 50 signals. Note that the TwoSpect pipeline reports a fixed upper-limit value for all non-detections.
  Details of this procedure are given
in the methods paper~\cite{TwoSpectMDCMethods2015}.}
\end{table}
\begin{table}
\caption{Comparison of $h_{0}$ estimation.\label{tab:h0res}}
\makeatletter{}\begin{tabular}{cc|cccc}
\hline
idx & $h_{0}\times 10^{25}$ & CrossCorr$^*$ & TwoSpect & Radiometer & Sideband \\
\hline
1 & 4.16 & $5.27 \pm 1.60$ & $4.59 \pm 1.71$ & $4.80 \pm 1.87$ & $4.79 \pm 1.60$ \\
2 & 4.04 & $4.80 \pm 1.46$ & $4.45 \pm 1.66$ & $4.38 \pm 1.72$ & $3.98 \pm 1.39$ \\
3 & 3.57 & $6.68 \pm 2.03$ & $6.04 \pm 2.25$ & $6.09 \pm 2.39$ & $5.32 \pm 1.77$ \\
5 & 1.25 & $2.19 \pm 0.67$ & $2.04 \pm 0.79$ & -- & -- \\
11 & 3.09 & $2.59 \pm 0.79$ & $2.83 \pm 1.07$ & -- & -- \\
14 & 2.04 & $2.41 \pm 0.73$ & $2.55 \pm 0.97$ & -- & -- \\
15 & 11.76 & $7.87 \pm 2.39$ & $8.22 \pm 3.05$ & $7.40 \pm 3.10$ & $6.73 \pm 2.19$ \\
17 & 3.47 & $2.60 \pm 0.79$ & $2.77 \pm 1.05$ & $3.07 \pm 1.33$ & -- \\
19 & 6.03 & $4.33 \pm 1.32$ & $4.58 \pm 1.71$ & $3.59 \pm 1.55$ & $4.34 \pm 1.59$ \\
20 & 9.71 & $5.89 \pm 1.79$ & $6.56 \pm 2.44$ & $5.43 \pm 2.33$ & $5.64 \pm 1.92$ \\
21 & 1.82 & $1.52 \pm 0.46$ & -- & -- & -- \\
23 & 2.97 & $3.14 \pm 0.96$ & $3.15 \pm 1.19$ & $3.35 \pm 1.48$ & -- \\
26 & 1.42 & $2.23 \pm 0.68$ & $2.69 \pm 1.02$ & -- & -- \\
29 & 4.27 & $5.99 \pm 1.82$ & $5.51 \pm 2.05$ & $5.49 \pm 2.41$ & $5.00 \pm 1.79$ \\
32 & 10.04 & $8.72 \pm 2.65$ & $8.87 \pm 3.29$ & $9.05 \pm 4.00$ & $6.73 \pm 2.24$ \\
35 & 16.40 & $17.75 \pm 5.40$ & $16.39 \pm 6.07$ & $16.80 \pm 7.49$ & $14.86 \pm 4.44$ \\
36 & 3.86 & $3.14 \pm 0.95$ & $3.33 \pm 1.25$ & $3.26 \pm 1.49$ & -- \\
41 & 1.56 & $1.42 \pm 0.43$ & -- & -- & -- \\
44 & 2.24 & $3.84 \pm 1.17$ & $3.62 \pm 1.36$ & $4.09 \pm 1.86$ & -- \\
47 & 4.88 & $4.05 \pm 1.23$ & $4.14 \pm 1.55$ & $4.25 \pm 1.94$ & -- \\
48 & 1.81 & $1.27 \pm 0.39$ & -- & -- & -- \\
50 & 1.09 & $1.96 \pm 0.60$ & -- & -- & -- \\
51 & 9.15 & $6.21 \pm 1.89$ & $6.81 \pm 2.53$ & $4.89 \pm 2.24$ & $5.66 \pm 2.02$ \\
52 & 2.79 & $2.81 \pm 0.85$ & -- & $3.12 \pm 1.47$ & -- \\
54 & 1.52 & $2.42 \pm 0.74$ & -- & -- & -- \\
57 & 1.58 & $1.47 \pm 0.45$ & -- & -- & -- \\
58 & 3.42 & $2.47 \pm 0.75$ & -- & -- & -- \\
59 & 8.83 & $9.33 \pm 2.84$ & $9.07 \pm 3.36$ & $7.71 \pm 3.52$ & $7.21 \pm 2.40$ \\
60 & 2.96 & $4.68 \pm 1.42$ & $4.49 \pm 1.68$ & $3.43 \pm 1.61$ & -- \\
61 & 6.06 & $4.50 \pm 1.37$ & $4.69 \pm 1.75$ & $4.62 \pm 2.12$ & -- \\
62 & 10.74 & $6.63 \pm 2.02$ & $7.32 \pm 2.72$ & $5.66 \pm 2.59$ & $6.33 \pm 2.21$ \\
63 & 1.12 & $1.44 \pm 0.44$ & -- & -- & -- \\
64 & 1.60 & $1.06 \pm 0.32$ & -- & -- & -- \\
65 & 8.47 & $8.29 \pm 2.52$ & $8.17 \pm 3.03$ & $6.02 \pm 2.75$ & $6.90 \pm 2.35$ \\
66 & 9.31 & $11.22 \pm 3.42$ & $10.57 \pm 3.92$ & $10.80 \pm 4.90$ & $9.33 \pm 2.97$ \\
67 & 4.58 & $2.81 \pm 0.86$ & $2.94 \pm 1.11$ & -- & -- \\
68 & 3.70 & $3.44 \pm 1.05$ & $3.56 \pm 1.34$ & $3.11 \pm 1.47$ & -- \\
69 & 2.89 & $1.95 \pm 0.59$ & -- & -- & -- \\
71 & 2.92 & $2.40 \pm 0.73$ & -- & -- & -- \\
72 & 1.25 & $1.56 \pm 0.48$ & -- & -- & -- \\
73 & 2.44 & $1.75 \pm 0.53$ & -- & -- & -- \\
75 & 7.68 & $7.84 \pm 2.38$ & $7.41 \pm 2.75$ & $6.48 \pm 2.90$ & $6.34 \pm 2.21$ \\
76 & 3.26 & $3.41 \pm 1.04$ & $3.34 \pm 1.26$ & $3.37 \pm 1.56$ & -- \\
79 & 4.68 & $3.19 \pm 0.97$ & $3.54 \pm 1.33$ & -- & -- \\
83 & 5.92 & $4.54 \pm 1.38$ & $4.63 \pm 1.78$ & $4.16 \pm 1.78$ & -- \\
84 & 11.61 & $13.78 \pm 4.19$ & $12.54 \pm 4.65$ & $12.80 \pm 5.32$ & $10.92 \pm 3.33$ \\
85 & 4.55 & $3.27 \pm 1.00$ & $3.74 \pm 1.47$ & -- & -- \\
90 & 0.68 & $0.97 \pm 0.30$ & -- & -- & -- \\
95 & 4.29 & $5.83 \pm 1.77$ & $5.51 \pm 2.09$ & $4.86 \pm 1.92$ & -- \\
98 & 5.40 & $3.65 \pm 1.11$ & -- & -- & -- \\
\hline
\end{tabular}
 
\caption*{The simulated signal strain amplitude $h_{0}$ compared
  to the estimated values and their 1--$\sigma$ uncertainties from
  each search algorithm.  The Polynomial algorithm does not return
  $h_{0}$ estimates.
}
\end{table}
\begin{table*}
\caption{Comparison of frequency estimation.\label{tab:f0res}}
\makeatletter{}\begin{tabular}{cc|ccccc}
\hline
index & $f_{0}$ (Hz) & CrossCorr$^*$ & TwoSpect & Radiometer & Sideband & Polynomial \\
\hline
1 & 54.498391 & $54.498397 \pm 0.000005$ & $54.4982 \pm 0.0003749$ & $54.5 \pm 0.125$ & $54.499 \pm 0.002$ & -- \\
2 & 64.411966 & $64.411972 \pm 0.000006$ & $64.4119 \pm 0.0003749$ & $64.5 \pm 0.125$ & $64.407 \pm 0.003$ & -- \\
3 & 73.795581 & $73.795575 \pm 0.000005$ & $73.7952 \pm 0.0003749$ & $73.75 \pm 0.125$ & $73.795 \pm 0.003$ & -- \\
5 & 93.909518 & $93.909525 \pm 0.000006$ & $93.9113 \pm 0.0003749$ & -- & -- & -- \\
11 & 154.916884 & $154.916878 \pm 0.000006$ & $154.917 \pm 0.0003749$ & -- & -- & -- \\
14 & 183.974917 & $183.974911 \pm 0.000007$ & $183.975 \pm 0.0003749$ & -- & -- & -- \\
15 & 191.580343 & $191.580343 \pm 0.000006$ & $191.58 \pm 0.0003749$ & $191.5 \pm 0.125$ & $191.578 \pm 0.008$ & $191.58 \pm 0.04$ \\
17 & 213.232194 & $213.232193 \pm 0.000006$ & $213.232 \pm 0.0003749$ & $213.25 \pm 0.125$ & -- & -- \\
19 & 233.432566 & $233.432561 \pm 0.000006$ & $233.433 \pm 0.0003749$ & $233.5 \pm 0.125$ & $233.436 \pm 0.01$ & -- \\
20 & 244.534698 & $244.534698 \pm 0.000006$ & $244.535 \pm 0.0003749$ & $244.5 \pm 0.125$ & $244.536 \pm 0.01$ & -- \\
21 & 254.415048 & $254.415051 \pm 0.000022$ & -- & -- & -- & -- \\
23 & 271.739908 & $271.739915 \pm 0.000006$ & $271.74 \pm 0.0003749$ & $271.75 \pm 0.125$ & -- & -- \\
26 & 300.590450 & $300.590443 \pm 0.000007$ & $300.591 \pm 0.0003749$ & -- & -- & -- \\
29 & 330.590358 & $330.590352 \pm 0.000006$ & $330.591 \pm 0.0003749$ & $330.5 \pm 0.125$ & $330.59 \pm 0.014$ & -- \\
32 & 362.990821 & $362.990816 \pm 0.000006$ & $362.99 \pm 0.0003749$ & $363 \pm 0.125$ & $362.984 \pm 0.015$ & $363 \pm 0.09$ \\
35 & 394.685590 & $394.685584 \pm 0.000005$ & $394.686 \pm 0.0003749$ & $394.75 \pm 0.125$ & $394.684 \pm 0.017$ & $394.69 \pm 0.02$ \\
36 & 402.721234 & $402.721231 \pm 0.000008$ & $402.721 \pm 0.0003749$ & $402.75 \pm 0.125$ & -- & -- \\
41 & 454.865249 & $454.865253 \pm 0.000017$ & -- & -- & -- & -- \\
44 & 483.519618 & $483.519625 \pm 0.000007$ & $483.519 \pm 0.0003749$ & $483.5 \pm 0.125$ & -- & -- \\
47 & 514.568400 & $514.568406 \pm 0.000010$ & $514.568 \pm 0.0003749$ & $514.5 \pm 0.125$ & -- & -- \\
48 & 520.177348 & $520.177354 \pm 0.000002$ & -- & -- & -- & -- \\
50 & 542.952477 & $542.952467 \pm 0.000019$ & -- & -- & -- & -- \\
51 & 552.120599 & $552.120596 \pm 0.000006$ & $552.121 \pm 0.0003749$ & $552 \pm 0.125$ & $552.116 \pm 0.023$ & -- \\
52 & 560.755049 & $560.755040 \pm 0.000014$ & -- & $560.75 \pm 0.125$ & -- & -- \\
54 & 593.663031 & $593.663041 \pm 0.000010$ & -- & -- & -- & -- \\
57 & 622.605388 & $622.605391 \pm 0.000017$ & -- & -- & -- & -- \\
58 & 641.491605 & $641.491605 \pm 0.000010$ & -- & -- & -- & -- \\
59 & 650.344231 & $650.344225 \pm 0.000006$ & $650.344 \pm 0.0003749$ & $650.25 \pm 0.125$ & $650.326 \pm 0.027$ & $650.31 \pm 0.15$ \\
60 & 664.611447 & $664.611440 \pm 0.000007$ & $664.611 \pm 0.0003749$ & $664.5 \pm 0.125$ & -- & -- \\
61 & 674.711568 & $674.711567 \pm 0.000007$ & $674.712 \pm 0.0003749$ & $674.75 \pm 0.125$ & -- & -- \\
62 & 683.436211 & $683.436214 \pm 0.000006$ & $683.436 \pm 0.0003749$ & $683.5 \pm 0.125$ & $683.447 \pm 0.029$ & -- \\
63 & 690.534688 & $690.534690 \pm 0.000017$ & -- & -- & -- & -- \\
64 & 700.866836 & $700.866835 \pm 0.000003$ & -- & -- & -- & -- \\
65 & 713.378002 & $713.377996 \pm 0.000006$ & $713.378 \pm 0.0003749$ & $713.25 \pm 0.125$ & $713.364 \pm 0.03$ & -- \\
66 & 731.006818 & $731.006813 \pm 0.000005$ & $731.007 \pm 0.0003749$ & $731 \pm 0.125$ & $731.014 \pm 0.03$ & $731.01 \pm 0.07$ \\
67 & 744.255708 & $744.255707 \pm 0.000009$ & $744.282 \pm 0.0003749$ & -- & -- & -- \\
68 & 754.435957 & $754.435962 \pm 0.000008$ & $754.436 \pm 0.0003749$ & $754.5 \pm 0.125$ & -- & -- \\
69 & 761.538797 & $761.538791 \pm 0.000019$ & -- & -- & -- & -- \\
71 & 804.231718 & $804.231723 \pm 0.000014$ & -- & -- & -- & -- \\
72 & 812.280741 & $812.280731 \pm 0.000022$ & -- & -- & -- & -- \\
73 & 824.988633 & $824.988636 \pm 0.000030$ & -- & -- & -- & -- \\
75 & 862.398935 & $862.398930 \pm 0.000006$ & $862.399 \pm 0.0003749$ & $862.5 \pm 0.125$ & $862.384 \pm 0.036$ & $862.4 \pm 0.05$ \\
76 & 882.747980 & $882.747971 \pm 0.000015$ & $882.747 \pm 0.0003749$ & $882.75 \pm 0.125$ & -- & -- \\
79 & 931.006000 & $931.006001 \pm 0.000011$ & $931.006 \pm 0.0003749$ & -- & -- & -- \\
83 & 1081.398956 & $1081.398954 \pm 0.000008$ & $1081.4 \pm 0.0007659$ & $1081.5 \pm 0.25$ & -- & -- \\
84 & 1100.906018 & $1100.906024 \pm 0.000006$ & $1100.91 \pm 0.0003749$ & $1101 \pm 0.25$ & $1100.89 \pm 0.046$ & $1100.92 \pm 0.04$ \\
85 & 1111.576832 & $1111.576830 \pm 0.000011$ & $1111.58 \pm 0.0007659$ & -- & -- & -- \\
90 & 1193.191891 & $1193.191898 \pm 0.000009$ & -- & -- & -- & -- \\
95 & 1324.567365 & $1324.567360 \pm 0.000007$ & $1324.57 \pm 0.0007659$ & $1324.5 \pm 0.25$ & -- & -- \\
98 & 1372.042155 & $1372.042158 \pm 0.000014$ & -- & -- & -- & -- \\
\hline
\end{tabular}
 
\caption*{The simulated signal intrinsic frequency $f_{0}$ compared with the
estimates provided by all search pipelines for their respective detected signals.}
\end{table*}
\begin{table}
\caption{Comparison of $\asini$ estimation.\label{tab:asinires}}
\makeatletter{}\begin{tabular}{cc|cc}
\hline
index & $\asini$ (sec) & CrossCorr$^*$ & TwoSpect \\
\hline
1 & 1.37952 & $1.37973 \pm 0.00055$ & $1.32004 \pm 0.01839$ \\
2 & 1.76461 & $1.76468 \pm 0.00053$ & $1.78418 \pm 0.01839$ \\
3 & 1.53460 & $1.53462 \pm 0.00040$ & $1.54807 \pm 0.01839$ \\
5 & 1.52018 & $1.52012 \pm 0.00055$ & $1.27114 \pm 0.01839$ \\
11 & 1.39229 & $1.39235 \pm 0.00038$ & $1.39849 \pm 0.01839$ \\
14 & 1.50970 & $1.50965 \pm 0.00034$ & $1.47490 \pm 0.01839$ \\
15 & 1.51814 & $1.51813 \pm 0.00017$ & $1.50757 \pm 0.01839$ \\
17 & 1.31021 & $1.31032 \pm 0.00029$ & $1.32593 \pm 0.01839$ \\
19 & 1.23123 & $1.23131 \pm 0.00018$ & $1.23230 \pm 0.01839$ \\
20 & 1.28442 & $1.28449 \pm 0.00015$ & $1.26879 \pm 0.01839$ \\
21 & 1.07219 & $1.07267 \pm 0.00132$ & -- \\
23 & 1.44287 & $1.44289 \pm 0.00019$ & $1.44599 \pm 0.01839$ \\
26 & 1.25869 & $1.25876 \pm 0.00023$ & $1.27430 \pm 0.01839$ \\
29 & 1.33070 & $1.33073 \pm 0.00011$ & $1.32816 \pm 0.01839$ \\
32 & 1.61109 & $1.61110 \pm 0.00009$ & $1.60622 \pm 0.01839$ \\
35 & 1.31376 & $1.31376 \pm 0.00007$ & $1.29794 \pm 0.01839$ \\
36 & 1.25484 & $1.25497 \pm 0.00027$ & $1.23518 \pm 0.01839$ \\
41 & 1.46578 & $1.46582 \pm 0.00057$ & -- \\
44 & 1.55221 & $1.55226 \pm 0.00019$ & $1.55747 \pm 0.01839$ \\
47 & 1.14021 & $1.14011 \pm 0.00026$ & $1.13354 \pm 0.01839$ \\
48 & 1.33669 & $1.33673 \pm 0.00008$ & -- \\
50 & 1.11915 & $1.11890 \pm 0.00053$ & -- \\
51 & 1.32783 & $1.32784 \pm 0.00011$ & $1.32385 \pm 0.01839$ \\
52 & 1.79214 & $1.79220 \pm 0.00034$ & -- \\
54 & 1.61276 & $1.61268 \pm 0.00024$ & -- \\
57 & 1.51329 & $1.51332 \pm 0.00041$ & -- \\
58 & 1.58443 & $1.58446 \pm 0.00022$ & -- \\
59 & 1.67711 & $1.67711 \pm 0.00010$ & $1.66654 \pm 0.01839$ \\
60 & 1.58262 & $1.58262 \pm 0.00011$ & $1.58219 \pm 0.01839$ \\
61 & 1.49937 & $1.49939 \pm 0.00012$ & $1.49017 \pm 0.01839$ \\
62 & 1.26951 & $1.26953 \pm 0.00008$ & $1.27473 \pm 0.01839$ \\
63 & 1.51824 & $1.51838 \pm 0.00036$ & -- \\
64 & 1.39993 & $1.39997 \pm 0.00007$ & -- \\
65 & 1.14577 & $1.14581 \pm 0.00010$ & $1.13298 \pm 0.01839$ \\
66 & 1.32179 & $1.32180 \pm 0.00006$ & $1.33204 \pm 0.01839$ \\
67 & 1.67774 & $1.67772 \pm 0.00016$ & $1.27351 \pm 0.01839$ \\
68 & 1.41389 & $1.41389 \pm 0.00013$ & $1.40005 \pm 0.01839$ \\
69 & 1.62613 & $1.62588 \pm 0.00037$ & -- \\
71 & 1.65203 & $1.65194 \pm 0.00024$ & -- \\
72 & 1.19649 & $1.19660 \pm 0.00039$ & -- \\
73 & 1.41715 & $1.41718 \pm 0.00056$ & -- \\
75 & 1.56703 & $1.56705 \pm 0.00007$ & $1.55329 \pm 0.01839$ \\
76 & 1.46249 & $1.46251 \pm 0.00025$ & $1.46132 \pm 0.01839$ \\
79 & 1.49171 & $1.49177 \pm 0.00015$ & $1.48842 \pm 0.01839$ \\
83 & 1.19854 & $1.19857 \pm 0.00010$ & $1.19267 \pm 0.01839$ \\
84 & 1.58972 & $1.58972 \pm 0.00004$ & $1.58362 \pm 0.01839$ \\
85 & 1.34479 & $1.34488 \pm 0.00013$ & $1.33880 \pm 0.01839$ \\
90 & 1.57513 & $1.57521 \pm 0.00008$ & -- \\
95 & 1.59168 & $1.59167 \pm 0.00006$ & $1.58786 \pm 0.01839$ \\
98 & 1.31510 & $1.31514 \pm 0.00015$ & -- \\
\hline
\end{tabular}
 
\caption*{The simulated signal projected orbital semi-major axis $\asini$
  compared to the estimates from the TwoSpect and CrossCorr pipelines.
Note that the TwoSpect pipeline reports a fixed $\asini$ uncertainty
 for all detections.
  Details of this procedure are given
in the methods paper~\cite{TwoSpectMDCMethods2015}.}
\end{table}
\begin{table}
\caption{Comparison of $\Tasc$ estimation.\label{tab:Tascres}}
\makeatletter{}\begin{tabular}{cc|c}
\hline
index & $\Tasc$ (GPS sec) & CrossCorr$^*$ \\
\hline
1 & 1245967666.0 & $1245967664.9 \pm 1.5$ \\
2 & 1245967593.0 & $1245967592.6 \pm 1.9$ \\
3 & 1245967461.3 & $1245967461.8 \pm 0.8$ \\
5 & 1245966927.9 & $1245966927.7 \pm 3.3$ \\
11 & 1245967560.0 & $1245967560.5 \pm 2.6$ \\
14 & 1245967551.0 & $1245967551.5 \pm 2.2$ \\
15 & 1245967298.5 & $1245967298.5 \pm 0.6$ \\
17 & 1245967522.5 & $1245967523.3 \pm 2.1$ \\
19 & 1245967331.1 & $1245967330.9 \pm 1.2$ \\
20 & 1245967111.0 & $1245967111.2 \pm 0.8$ \\
21 & 1245967346.4 & $1245967360.8 \pm 12.4$ \\
23 & 1245967302.3 & $1245967302.0 \pm 1.2$ \\
26 & 1245967177.5 & $1245967177.1 \pm 1.8$ \\
29 & 1245967520.8 & $1245967521.8 \pm 0.6$ \\
32 & 1245967585.6 & $1245967585.6 \pm 0.3$ \\
35 & 1245967198.0 & $1245967197.5 \pm 0.2$ \\
36 & 1245967251.3 & $1245967249.4 \pm 2.2$ \\
41 & 1245967225.8 & $1245967220.8 \pm 4.0$ \\
44 & 1245967397.9 & $1245967398.0 \pm 1.2$ \\
47 & 1245967686.8 & $1245967686.4 \pm 2.4$ \\
48 & 1245967675.3 & $1245967674.0 \pm 0.7$ \\
50 & 1245967927.5 & $1245967930.1 \pm 4.9$ \\
51 & 1245967589.5 & $1245967590.2 \pm 0.8$ \\
52 & 1245967377.2 & $1245967379.4 \pm 2.1$ \\
54 & 1245967624.5 & $1245967623.9 \pm 1.6$ \\
57 & 1245967203.2 & $1245967202.6 \pm 2.8$ \\
58 & 1245967257.7 & $1245967256.8 \pm 1.4$ \\
59 & 1245967829.9 & $1245967829.5 \pm 0.6$ \\
60 & 1245967612.3 & $1245967610.8 \pm 0.7$ \\
61 & 1245967003.3 & $1245967003.2 \pm 0.8$ \\
62 & 1245967454.0 & $1245967454.3 \pm 0.6$ \\
63 & 1245967419.4 & $1245967418.1 \pm 2.6$ \\
64 & 1245967596.1 & $1245967595.0 \pm 0.6$ \\
65 & 1245967094.6 & $1245967095.1 \pm 0.8$ \\
66 & 1245967576.5 & $1245967576.4 \pm 0.3$ \\
67 & 1245967084.3 & $1245967084.0 \pm 1.0$ \\
68 & 1245967538.7 & $1245967538.9 \pm 1.0$ \\
69 & 1245966821.5 & $1245966819.8 \pm 2.5$ \\
71 & 1245967156.5 & $1245967157.2 \pm 1.6$ \\
72 & 1245967159.1 & $1245967158.6 \pm 3.6$ \\
73 & 1245967876.8 & $1245967876.5 \pm 4.1$ \\
75 & 1245967346.3 & $1245967346.9 \pm 0.5$ \\
76 & 1245966753.2 & $1245966751.6 \pm 1.9$ \\
79 & 1245967290.1 & $1245967290.1 \pm 1.1$ \\
83 & 1245967313.9 & $1245967314.6 \pm 0.8$ \\
84 & 1245967204.1 & $1245967204.9 \pm 0.2$ \\
85 & 1245967049.3 & $1245967050.4 \pm 1.0$ \\
90 & 1245966914.3 & $1245966916.5 \pm 0.8$ \\
95 & 1245967424.8 & $1245967424.6 \pm 0.4$ \\
98 & 1245966869.9 & $1245966871.6 \pm 1.2$ \\
\hline
\end{tabular}
 
\caption*{The simulated signal time of passage of the ascending node
  compared to the estimate from the CrossCorr pipeline, which was
  the only search to estimate this parameter.}
\end{table}

\section{Radiometer technical details\label{sec:radiometertech}}

\subsection{Re-expressing the Radiometer results\label{sec:radiometerintro}}
Unlike the other search methods presented in this paper, the Radiometer algorithm grew out of the search for an anisotropic stochastic \ac{GW} background where the strongest sources dominated~\cite{Ballmer2006CQG,2007PhRvD..76h2003A,2011PhRvL.107A1102A}.  This focus shaped how the Radiometer results were reported.  In order to accurately compare the different search algorithms, the Radiometer search has converted its results to be in the format used in this paper.  The changes applied are for converting from strain power to strain amplitude, for generalizing the assumption of a circularly polarized signal with spin axis is aligned with the line of sight to a random polarization, and for applying a factor to account for signals spanning multiple frequency bins.

\subsection{Converting strain power to strain amplitude\label{sec:radiometerh0}}
The radiometer algorithm is normalized such that its $Y$-statistic equals the strain power.  Under the assumption of circular polarization, conversion from ${\hat{Y}_{\rm tot}}$ to the strain amplitude $h_0$ is straightforward:
\begin{equation}
h_0=\sqrt{{\hat{Y}_{\rm tot}}df}
\end{equation}
where $df$ is the width of the frequency bin.

\subsection{Converting from circular to random polarization\label{sec:radiometercp}}
In its current form, the Radiometer search assumes a circularly polarized signal.  If a signal is exactly circularly polarized, the estimate of $h_0$ should be unbiased.  However, the closer a signal is to being linearly polarized, the greater the underestimate of $h_0$.  This is the same as the TwoSpect algorithm and the same method for characterizing the effect as described in Appendix~\ref{app:TwoSpectambiguity} is used here, resulting in an average circular polarization correction factor of $C_{\rm cp}=1.74\pm0.37$ on $h_0$.

\subsection{Signals spanning multiple frequency bins\label{sec:radiometermfb}}
The current implementation of the Radiometer search is not tuned to \ac{ScoX1}.  One aspect of this is that the search uses 0.25~Hz bins rather than bin size based on twice the modulation depth of $(2\pi f_0\,\asini)/\Porb$ and on the Earth's annual motion around the Sun.  Another aspect is that frequency bins are non-overlapping even though \ac{ScoX1}'s frequency $f_0$ is unknown.  Both of these situations lead to the fact that signals can span more than one frequency bin.  Below 538~Hz, signals can span two bins if the signal frequency is sufficiently close to the boundary between frequency bins.  Signals span 2-3 frequency bins above 538~Hz and 3-4 bins above 1076~Hz depending on where the signal frequency is relative to the frequency bin borders.  When signals span more than one bin, the signal \ac{SNR} is lessened in each individual bin.  This causes the Radiometer, statistically speaking, to underestimate $h_0$ and to potentially miss borderline detections.

To account for the $h_0$ underestimate, we statistically calculate a frequency-dependent correction factor.  We start by determining the correction factor for a single frequency bin at a particular frequency ('the trial correction factor') by simulating $10^4$ trials, where each trial has a single signal injection in it.  The set of signal injections are uniformly spaced across the bin.  For each trial, we assume that the signal is uniform over the frequency span and that the center value of the bin frequency for the calculation of $\Delta f_{\rm obs}$ (at most a 1\% error).

\begin{figure}
\includegraphics[width=\columnwidth]{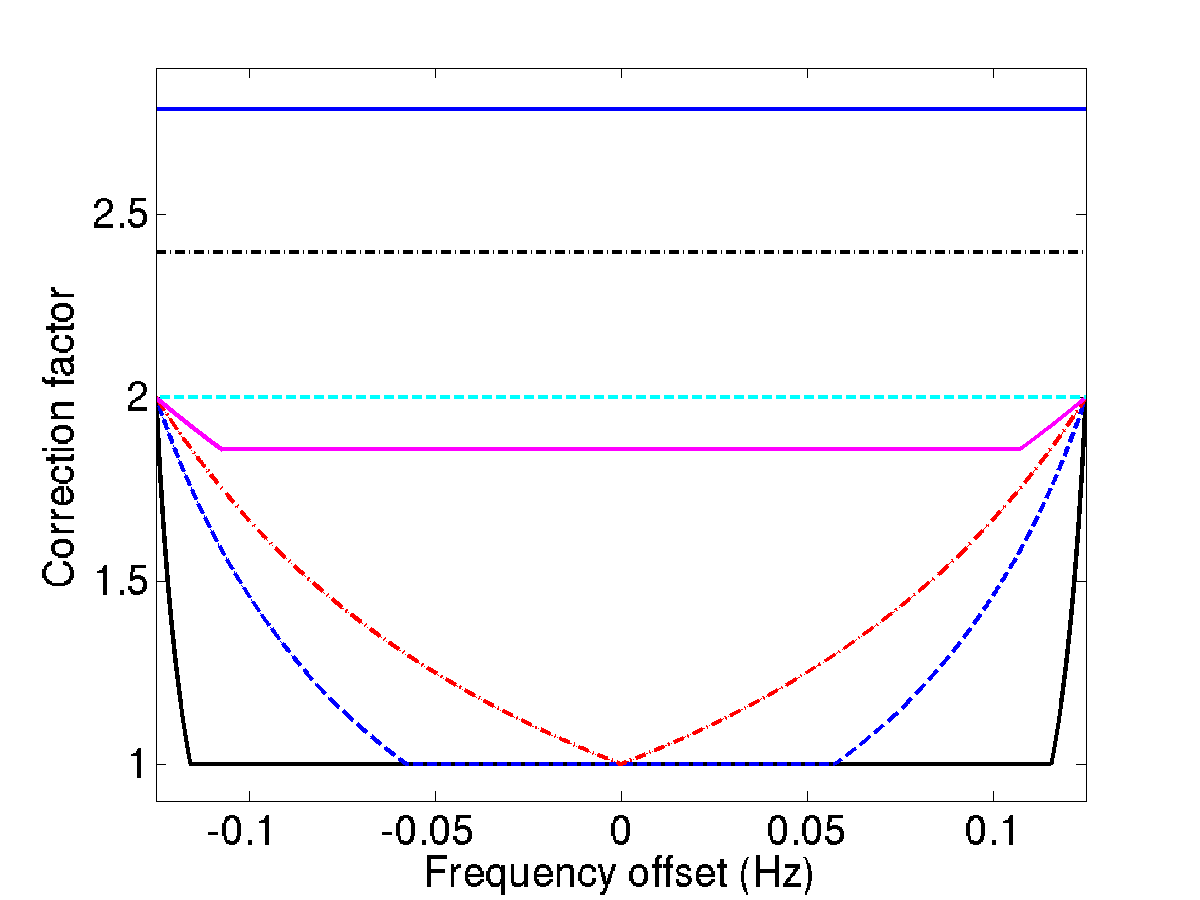}
\caption{ Simulated correction of $\hat{Y}_{\rm tot}$ for the Radiometer search across the standard 0.25~Hz-wide bin for a variety of frequencies.  Bin center falls at the frequency offset of 0~Hz.  The traces, from bottom to top, display the simulation for 40 (black, solid), 290 (blue, dashed), 538 (red, dot-dashed), 1000 (magenta, solid), 1076 (cyan, dashed), 1290 (black, dot-dashed), and 1500~Hz (blue, solid).  Below 538~Hz note that at the bin boundary, where only half of the \ac{SNR} is in this bin, the correction factor is two.  The correction factor is one when the signal is completely within the bin.  $\Delta f$ is larger for 290~Hz than for 40~Hz, thus the frequencies at which the injection is completely within the chosen bin (and thus has a correction factor of one) are fewer.  Above 538~Hz, it is not possible to have the entire injection within a single frequency bin $df$.  At most, $df/{\Delta f}$ of the injection can be within a single bin, resulting in a correction factor that is always greater than one.  Above 1076~Hz, $\hat{Y}_{\rm tot}$ is always $df/{\Delta f}$ of the injection.
\label{fig:cfVsF_combo}} 
\end{figure}

At or below 538~Hz, $\Delta f _{\rm obs} \leq df$ and there are three separate regimes that need to be considered: 1) when the trial injected signal is divided between the chosen frequency bin and the adjacent bin lower in frequency (while still having a maximum \ac{SNR} in the chosen bin), 2) when the signal falls completely within the chosen bin, and 3) when the signal is divided between the chosen frequency bin and the adjacent bin higher in frequency (while still having a maximum \ac{SNR} in the chosen bin).  If the injected signal falls exactly on the border between bins, the recovered $\hat{Y}_{\rm tot}$ will be half the value expected from the injection.  If the injected signal is completely within the chosen bin, the recovered $\hat{Y}_{\rm tot}$ will be the same as expected from the injected value.  In case 1), the trial correction factor on $\hat{Y}_{\rm tot}$ (for a single frequency bin) is given by
\begin{equation}
\frac{1}{C_{\rm mfb\_{trial}}}=\frac{f_{\rm inj}}{\Delta f}+0.5-\frac{df_{\rm low}}{\Delta f}
\label{radiometer_eq2}
\end{equation}
where $f_{\rm inj}$ is the central frequency of the injection, $\Delta f$ is the frequency width of the injection, and $df_{\rm low}$ is the low frequency border of the chosen bin.  In case 2), $C_{\rm mfb\_{trial}}=1$ since no correction is necessary.  In case 3),
\begin{equation}
\frac{1}{C_{\rm mfb\_{trial}}}=-\frac{f_{\rm inj}}{\Delta f}+0.5+\frac{df_{\rm high}}{\Delta f}
\label{radiometer_eq3}
\end{equation}
where $df_{\rm high}$ is the high frequency border of the chosen bin.  The trial correction factor as a function of frequency for the 40, 290, and 538~Hz frequency bins are shown in Figure~\ref{fig:cfVsF_combo}. 

Above 538~Hz and at or below 1076~Hz, $df < \Delta f_{\rm obs} \leq 2*df$ and there are again three regimes to be considered: 1) when the trial injected signal is divided between the chosen frequency bin and the adjacent bin lower in frequency (while still having a maximum \ac{SNR} in the chosen bin), 2) when the signal completely fills the chosen bin, and 3) when the signal is divided between the chosen frequency bin and the adjacent bin higher in frequency (while still having a maximum \ac{SNR} in the chosen bin).  Similar to at lower frequencies, if the injected signal falls exactly on the bin border, the recovered $\hat{Y}_{\rm tot}$ will be half the value expected from the injection and thus requires a correction factor of two.  If the signal completely fills the chosen bin, the recovered $\hat{Y}_{\rm tot}$ will be the fraction of the injected signal it spans.  Above 538~Hz and below 1076~Hz, Eq.'s~\ref{radiometer_eq2} and \ref{radiometer_eq3} still apply to cases 1) and 3) respectively.  However case 2) becomes
\begin{equation}
\frac{1}{C_{\rm mfb\_{trial}}}=\frac{df}{{\Delta f}}.
\label{radiometer_eq4}
\end{equation}
The trial correction factor as a function of frequency for the 1000 and 1076~Hz frequency bins are shown in Figure~\ref{fig:cfVsF_combo}.

Above 1076~Hz, $\Delta f_{\rm obs} > 2*df$.  Here, the trial injected signal only has a maximum \ac{SNR} in the chosen bin when the signal completely fills the bin.  The correction factor is then always Eq.~\ref{radiometer_eq4}.  
The trail correction factor as a function of frequency for the 1290 and 1500~Hz bins are shown in Figure~\ref{fig:cfVsF_combo}.

\begin{figure}
\includegraphics[width=\columnwidth]{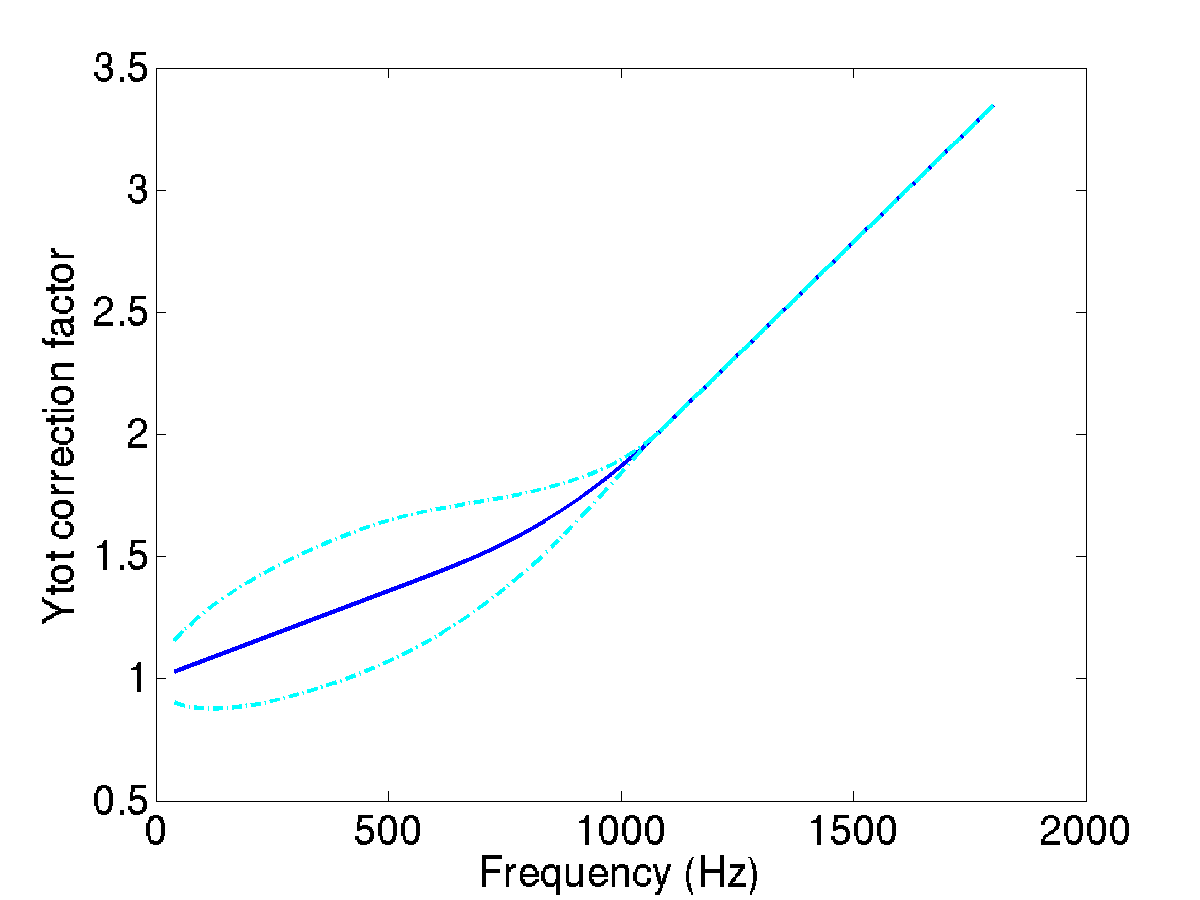}
\caption{ Frequency bin correction factor necessary to account for signals spanning multiple bins in the Radiometer search.  The solid blue line denotes the correction factor and the dashed cyan lines denote its $1\sigma$ uncertainty.  At low frequencies, it is statistically infrequent for a signal to span multiple frequency bins, resulting in a correction factor close to one.  At high frequencies, signals always span at least two frequency bins which is reflected by the larger correction factor.  The increasing uncertainty with increasing frequency below 538~Hz reflects that increasing $\Delta f$ results in fewer occurrences of a simulation falling completely within a frequency bin which results in a broadened distribution of the simulated trials.  The decreasing uncertainty with increasing frequency above 538~Hz is affected by two factors.  One is that increasing $\Delta f$ increases the occurrences among the simulated trials of completely spanning the frequency bin.  The other factor is that the range of correction factors for a particular frequency bin lessens with increasing frequency (and hence increasing $\Delta f$).
\label{fig:signalSpanningMultibinCF}} 
\end{figure}

In the above discussion, the trial correction factor is exactly determined for a large number of frequencies within an individual frequency bin.  In order to determine a single corrected value of $\hat{Y}_{\rm tot}$ for each frequency bin, we calculate
\begin{equation}
\hat{Y}_{\rm tot  \_ BCFcorr}=\hat{Y}_{\rm tot \_ meas} \times C_{\rm mbf}
\label{radiometer_eq5}
\end{equation}
where $\hat{Y}_{\rm tot \_ meas}$ is the (uncorrected) measured value of $\hat{Y}_{\rm tot}$ for a particular frequency bin and $C_{\rm mbf}$ is the expectation value of $C_{\rm mfb\_{trial}}$, the set of correction factors from the simulated trials.  Figure~\ref{fig:signalSpanningMultibinCF} shows the correction factor and associated $1\sigma$ uncertainty for each frequency bin over a large range of frequency bins. 

\subsection{The combined correction factor\label{sec:radiometerccf}}

The measured ${\hat{Y}_{\rm tot}}$, when accounting for the circular polarization assumption and for a signal spanning multiple frequency bins, is converted to $h_0$ by
\begin{equation}
h_0=C_{\rm cp}\times \sqrt{C_{\rm mbf}\times {\hat{Y}_{\rm tot}}df}
\end{equation}
and uncertainties on this value are determined by standard propagation of error techniques for uncorrelated variables as well as the Goodman expression for exact variance \cite{Goodman}.  Bayesian upper limits are calculated for the MDC's 5~Hz injection search bands in which the Radiometer made no detection, using the frequency bin within the search band with the largest measured $h_0$ (after application of correction factors).

\begin{figure}
\includegraphics[width=\columnwidth]{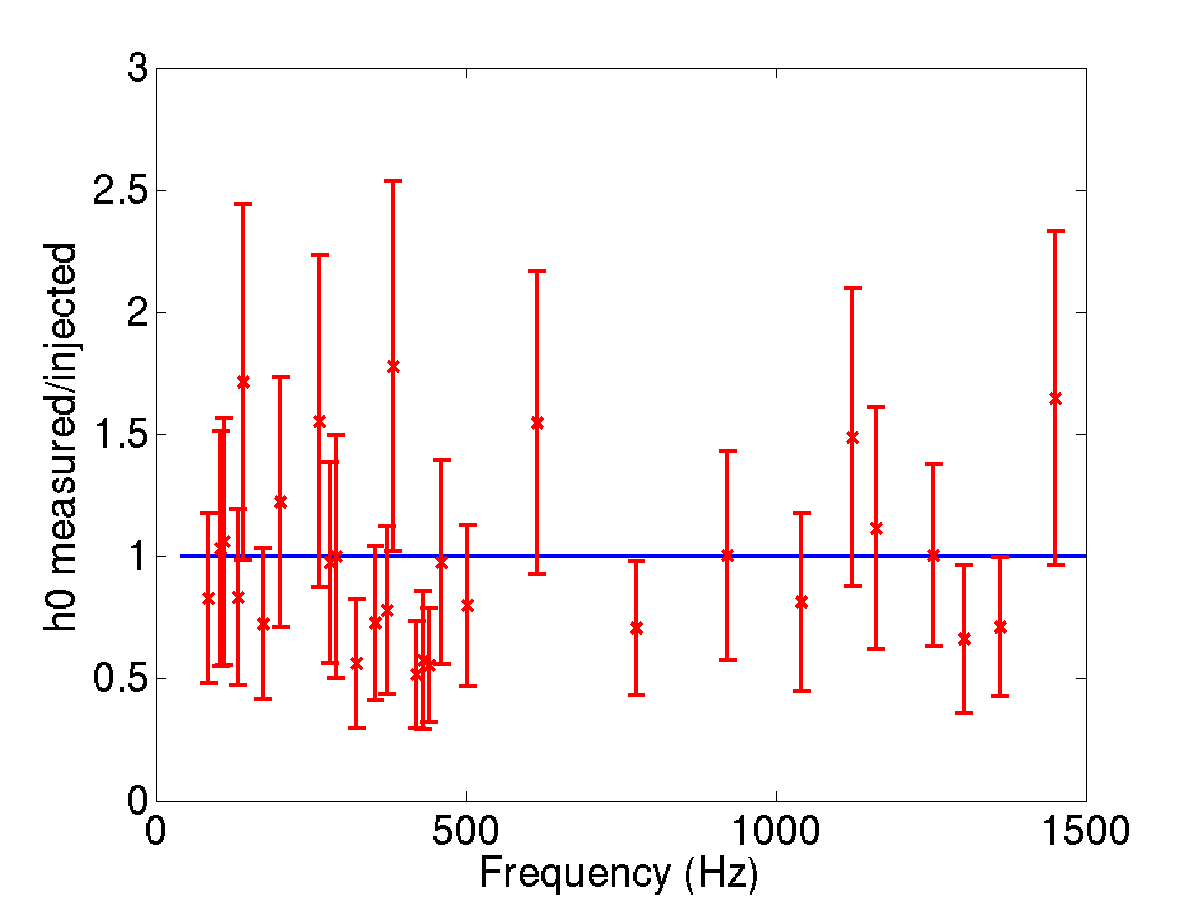}
\caption{ Ratio of measured and corrected to injected $h_0$ for the Radiometer search for the open signals.  An extra empirical factor of $\sqrt{3}$ has been included in the uncertainties.  When the measured and the injected $h_0$ agree to within one standard deviation, the error bars for the data (red) should intersect the blue line which has a value of one.
\label{fig:h0MeasToInj_openSig_sqrt3Factor}} 
\end{figure}

The conversion factors were verified with the open signals.  Without correction factors applied, the average ratio of detected and measured to injected $h_0$ is $0.48$ and none of the 29 detected and measured $h_0$ agree with the injected $h_0$ to within one standard deviation.  After applying the conversion factors, the average ratio of detected and measured to injected $h_0$ is $1.00$.  Independent of frequency, the uncertainties were found to be too small with only 12 of the 29 detections agreeing to within one standard deviation with the injection.  Work is in progress to identify the source of this discrepancy, but currently the open signals have been used to establish a factor by which to increase the error bars.  Choosing a factor of $\sqrt{3}$ causes 21 of the 29 detections to agree with the injection to within one standard deviation (and all to agree within two standard deviations).  
The ratio of the measured and corrected to injected $h_0$ is shown in Figure~\ref{fig:h0MeasToInj_openSig_sqrt3Factor}.

\section{TwoSpect technical details\label{sec:TwoSpecttech}}

Addition details on TwoSpect are presented in a forthcoming methods paper~\cite{TwoSpectMDCMethods2015}.

In this~\ac{MDC}, TwoSpect used Gaussian noise data from the open signal
to calibrate the false alarm probability to 0.01 or better.
Outliers in $R$-statistic and $p$-value were coincidence-tested;
detections were required to be present in at least one interferometer pair.
Initially, a threshold of $\log_{10} p = -7.75$  yielded the desired
false alarm probability in bands using 840-s \acp{SFT}, but $\log_{10} p = -12.0$ was
needed for 360-s \ac{SFT} bands. 
This was found, after the deadline,
 to be caused by some non-Gaussian signal having been
sampled in our data.
If this had been known, the threshold for 360-s \ac{SFT} bands could have been
lower, $\log_{10} p = -8.80$, in accordance with the expectation that
$p$-value thresholds should be independent of coherence time.
At most one detection was lost due to this mistake.

\subsection{Detection criteria}

\begin{itemize}
\item Single-IFO candidates are the up-to-200 most extreme $p$-value 
outliers in a 5-Hz band that had a $\log_{10}p \leq$ threshold, where 
threshold = $-7.75$ if $f <$ 360.0 Hz (those that used 840-s \acp{SFT}) or 
$-12.0$ if $f \geq$ 360.0 Hz (those that used 360-s \acp{SFT}).
\item Each candidate must survive at least one double-IFO coincidence test, 
involving a pairwise comparison of single-IFO candidates to see whether 
they are within 1/$T_{\textup{SFT}}$ in both frequency ($f$) and 
modulation depth ($\Delta f_\textup{obs}$, also known as $df$).
\end{itemize}

If there is any candidate surviving these criteria in a 5 Hz band, 
we mark detected, else not detected. Note that the coincidence
requirement achieves the desired false alarm probability, but it
means that calculating an accurate joint $p$-value for the 
detection would be require additional studies with computing
cost beyond the scope of the \ac{MDC}.

\subsection{Parameter Estimation\label{app:TwoSpectPE}}

Open signals allowed the calibration of parameter estimation methods. 
Signal parameters were estimated with a standard deviation according to the error in open signals,
using the extremal $p$-value for a coincident signal to read the estimated parameters.
In addition to the uncertainty inherent in this procedure, there is also a systematic uncertainty
due to $\cos \iota$, because the pipeline is instead sensitive to the circularly-polarized-equivalent,
$h_0^{\text{eff}}$, defined in \eqref{eq:h0eff}
This systematic error is also included in the uncertainties for $h_0$.

Upper limits and detection efficiency estimates were made using open pulsar data.

\subsection{Ambiguity between \texorpdfstring{$h_{0}$ and $\cos\iota$}{h0 and (cosiota)}}
\label{app:TwoSpectambiguity}
The cosine of the inclination angle of the pulsar, $\cos \iota$, 
casts an ambiguity over the determination of $h_0$. 
For TwoSpect, which assumes circular polarization, 
the approximate true value of $h_0$ will indeed be 
as reported if $|\cos \iota|$ = 1, but will be greater for smaller 
$| \cos \iota |$ is less (i.e., the \ac{GW} is elliptically polarized). 
In the case of linear polarization, $h_0$ will be about 
$2^{3/2} \approx 2.83$ times larger than reported.
A simulation of $2\times10^6$ pulsars, generated uniformly in $1/h_0$ for $h_0$ between 
$3\times 10^{-26}$ and $3\times 10^{-24}$, demonstrated that the \textit{average} factor is
1.74 with $1\sigma$-uncertainty of $\pm 0.37$.

We validated the fraction of open analyses that 
estimated $h_0$, $f$ and $a \sin i$ within 
their 1-$\sigma$ error bars:
\begin{itemize}
\item $h_0$: 77.4\%
\item $f$: 74.2\%
\item $asini$: 67.7\%
\item Period: 100\% with only 68023.8259 s tried
\end{itemize}

Because these percentages are larger than fraction expected in 1-$\sigma$,
the parameter estimation uncertainties for the open data set were conservative.

\section{CrossCorr technical details\label{app:CrossCorrtech}}

Complete details of the CrossCorr analysis as implemented for the
\ac{MDC} will appear in \cite{CrossCorrMDC}.  This appendix briefly
describes the most important aspects.

\subsection{Choice of \texorpdfstring{$\Tmax$}{Tmax} parameter}
\label{app:CrossCorrTmax}

\begin{table}[tbp]
  \caption{Choice of \ac{SFT} durations and coherence times
    for the cross-correlation search.}
  \label{tab:sftparams}
  \begin{tabular}{c|c|cc}
    \hline
    Freq range & $\Tsft$(s)
    & $T_{\text{max,inner}}$(s) & $T_{\text{max,outer}}$ (s)
    \\
    \hline
$50$--$100$ & $900$ & $5400$ & $3600$
\\
$100$--$200$ & $600$ & $2400$ & $1200$
\\
$200$--$400$ & $420$ & $2100$ & $840$
\\
$400$--$800$ & $300$ & $1140$ & $840$
\\
$800$--$1375$ & $240$ & $780$ & $540$
\\
    \hline
  \end{tabular}
  \caption*{As described in Sec.~\ref{app:CrossCorrTmax}, the
    cross-correlation search used a different duration $\Tsft$
    for the Fourier transforms of the data as a function of frequency.
    It also used two different values of maximum allowed time offset
    $\Tmax$ between correlated \acp{SFT} for the two regions of
    parameter space shown in Fig.~\ref{fig:crosscorrlags}.  These
    choices were made to balance computing cost and sensitivity.}
\end{table}

\begin{figure}
\includegraphics[width=\columnwidth]{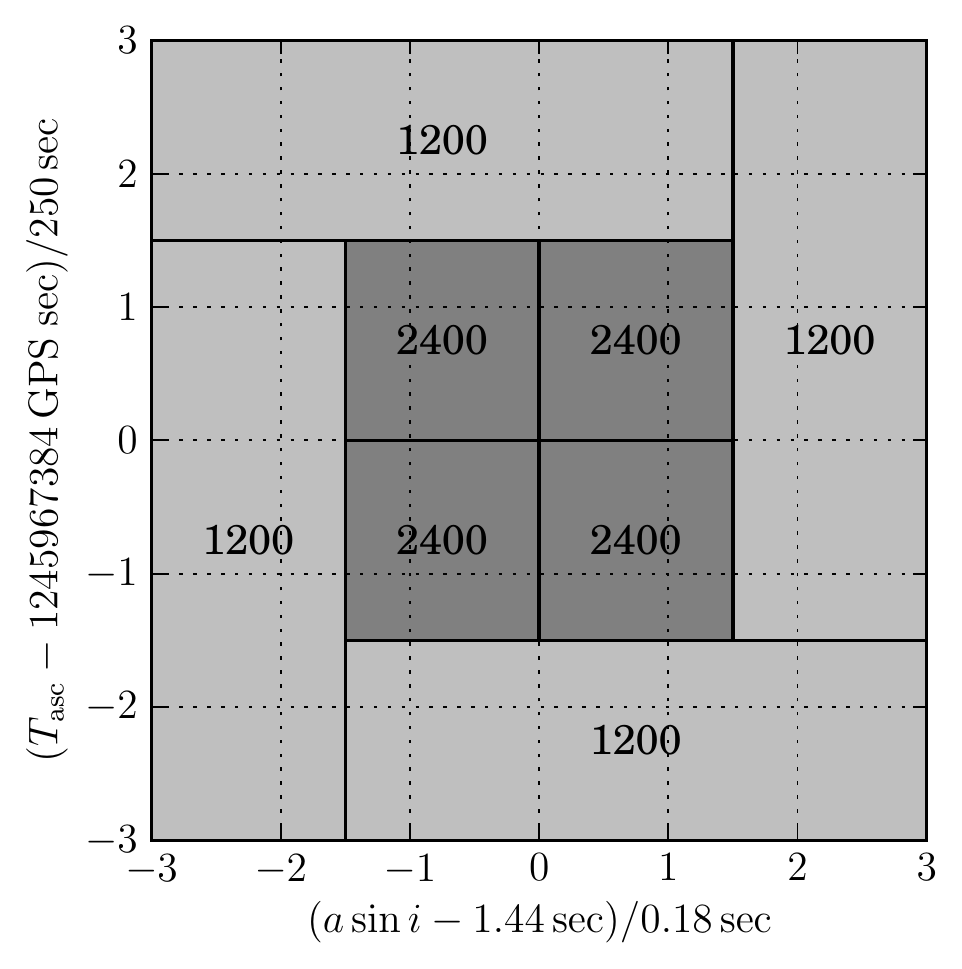}
\caption{Example of $\Tmax$ values for the CrossCorr search divided
  according to orbital parameter space.  The ``inner regions'', which
  are more likely to contain the signal parameters, use a longer
  coherence length $\Tmax$.\label{fig:crosscorrlags} }
\end{figure}

The primary determining factor in both the sensitivity and
computational cost of the CrossCorr search is the coherence
time $\Tmax$.  This is the maximum time offset allowed between pairs
of \acp{SFT} to be included in the CrossCorr statistic.  Use of a
single $\Tmax$ value for the entire MDC would not have been ideal for
two reasons:
\begin{itemize}
\item The density of templates in each of the orbital parameter space
  directions ($\asini$ and $\Tasc$) grows with frequency.
  Additionally, since the method treats the Doppler-shifted frequency
  as constant over an \ac{SFT}, the \ac{SFT} length needs to be
  decreased with frequency in order to avoid loss of \ac{SNR} due to
  unmodelled Doppler acceleration.  These two effects mean a search
  with the same $\Tmax$ will be more computationally expensive at
  higher frequencies.
\item If the prior uncertainties on orbital parameters are assumed to
  be Gaussian distributed (as they were for the \ac{MDC}), the
  ``inner'' regions of parameter space (close to the most likely
  values) are more likely to contain the signal parameters than the
  ``outer'' regions.
\end{itemize}
Since it is impractical to have the coherence time $\Tmax$ (and the
\ac{SFT} duration $\Tsft$) vary continuously with frequency, the
CrossCorr search was implemented with a single setup for all of the
frequency bands lying in a particular ``octave'', as described in
Table~\ref{tab:sftparams}.
The setups were chosen to have roughly
constant computing cost (estimated as proportional to the number of
parameter space templates times the number of \ac{SFT} pairs) for each
5~Hz band.  To determine the ideal coherence time $\Tsft$ within this
constraint, the orbital parameter space was divided into ``inner'' and
``outer'' regions, as shown in Figure~\ref{fig:crosscorrlags}.  The
``inner'' regions, with both $\asini$ and $\Tasc$ within 1.5 standard
deviations ($1.5\sigma$) of their a priori most likely values, had,
according to the assumed Gaussian prior distribution, a $75.1\%$
probability of containing any given signal.  The outer regions, where
both parameters were within $3\sigma$ of their most likely value, but
at least one was more than $1.5\sigma$ away, constituted three times
the area in parameter space, but only had a $24.4\%$ chance of
containing the signal.  Different combinations of
$T_{\text{max,inner}}$ and $T_{\text{max,outer}}$ which fit within the
computing budget were considered, and the ones which provided the
greatest overall likelihood of producing a detection (considering the
probability of the signal lying in each region and the conditional
probability of observing a high enough statistic value with the chosen
$\Tmax$, if the signal was in that region) were chosen.  These are
summarized in Table~\ref{tab:sftparams}.

\subsection{Detection confidence}
\label{app:CrossCorrPval}

A na\"{\i}ve estimate of significance for the CrossCorr search, which
constructs a statistic which in the absence of a signal has zero mean
and unit variance, is to assume the distribution to be Gaussian.  The
single-template false-alarm probability corresponding to a value
$\rho$ for the statistic defined in \eqref{e:CCrhodef} is then
\begin{equation}
  \label{e:CCpvalnaive}
  p_1 = \frac{1}{2}\erfc\left(\frac{\rho}{\sqrt{2}}\right)
  \approx \frac{e^{-\rho^2/2}}{\rho\sqrt{2\pi}} (1-\rho^{-2})
\end{equation}
where the asymptotic form of the complementary error function should
be used to avoid underflow when $\rho>20$.  To estimate a combined
$p$-value of a search using $N$ templates, these were assumed to be
$N$ independent trials and constructs the per-band false alarm
probability corresponding to the maximum $\rho$ over the band:
\begin{equation}
  p_N = 1 - (1-p_1)^N \approx 1 - e^{Np_1} \approx Np_1
\end{equation}
where the first approximation is valid when $N$ is large and $p_1$ is
small, and the second is valid when $Np_1$ is also small.  The nominal
detection threshold would be $p_N<10^{-2}$, however with $N\sim 10^8$
templates per band, the Gaussian approximation \eqref{e:CCpvalnaive}
will be invalid at the required small values of $p_1\lesssim10^{-10}$.
Since the methods described in \cite{LMXBCrossCorr} for estimating the
false-alarm probability more accurately were not implemented at the
time of the analysis, the na\"{\i}ve value $p_N$ was calculated.
Although the resulting values do not represent a realistic
quantification of the false-alarm probability (and are therefore not
shown in Figure\ref{fig:detoverview}), they can be compared to the
results of 49 searches over signal-free bands similar to the closed
signal bands of the MDC.  The lowest na\"{\i}ve $p$-value from those
searches, which wa s $p_N=7.5\times 10^{-5}$, provides an empirical
estimate of an actual $p$-value of $1.02\%$.  In comparison, the highest
na\"{\i}ve $p$-value for any closed signal was $p_N=3.4\times
10^{-129}$, so it is possible to comfortably declare all of them to be
confident detections, even without reliable $p$-values.

\subsection{Parameter estimation and errorbars}

The CrossCorr search is performed over a grid of templates in $f$,
$\asini$ and $\Tasc$, whose spacing is determined by the metric given
in \cite{LMXBCrossCorr}.  The spacing in each direction scales (in
cases where $\Tmax$ is small compared to the orbital period, as it was
for the \ac{MDC} analysis) as $\sqrt{m}/\Tmax$, where $m$ is the
chosen mismatch and $\Tmax$ is the maximum time separation allowed
between pairs of \ac{SFT}s to be included in the cross-correlation.
Additionally, the spacing in $\asini$ and $\Tasc$ each scale as
$1/f_0$ and the spacing in $\Tasc$ scales as $1/(\asini)$.  The
initial grid spacing was chosen to produce a mismatch of around
$0.25$.  A refined grid of $13\times13\times13$ points, with one-third
the original spacing, was then generated around the loudest candidate
signal in the band.  For some quieter signals there was a further
followup with a larger value of $\Tmax$ to produce an even finer
$13\times13\times13$ grid centered on the loudest candidate in the
refined grid.  For parameter estimation purposes, an interpolation
procedure was used, where the statistic values in a $3\times3\times3$
subgrid of the final fine grid, centered on the loudest value, were
fit to a multivariate quadratic.  The peak of this quadratic function
was chosen as the best estimate of the parameters, allowing them to be
estimated more accurately than the spacing of the final grid.  The
errorbars quoted for the CrossCorr results are a quadrature
combination of three effects: the usual statistical uncertainty
associated with the unknown noise realization (which scales inversely
with the observed $\rho$ value), an interpolation error estimated
using the residuals of the quadratic fit, and a systematic uncertainty
associated with the unknown value of $\cos\iota$.  Parameter estimates
were produced using this method for the open signals and compared to
the actual parameter values.  The actual offsets between estimated and
true parameters were mostly consistent with the estimated
uncertainties, aside from two observed effects described in the
following sections.  Analysis of this effects led to an
empirically-determined modifications to the procedure, and this
modified procedure was used to obtain the parameter estimates and
errorbars for the closed data.

\subsubsection{Empirical adjustment of \texorpdfstring{$\asini$}{asini} values}

Analysis of the open signals indicated that the $\asini$ estimates
were systematically lower than the true values, with the qualitative
feature that the underestimates were larger at lower frequencies.  The
explanation for this is unknown, but the dependence was assumed on
dimensional grounds to be inversely proportional to frequency.  The
proportionally constant was estimated from the open signals, and the
adjustment was made to replace the old estimate with
\begin{equation}
  (\asini)_{\text{est,new}} = (\asini)_{\text{est,old}} + \frac{0.028}{f_0}
\end{equation}
In order to produce conservative estimates of our parameter
uncertainties, this offset amount $0.028/f_0$ was combined in
quadrature with the other three contributions to the errorbars
reported for $\asini$.

\subsubsection{Empirical adjustment of statistical errorbars for loud
  signals}

A second discrepancy seen in the recovered parameters of the open
signals was that some of the loudest recovered signals ($\rho\gtrsim
300$) had recovered parameters (especially $\Tasc$) which were
significantly larger than the calculated erorrbars would suggest.  A
conjectured explanation is that the expressions used for generating
the statistical errorbars neglected higher order terms in the signal
amplitude $h_0$.  Scaling arguments indicate that the relative size of
such terms should be $\sim \rho/\sqrt{\Npair}$.  The statistical
errorbars for all parameters were thus increased by a factor of
$1+150\frac{\rho}{\sqrt{\Npair}}$, where the coefficient $150$ was
empirically determined to make the $\Tasc$ errorbars calculated from
the open data consistent with the actual parameter offsets.


\begin{thebibliography}{69}\makeatletter
\providecommand \@ifxundefined [1]{ \@ifx{#1\undefined}
}\providecommand \@ifnum [1]{ \ifnum #1\expandafter \@firstoftwo
 \else \expandafter \@secondoftwo
 \fi
}\providecommand \@ifx [1]{ \ifx #1\expandafter \@firstoftwo
 \else \expandafter \@secondoftwo
 \fi
}\providecommand \natexlab [1]{#1}\providecommand \enquote  [1]{``#1''}\providecommand \bibnamefont  [1]{#1}\providecommand \bibfnamefont [1]{#1}\providecommand \citenamefont [1]{#1}\providecommand \href@noop [0]{\@secondoftwo}\providecommand \href [0]{\begingroup \@sanitize@url \@href}\providecommand \@href[1]{\@@startlink{#1}\@@href}\providecommand \@@href[1]{\endgroup#1\@@endlink}\providecommand \@sanitize@url [0]{\catcode `\\12\catcode `\$12\catcode
  `\&12\catcode `\#12\catcode `\^12\catcode `\_12\catcode `\%12\relax}\providecommand \@@startlink[1]{}\providecommand \@@endlink[0]{}\providecommand \url  [0]{\begingroup\@sanitize@url \@url }\providecommand \@url [1]{\endgroup\@href {#1}{\urlprefix }}\providecommand \urlprefix  [0]{URL }\providecommand \Eprint [0]{\href }\providecommand \doibase [0]{http://dx.doi.org/}\providecommand \selectlanguage [0]{\@gobble}\providecommand \bibinfo  [0]{\@secondoftwo}\providecommand \bibfield  [0]{\@secondoftwo}\providecommand \translation [1]{[#1]}\providecommand \BibitemOpen [0]{}\providecommand \bibitemStop [0]{}\providecommand \bibitemNoStop [0]{.\EOS\space}\providecommand \EOS [0]{\spacefactor3000\relax}\providecommand \BibitemShut  [1]{\csname bibitem#1\endcsname}\let\auto@bib@innerbib\@empty
\bibitem [{\citenamefont {Aasi}\ \emph {et~al.}(2015)\citenamefont {Aasi} \emph
  {et~al.}}]{aligo}  \BibitemOpen
  \bibfield  {author} {\bibinfo {author} {\bibfnamefont {J.}~\bibnamefont
  {Aasi}} \emph {et~al.} (\bibinfo {collaboration} {LIGO Scientific
  Collaboration}),\ }\bibfield  {title} {\enquote {\bibinfo {title} {{Advanced
  LIGO}},}\ }\href {\doibase 10.1088/0264-9381/32/7/074001} {\bibfield
  {journal} {\bibinfo  {journal} {Class.\ Quant.\ Grav.}\ }\textbf {\bibinfo
  {volume} {32}},\ \bibinfo {pages} {074001} (\bibinfo {year} {2015})},\
  \Eprint {http://arxiv.org/abs/1411.4547} {arXiv:1411.4547} \BibitemShut
  {NoStop}\bibitem [{\citenamefont {Acernese}\ \emph
  {et~al.}(2015{\natexlab{a}})\citenamefont {Acernese} \emph {et~al.}}]{adv}  \BibitemOpen
  \bibfield  {author} {\bibinfo {author} {\bibfnamefont {F.}~\bibnamefont
  {Acernese}} \emph {et~al.} (\bibinfo {collaboration} {Virgo Collaboration}),\
  }\bibfield  {title} {\enquote {\bibinfo {title} {{Advanced Virgo: a
  second-generation interferometric gravitational wave detector}},}\ }\href
  {\doibase 10.1088/0264-9381/32/2/024001} {\bibfield  {journal} {\bibinfo
  {journal} {Class.\ Quant.\ Grav.}\ }\textbf {\bibinfo {volume} {32}},\
  \bibinfo {pages} {024001} (\bibinfo {year} {2015}{\natexlab{a}})},\ \Eprint
  {http://arxiv.org/abs/1408.3978} {arXiv:1408.3978} \BibitemShut {NoStop}\bibitem [{\citenamefont {{Papaloizou}}\ and\ \citenamefont
  {{Pringle}}(1978)}]{1978MNRAS.184..501P}  \BibitemOpen
  \bibfield  {author} {\bibinfo {author} {\bibfnamefont {J.}~\bibnamefont
  {{Papaloizou}}}\ and\ \bibinfo {author} {\bibfnamefont {J.~E.}\ \bibnamefont
  {{Pringle}}},\ }\bibfield  {title} {\enquote {\bibinfo {title}
  {{Gravitational radiation and the stability of rotating stars}},}\ }\href
  {\doibase 10.1093/mnras/184.3.501} {\bibfield  {journal} {\bibinfo  {journal}
  {Mon.\ Not.\ R.\ Astron.\ Soc.}\ }\textbf {\bibinfo {volume} {184}},\
  \bibinfo {pages} {501--508} (\bibinfo {year} {1978})}\BibitemShut {NoStop}\bibitem [{\citenamefont {{Wagoner}}(1984)}]{1984ApJ...278..345W}  \BibitemOpen
  \bibfield  {author} {\bibinfo {author} {\bibfnamefont {R.~V.}\ \bibnamefont
  {{Wagoner}}},\ }\bibfield  {title} {\enquote {\bibinfo {title}
  {{Gravitational radiation from accreting neutron stars}},}\ }\href {\doibase
  10.1086/161798} {\bibfield  {journal} {\bibinfo  {journal} {Astrophys.\ J.}\
  }\textbf {\bibinfo {volume} {278}},\ \bibinfo {pages} {345--348} (\bibinfo
  {year} {1984})}\BibitemShut {NoStop}\bibitem [{\citenamefont {{Bildsten}}(1998)}]{1998ApJ...501L..89B}  \BibitemOpen
  \bibfield  {author} {\bibinfo {author} {\bibfnamefont {L.}~\bibnamefont
  {{Bildsten}}},\ }\bibfield  {title} {\enquote {\bibinfo {title}
  {{Gravitational Radiation and Rotation of Accreting Neutron Stars}},}\ }\href
  {\doibase 10.1086/311440} {\bibfield  {journal} {\bibinfo  {journal}
  {Astrophys.\ J.\ Lett.}\ }\textbf {\bibinfo {volume} {501}},\ \bibinfo
  {pages} {L89--L93} (\bibinfo {year} {1998})},\ \Eprint
  {http://arxiv.org/abs/astro-ph/9804325} {astro-ph/9804325} \BibitemShut
  {NoStop}\bibitem [{\citenamefont {{Ushomirsky}}\ \emph {et~al.}(2000)\citenamefont
  {{Ushomirsky}}, \citenamefont {{Cutler}},\ and\ \citenamefont
  {{Bildsten}}}]{2000MNRAS.319..902U}  \BibitemOpen
  \bibfield  {author} {\bibinfo {author} {\bibfnamefont {G.}~\bibnamefont
  {{Ushomirsky}}}, \bibinfo {author} {\bibfnamefont {C.}~\bibnamefont
  {{Cutler}}}, \ and\ \bibinfo {author} {\bibfnamefont {L.}~\bibnamefont
  {{Bildsten}}},\ }\bibfield  {title} {\enquote {\bibinfo {title}
  {{Deformations of accreting neutron star crusts and gravitational wave
  emission}},}\ }\href {\doibase 10.1046/j.1365-8711.2000.03938.x} {\bibfield
  {journal} {\bibinfo  {journal} {Mon.\ Not.\ R.\ Astron.\ Soc.}\ }\textbf
  {\bibinfo {volume} {319}},\ \bibinfo {pages} {902--932} (\bibinfo {year}
  {2000})},\ \Eprint {http://arxiv.org/abs/astro-ph/0001136} {astro-ph/0001136}
  \BibitemShut {NoStop}\bibitem [{\citenamefont {{Cutler}}(2002)}]{2002PhRvD..66h4025C}  \BibitemOpen
  \bibfield  {author} {\bibinfo {author} {\bibfnamefont {C.}~\bibnamefont
  {{Cutler}}},\ }\bibfield  {title} {\enquote {\bibinfo {title} {{Gravitational
  waves from neutron stars with large toroidal B fields}},}\ }\href {\doibase
  10.1103/PhysRevD.66.084025} {\bibfield  {journal} {\bibinfo  {journal}
  {Phys.\ Rev.\ D}\ }\textbf {\bibinfo {volume} {66}},\ \bibinfo {eid} {084025}
  (\bibinfo {year} {2002})},\ \Eprint {http://arxiv.org/abs/gr-qc/0206051}
  {gr-qc/0206051} \BibitemShut {NoStop}\bibitem [{\citenamefont {Premachandra}(2015)}]{PremachandraThesis2015}  \BibitemOpen
  \bibfield  {author} {\bibinfo {author} {\bibfnamefont {S.}~\bibnamefont
  {Premachandra}},\ }\emph {\bibinfo {title} {Precision ephemerides of neutron
  star binaries to assist gravitational wave searches: {Sco X-1} \& {Cyg
  X-2}}},\ \href@noop {} {Ph.D. thesis},\ \bibinfo  {school} {Monash} (\bibinfo
  {year} {2015})\BibitemShut {NoStop}\bibitem [{\citenamefont {Abbott}\ \emph
  {et~al.}(2007{\natexlab{a}})\citenamefont {Abbott} \emph
  {et~al.}}]{2007PhRvD..76h2003A}  \BibitemOpen
  \bibfield  {author} {\bibinfo {author} {\bibfnamefont {B.}~\bibnamefont
  {Abbott}} \emph {et~al.} (\bibinfo {collaboration} {LIGO Scientific
  Collaboration}),\ }\bibfield  {title} {\enquote {\bibinfo {title} {{Upper
  limit map of a background of gravitational waves}},}\ }\href {\doibase
  10.1103/PhysRevD.76.082003} {\bibfield  {journal} {\bibinfo  {journal}
  {Phys.\ Rev.\ D}\ }\textbf {\bibinfo {volume} {76}},\ \bibinfo {eid} {082003}
  (\bibinfo {year} {2007}{\natexlab{a}})},\ \Eprint
  {http://arxiv.org/abs/astro-ph/0703234} {astro-ph/0703234} \BibitemShut
  {NoStop}\bibitem [{\citenamefont {Abbott}\ \emph
  {et~al.}(2007{\natexlab{b}})\citenamefont {Abbott} \emph
  {et~al.}}]{2007PhRvD..76h2001A}  \BibitemOpen
  \bibfield  {author} {\bibinfo {author} {\bibfnamefont {B.}~\bibnamefont
  {Abbott}} \emph {et~al.} (\bibinfo {collaboration} {LIGO Scientific
  Collaboration}),\ }\bibfield  {title} {\enquote {\bibinfo {title} {{Searches
  for periodic gravitational waves from unknown isolated sources and Scorpius
  X-1: Results from the second LIGO science run}},}\ }\href {\doibase
  10.1103/PhysRevD.76.082001} {\bibfield  {journal} {\bibinfo  {journal}
  {Phys.\ Rev.\ D}\ }\textbf {\bibinfo {volume} {76}},\ \bibinfo {eid} {082001}
  (\bibinfo {year} {2007}{\natexlab{b}})},\ \Eprint
  {http://arxiv.org/abs/gr-qc/0605028} {gr-qc/0605028} \BibitemShut {NoStop}\bibitem [{\citenamefont {Abbott}\ \emph {et~al.}(2011)\citenamefont {Abbott}
  \emph {et~al.}}]{2011PhRvL.107A1102A}  \BibitemOpen
  \bibfield  {author} {\bibinfo {author} {\bibfnamefont {B.~P.}\ \bibnamefont
  {Abbott}} \emph {et~al.} (\bibinfo {collaboration} {LIGO Scientific
  Collaboration and Virgo Collaboration}),\ }\bibfield  {title} {\enquote
  {\bibinfo {title} {Directional limits on persistent gravitational waves using
  ligo s5 science data},}\ }\href {\doibase 10.1103/PhysRevLett.107.271102}
  {\bibfield  {journal} {\bibinfo  {journal} {Phys.\ Rev.\ Lett.}\ }\textbf
  {\bibinfo {volume} {107}},\ \bibinfo {pages} {271102} (\bibinfo {year}
  {2011})},\ \Eprint {http://arxiv.org/abs/1109.1809} {arXiv:1109.1809}
  \BibitemShut {NoStop}\bibitem [{\citenamefont {Aasi}\ \emph
  {et~al.}(2014{\natexlab{a}})\citenamefont {Aasi} \emph
  {et~al.}}]{PhysRevD.90.062010}  \BibitemOpen
  \bibfield  {author} {\bibinfo {author} {\bibfnamefont {J.}~\bibnamefont
  {Aasi}} \emph {et~al.} (\bibinfo {collaboration} {LIGO Scientific
  Collaboration and Virgo Collaboration}),\ }\bibfield  {title} {\enquote
  {\bibinfo {title} {First all-sky search for continuous gravitational waves
  from unknown sources in binary systems},}\ }\href {\doibase
  10.1103/PhysRevD.90.062010} {\bibfield  {journal} {\bibinfo  {journal}
  {Phys.\ Rev.\ D}\ }\textbf {\bibinfo {volume} {90}},\ \bibinfo {pages}
  {062010} (\bibinfo {year} {2014}{\natexlab{a}})}\BibitemShut {NoStop}\bibitem [{\citenamefont {Aasi}\ \emph
  {et~al.}(2014{\natexlab{b}})\citenamefont {Aasi} \emph
  {et~al.}}]{S5sideband}  \BibitemOpen
  \bibfield  {author} {\bibinfo {author} {\bibfnamefont {J.}~\bibnamefont
  {Aasi}} \emph {et~al.} (\bibinfo {collaboration} {LIGO Scientific
  Collaboration and Virgo Collaboration}),\ }\bibfield  {title} {\enquote
  {\bibinfo {title} {{A directed search for gravitational waves from Scorpius
  X-1 with initial LIGO}},}\ }\href@noop {} {\  (\bibinfo {year}
  {2014}{\natexlab{b}})},\ \Eprint {http://arxiv.org/abs/1412.0605}
  {arXiv:1412.0605} \BibitemShut {NoStop}\bibitem [{\citenamefont {{The LIGO Scientific Collaboration}}\ \emph
  {et~al.}(2015)\citenamefont {{The LIGO Scientific Collaboration}},
  \citenamefont {{Aasi}}, \citenamefont {{Abbott}}, \citenamefont {{Abbott}},
  \citenamefont {{Abbott}}, \citenamefont {{Abernathy}}, \citenamefont
  {{Ackley}}, \citenamefont {{Adams}}, \citenamefont {{Adams}}, \citenamefont
  {{Addesso}},\ and\ \citenamefont {et~al.}}]{2015CQGra..32g4001T}  \BibitemOpen
  \bibfield  {author} {\bibinfo {author} {\bibnamefont {{The LIGO Scientific
  Collaboration}}}, \bibinfo {author} {\bibfnamefont {J.}~\bibnamefont
  {{Aasi}}}, \bibinfo {author} {\bibfnamefont {B.~P.}\ \bibnamefont
  {{Abbott}}}, \bibinfo {author} {\bibfnamefont {.}~\bibnamefont {{Abbott}}},
  \bibinfo {author} {\bibfnamefont {T.}~\bibnamefont {{Abbott}}}, \bibinfo
  {author} {\bibfnamefont {M.~R.}\ \bibnamefont {{Abernathy}}}, \bibinfo
  {author} {\bibfnamefont {K.}~\bibnamefont {{Ackley}}}, \bibinfo {author}
  {\bibfnamefont {C.}~\bibnamefont {{Adams}}}, \bibinfo {author} {\bibfnamefont
  {T.}~\bibnamefont {{Adams}}}, \bibinfo {author} {\bibfnamefont
  {P.}~\bibnamefont {{Addesso}}}, \ and\ \bibinfo {author} {\bibnamefont
  {et~al.}},\ }\bibfield  {title} {\enquote {\bibinfo {title} {{Advanced
  LIGO}},}\ }\href {\doibase 10.1088/0264-9381/32/7/074001} {\bibfield
  {journal} {\bibinfo  {journal} {Classical and Quantum Gravity}\ }\textbf
  {\bibinfo {volume} {32}},\ \bibinfo {eid} {074001} (\bibinfo {year}
  {2015})},\ \Eprint {http://arxiv.org/abs/1411.4547} {arXiv:1411.4547 [gr-qc]}
  \BibitemShut {NoStop}\bibitem [{\citenamefont {{LIGO Scientific Collaboration}}\ \emph
  {et~al.}(2013)\citenamefont {{LIGO Scientific Collaboration}}, \citenamefont
  {{Virgo Collaboration}}, \citenamefont {{Aasi}}, \citenamefont {{Abadie}},
  \citenamefont {{Abbott}}, \citenamefont {{Abbott}}, \citenamefont {{Abbott}},
  \citenamefont {{Abernathy}}, \citenamefont {{Accadia}}, \citenamefont
  {{Acernese}},\ and\ \citenamefont {et~al.}}]{2013arXiv1304.0670L}  \BibitemOpen
  \bibfield  {author} {\bibinfo {author} {\bibnamefont {{LIGO Scientific
  Collaboration}}}, \bibinfo {author} {\bibnamefont {{Virgo Collaboration}}},
  \bibinfo {author} {\bibfnamefont {J.}~\bibnamefont {{Aasi}}}, \bibinfo
  {author} {\bibfnamefont {J.}~\bibnamefont {{Abadie}}}, \bibinfo {author}
  {\bibfnamefont {B.~P.}\ \bibnamefont {{Abbott}}}, \bibinfo {author}
  {\bibfnamefont {R.}~\bibnamefont {{Abbott}}}, \bibinfo {author}
  {\bibfnamefont {T.~D.}\ \bibnamefont {{Abbott}}}, \bibinfo {author}
  {\bibfnamefont {M.}~\bibnamefont {{Abernathy}}}, \bibinfo {author}
  {\bibfnamefont {T.}~\bibnamefont {{Accadia}}}, \bibinfo {author}
  {\bibfnamefont {F.}~\bibnamefont {{Acernese}}}, \ and\ \bibinfo {author}
  {\bibnamefont {et~al.}},\ }\bibfield  {title} {\enquote {\bibinfo {title}
  {{Prospects for Localization of Gravitational Wave Transients by the Advanced
  LIGO and Advanced Virgo Observatories}},}\ }\href@noop {} {\bibfield
  {journal} {\bibinfo  {journal} {ArXiv e-prints}\ } (\bibinfo {year}
  {2013})},\ \Eprint {http://arxiv.org/abs/1304.0670} {arXiv:1304.0670 [gr-qc]}
  \BibitemShut {NoStop}\bibitem [{\citenamefont {Harry}(2010)}]{Harry:2010zz}  \BibitemOpen
  \bibfield  {author} {\bibinfo {author} {\bibfnamefont {Gregory~M.}\
  \bibnamefont {Harry}} (\bibinfo {collaboration} {LIGO Scientific}),\
  }\bibfield  {title} {\enquote {\bibinfo {title} {{Advanced LIGO: The next
  generation of gravitational wave detectors}},}\ }\href {\doibase
  10.1088/0264-9381/27/8/084006} {\bibfield  {journal} {\bibinfo  {journal}
  {Class.Quant.Grav.}\ }\textbf {\bibinfo {volume} {27}},\ \bibinfo {pages}
  {084006} (\bibinfo {year} {2010})}\BibitemShut {NoStop}\bibitem [{\citenamefont {Acernese}\ \emph
  {et~al.}(2015{\natexlab{b}})\citenamefont {Acernese} \emph
  {et~al.}}]{TheVirgo:2014hva}  \BibitemOpen
  \bibfield  {author} {\bibinfo {author} {\bibfnamefont {F.}~\bibnamefont
  {Acernese}} \emph {et~al.} (\bibinfo {collaboration} {VIRGO}),\ }\bibfield
  {title} {\enquote {\bibinfo {title} {{Advanced Virgo: a second-generation
  interferometric gravitational wave detector}},}\ }\href {\doibase
  10.1088/0264-9381/32/2/024001} {\bibfield  {journal} {\bibinfo  {journal}
  {Class.Quant.Grav.}\ }\textbf {\bibinfo {volume} {32}},\ \bibinfo {pages}
  {024001} (\bibinfo {year} {2015}{\natexlab{b}})},\ \Eprint
  {http://arxiv.org/abs/1408.3978} {arXiv:1408.3978 [gr-qc]} \BibitemShut
  {NoStop}\bibitem [{\citenamefont {{Roy}}\ \emph {et~al.}(2015)\citenamefont {{Roy}}
  \emph {et~al.}}]{2015ApJ...800L..12R}  \BibitemOpen
  \bibfield  {author} {\bibinfo {author} {\bibfnamefont {J.}~\bibnamefont
  {{Roy}}} \emph {et~al.},\ }\bibfield  {title} {\enquote {\bibinfo {title}
  {{Discovery of Psr J1227-4853: A Transition from a Low-mass X-Ray Binary to a
  Redback Millisecond Pulsar}},}\ }\href {\doibase 10.1088/2041-8205/800/1/L12}
  {\bibfield  {journal} {\bibinfo  {journal} {Astrophys.\ J.\ Lett.}\ }\textbf
  {\bibinfo {volume} {800}},\ \bibinfo {eid} {L12} (\bibinfo {year} {2015})},\
  \Eprint {http://arxiv.org/abs/1412.4735} {arXiv:1412.4735} \BibitemShut
  {NoStop}\bibitem [{\citenamefont {{Deller}}\ \emph {et~al.}(2014)\citenamefont
  {{Deller}} \emph {et~al.}}]{2014arXiv1412.5155D}  \BibitemOpen
  \bibfield  {author} {\bibinfo {author} {\bibfnamefont {A.~T.}\ \bibnamefont
  {{Deller}}} \emph {et~al.},\ }\bibfield  {title} {\enquote {\bibinfo {title}
  {{Radio imaging observations of PSR J1023+0038 in an LMXB state}},}\
  }\href@noop {} {\  (\bibinfo {year} {2014})},\ \Eprint
  {http://arxiv.org/abs/1412.5155} {arXiv:1412.5155} \BibitemShut {NoStop}\bibitem [{\citenamefont {{Tendulkar}}\ \emph {et~al.}(2014)\citenamefont
  {{Tendulkar}} \emph {et~al.}}]{2014ApJ...791...77T}  \BibitemOpen
  \bibfield  {author} {\bibinfo {author} {\bibfnamefont {S.~P.}\ \bibnamefont
  {{Tendulkar}}} \emph {et~al.},\ }\bibfield  {title} {\enquote {\bibinfo
  {title} {{NuSTAR Observations of the State Transition of Millisecond Pulsar
  Binary PSR J1023+0038}},}\ }\href {\doibase 10.1088/0004-637X/791/2/77}
  {\bibfield  {journal} {\bibinfo  {journal} {Astrophys.\ J.}\ }\textbf
  {\bibinfo {volume} {791}},\ \bibinfo {eid} {77} (\bibinfo {year} {2014})},\
  \Eprint {http://arxiv.org/abs/1406.7009} {arXiv:1406.7009} \BibitemShut
  {NoStop}\bibitem [{\citenamefont {{Bogdanov}}\ \emph {et~al.}(2014)\citenamefont
  {{Bogdanov}} \emph {et~al.}}]{2014arXiv1412.5145B}  \BibitemOpen
  \bibfield  {author} {\bibinfo {author} {\bibfnamefont {S.}~\bibnamefont
  {{Bogdanov}}} \emph {et~al.},\ }\bibfield  {title} {\enquote {\bibinfo
  {title} {{Coordinated X-ray, Ultraviolet, Optical, and Radio Observations of
  the PSR J1023+0038 System in a Low-mass X-ray Binary State}},}\ }\href@noop
  {} {\  (\bibinfo {year} {2014})},\ \Eprint {http://arxiv.org/abs/1412.5145}
  {arXiv:1412.5145} \BibitemShut {NoStop}\bibitem [{\citenamefont {{Steeghs}}\ and\ \citenamefont
  {{Casares}}(2002)}]{2002ApJ...568..273S}  \BibitemOpen
  \bibfield  {author} {\bibinfo {author} {\bibfnamefont {D.}~\bibnamefont
  {{Steeghs}}}\ and\ \bibinfo {author} {\bibfnamefont {J.}~\bibnamefont
  {{Casares}}},\ }\bibfield  {title} {\enquote {\bibinfo {title} {{The Mass
  Donor of Scorpius X-1 Revealed}},}\ }\href {\doibase 10.1086/339224}
  {\bibfield  {journal} {\bibinfo  {journal} {Astrophys.\ J.}\ }\textbf
  {\bibinfo {volume} {568}},\ \bibinfo {pages} {273--278} (\bibinfo {year}
  {2002})},\ \Eprint {http://arxiv.org/abs/arXiv:astro-ph/0107343}
  {arXiv:astro-ph/0107343} \BibitemShut {NoStop}\bibitem [{\citenamefont {{Galloway}}\ \emph {et~al.}(2014)\citenamefont
  {{Galloway}}, \citenamefont {{Premachandra}}, \citenamefont {{Steeghs}},
  \citenamefont {{Marsh}}, \citenamefont {{Casares}},\ and\ \citenamefont
  {{Cornelisse}}}]{2014ApJ...781...14G}  \BibitemOpen
  \bibfield  {author} {\bibinfo {author} {\bibfnamefont {D.~K.}\ \bibnamefont
  {{Galloway}}}, \bibinfo {author} {\bibfnamefont {S.}~\bibnamefont
  {{Premachandra}}}, \bibinfo {author} {\bibfnamefont {D.}~\bibnamefont
  {{Steeghs}}}, \bibinfo {author} {\bibfnamefont {T.}~\bibnamefont {{Marsh}}},
  \bibinfo {author} {\bibfnamefont {J.}~\bibnamefont {{Casares}}}, \ and\
  \bibinfo {author} {\bibfnamefont {R.}~\bibnamefont {{Cornelisse}}},\
  }\bibfield  {title} {\enquote {\bibinfo {title} {{Precision Ephemerides for
  Gravitational-wave Searches. I. Sco X-1}},}\ }\href {\doibase
  10.1088/0004-637X/781/1/14} {\bibfield  {journal} {\bibinfo  {journal}
  {Astrophys.\ J.}\ }\textbf {\bibinfo {volume} {781}},\ \bibinfo {eid} {14}
  (\bibinfo {year} {2014})},\ \Eprint {http://arxiv.org/abs/1311.6246}
  {arXiv:1311.6246} \BibitemShut {NoStop}\bibitem [{\citenamefont {{Watts}}\ \emph {et~al.}(2008)\citenamefont
  {{Watts}}, \citenamefont {{Krishnan}}, \citenamefont {{Bildsten}},\ and\
  \citenamefont {{Schutz}}}]{2008MNRAS.389..839W}  \BibitemOpen
  \bibfield  {author} {\bibinfo {author} {\bibfnamefont {A.~L.}\ \bibnamefont
  {{Watts}}}, \bibinfo {author} {\bibfnamefont {B.}~\bibnamefont {{Krishnan}}},
  \bibinfo {author} {\bibfnamefont {L.}~\bibnamefont {{Bildsten}}}, \ and\
  \bibinfo {author} {\bibfnamefont {B.~F.}\ \bibnamefont {{Schutz}}},\
  }\bibfield  {title} {\enquote {\bibinfo {title} {{Detecting gravitational
  wave emission from the known accreting neutron stars}},}\ }\href {\doibase
  10.1111/j.1365-2966.2008.13594.x} {\bibfield  {journal} {\bibinfo  {journal}
  {Mon.\ Not.\ R.\ Astron.\ Soc.}\ }\textbf {\bibinfo {volume} {389}},\
  \bibinfo {pages} {839--868} (\bibinfo {year} {2008})},\ \Eprint
  {http://arxiv.org/abs/0803.4097} {arXiv:0803.4097} \BibitemShut {NoStop}\bibitem [{\citenamefont {{Haskell}}\ \emph {et~al.}(2015)\citenamefont
  {{Haskell}}, \citenamefont {{Priymak}}, \citenamefont {{Patruno}},
  \citenamefont {{Oppenoorth}}, \citenamefont {{Melatos}},\ and\ \citenamefont
  {{Lasky}}}]{2015arXiv150106039H}  \BibitemOpen
  \bibfield  {author} {\bibinfo {author} {\bibfnamefont {B.}~\bibnamefont
  {{Haskell}}}, \bibinfo {author} {\bibfnamefont {M.}~\bibnamefont
  {{Priymak}}}, \bibinfo {author} {\bibfnamefont {A.}~\bibnamefont
  {{Patruno}}}, \bibinfo {author} {\bibfnamefont {M.}~\bibnamefont
  {{Oppenoorth}}}, \bibinfo {author} {\bibfnamefont {A.}~\bibnamefont
  {{Melatos}}}, \ and\ \bibinfo {author} {\bibfnamefont {P.}~\bibnamefont
  {{Lasky}}},\ }\bibfield  {title} {\enquote {\bibinfo {title} {{Detecting
  gravitational waves from mountains on neutron stars in the Advanced Detector
  Era}},}\ }\href@noop {} {\  (\bibinfo {year} {2015})},\ \Eprint
  {http://arxiv.org/abs/1501.06039} {arXiv:1501.06039} \BibitemShut {NoStop}\bibitem [{\citenamefont {Bradshaw}\ \emph {et~al.}(1999)\citenamefont
  {Bradshaw}, \citenamefont {Fomalont},\ and\ \citenamefont
  {Geldzahler}}]{1538-4357-512-2-L121}  \BibitemOpen
  \bibfield  {author} {\bibinfo {author} {\bibfnamefont {C.~F.}\ \bibnamefont
  {Bradshaw}}, \bibinfo {author} {\bibfnamefont {E.~B.}\ \bibnamefont
  {Fomalont}}, \ and\ \bibinfo {author} {\bibfnamefont {B.~J.}\ \bibnamefont
  {Geldzahler}},\ }\href {\doibase 10.1086/311889} {\bibfield  {journal}
  {\bibinfo  {journal} {Astrophys.\ J.\ Lett.}\ }\textbf {\bibinfo {volume}
  {512}},\ \bibinfo {pages} {L121} (\bibinfo {year} {1999})}\BibitemShut
  {NoStop}\bibitem [{\citenamefont {{Liu}}\ \emph {et~al.}(2007)\citenamefont {{Liu}},
  \citenamefont {{van Paradijs}},\ and\ \citenamefont {{van den
  Heuvel}}}]{lmxb07}  \BibitemOpen
  \bibfield  {author} {\bibinfo {author} {\bibfnamefont {Q.~Z.}\ \bibnamefont
  {{Liu}}}, \bibinfo {author} {\bibfnamefont {J.}~\bibnamefont {{van
  Paradijs}}}, \ and\ \bibinfo {author} {\bibfnamefont {E.~P.~J.}\ \bibnamefont
  {{van den Heuvel}}},\ }\bibfield  {title} {\enquote {\bibinfo {title} {{A
  catalogue of low-mass X-ray binaries in the Galaxy, LMC, and SMC (Fourth
  edition)}},}\ }\href {\doibase 10.1051/0004-6361:20077303} {\bibfield
  {journal} {\bibinfo  {journal} {Astron.\ Astrophys.}\ }\textbf {\bibinfo
  {volume} {469}},\ \bibinfo {pages} {807--810} (\bibinfo {year} {2007})},\
  \Eprint {http://arxiv.org/abs/arXiv:0707.0544} {arXiv:0707.0544} \BibitemShut
  {NoStop}\bibitem [{\citenamefont {Leaci}\ and\ \citenamefont
  {Prix}(2015)}]{2015arXiv150200914L}  \BibitemOpen
  \bibfield  {author} {\bibinfo {author} {\bibfnamefont {Paola}\ \bibnamefont
  {Leaci}}\ and\ \bibinfo {author} {\bibfnamefont {Reinhard}\ \bibnamefont
  {Prix}},\ }\bibfield  {title} {\enquote {\bibinfo {title} {{Directed searches
  for continuous gravitational waves from binary systems: parameter-space
  metrics and optimal Scorpius X-1 sensitivity}},}\ }\href@noop {} {\
  (\bibinfo {year} {2015})},\ \Eprint {http://arxiv.org/abs/1502.00914}
  {arXiv:1502.00914} \BibitemShut {NoStop}\bibitem [{\citenamefont {{Middleditch}}\ and\ \citenamefont
  {{Priedhorsky}}(1986)}]{1986ApJ...306..230M}  \BibitemOpen
  \bibfield  {author} {\bibinfo {author} {\bibfnamefont {J.}~\bibnamefont
  {{Middleditch}}}\ and\ \bibinfo {author} {\bibfnamefont {W.~C.}\ \bibnamefont
  {{Priedhorsky}}},\ }\bibfield  {title} {\enquote {\bibinfo {title}
  {{Discovery of rapid quasi-periodic oscillations in Scorpius X-1}},}\ }\href
  {\doibase 10.1086/164335} {\bibfield  {journal} {\bibinfo  {journal}
  {Astrophys.\ J.}\ }\textbf {\bibinfo {volume} {306}},\ \bibinfo {pages}
  {230--237} (\bibinfo {year} {1986})}\BibitemShut {NoStop}\bibitem [{\citenamefont {{Wood}}\ \emph {et~al.}(1991)\citenamefont {{Wood}}
  \emph {et~al.}}]{1991ApJ...379..295W}  \BibitemOpen
  \bibfield  {author} {\bibinfo {author} {\bibfnamefont {K.~S.}\ \bibnamefont
  {{Wood}}} \emph {et~al.},\ }\bibfield  {title} {\enquote {\bibinfo {title}
  {{Searches for millisecond pulsations in low-mass X-ray binaries}},}\ }\href
  {\doibase 10.1086/170505} {\bibfield  {journal} {\bibinfo  {journal}
  {Astrophys.\ J.}\ }\textbf {\bibinfo {volume} {379}},\ \bibinfo {pages}
  {295--309} (\bibinfo {year} {1991})}\BibitemShut {NoStop}\bibitem [{\citenamefont {{Hertz}}\ \emph {et~al.}(1992)\citenamefont {{Hertz}}
  \emph {et~al.}}]{1992ApJ...396..201H}  \BibitemOpen
  \bibfield  {author} {\bibinfo {author} {\bibfnamefont {P.}~\bibnamefont
  {{Hertz}}} \emph {et~al.},\ }\bibfield  {title} {\enquote {\bibinfo {title}
  {{X-ray variability of Scorpius X-1 during a multiwavelength campaign}},}\
  }\href {\doibase 10.1086/171710} {\bibfield  {journal} {\bibinfo  {journal}
  {Astrophys.\ J.}\ }\textbf {\bibinfo {volume} {396}},\ \bibinfo {pages}
  {201--218} (\bibinfo {year} {1992})}\BibitemShut {NoStop}\bibitem [{\citenamefont {{Vaughan}}\ \emph {et~al.}(1994)\citenamefont
  {{Vaughan}} \emph {et~al.}}]{1994ApJ...435..362V}  \BibitemOpen
  \bibfield  {author} {\bibinfo {author} {\bibfnamefont {B.~A.}\ \bibnamefont
  {{Vaughan}}} \emph {et~al.},\ }\bibfield  {title} {\enquote {\bibinfo {title}
  {{Searches for millisecond pulsations in low-mass X-ray binaries, 2}},}\
  }\href {\doibase 10.1086/174818} {\bibfield  {journal} {\bibinfo  {journal}
  {Astrophys.\ J.}\ }\textbf {\bibinfo {volume} {435}},\ \bibinfo {pages}
  {362--371} (\bibinfo {year} {1994})}\BibitemShut {NoStop}\bibitem [{\citenamefont {{Jahoda}}\ \emph {et~al.}(1996)\citenamefont
  {{Jahoda}} \emph {et~al.}}]{xte96}  \BibitemOpen
  \bibfield  {author} {\bibinfo {author} {\bibfnamefont {K.}~\bibnamefont
  {{Jahoda}}} \emph {et~al.},\ }\bibfield  {title} {\enquote {\bibinfo {title}
  {In-orbit performance and calibration of the \protect{Rossi X-ray Timing
  Explorer ({\it RXTE}\/) Proportional Counter Array (PCA)}},}\ }\href
  {http://adsabs.harvard.edu/cgi-bin/nph-bib_query?bibcode=1996SPIE.2808...59J&db_key=AST}
  {\bibfield  {journal} {\bibinfo  {journal} {Proc.\ SPIE}\ }\textbf {\bibinfo
  {volume} {2808}},\ \bibinfo {pages} {59--70} (\bibinfo {year}
  {1996})}\BibitemShut {NoStop}\bibitem [{\citenamefont {{van der Klis}}\ \emph {et~al.}(1996)\citenamefont
  {{van der Klis}} \emph {et~al.}}]{1996ApJ...469L...1V}  \BibitemOpen
  \bibfield  {author} {\bibinfo {author} {\bibfnamefont {M.}~\bibnamefont {{van
  der Klis}}} \emph {et~al.},\ }\bibfield  {title} {\enquote {\bibinfo {title}
  {{Discovery of Submillisecond Quasi-periodic Oscillations in the X-Ray Flux
  of Scorpius X-1}},}\ }\href {\doibase 10.1086/310251} {\bibfield  {journal}
  {\bibinfo  {journal} {Astrophys.\ J.\ Lett.}\ }\textbf {\bibinfo {volume}
  {469}},\ \bibinfo {pages} {L1} (\bibinfo {year} {1996})},\ \Eprint
  {http://arxiv.org/abs/astro-ph/9607047} {astro-ph/9607047} \BibitemShut
  {NoStop}\bibitem [{\citenamefont {{van der Klis}}\ \emph {et~al.}(1997)\citenamefont
  {{van der Klis}}, \citenamefont {{Wijnands}}, \citenamefont {{Horne}},\ and\
  \citenamefont {{Chen}}}]{1997ApJ...481L..97V}  \BibitemOpen
  \bibfield  {author} {\bibinfo {author} {\bibfnamefont {M.}~\bibnamefont {{van
  der Klis}}}, \bibinfo {author} {\bibfnamefont {R.~A.~D.}\ \bibnamefont
  {{Wijnands}}}, \bibinfo {author} {\bibfnamefont {K.}~\bibnamefont {{Horne}}},
  \ and\ \bibinfo {author} {\bibfnamefont {W.}~\bibnamefont {{Chen}}},\
  }\bibfield  {title} {\enquote {\bibinfo {title} {{Kilohertz Quasi-Periodic
  Oscillation Peak Separation Is Not Constant in Scorpius X-1}},}\ }\href
  {\doibase 10.1086/310656} {\bibfield  {journal} {\bibinfo  {journal}
  {Astrophys.\ J.\ Lett.}\ }\textbf {\bibinfo {volume} {481}},\ \bibinfo
  {pages} {L97--L100} (\bibinfo {year} {1997})},\ \Eprint
  {http://arxiv.org/abs/astro-ph/9703025} {astro-ph/9703025} \BibitemShut
  {NoStop}\bibitem [{\citenamefont {{M{\'e}ndez}}\ and\ \citenamefont {{van der
  Klis}}(2000)}]{2000MNRAS.318..938M}  \BibitemOpen
  \bibfield  {author} {\bibinfo {author} {\bibfnamefont {M.}~\bibnamefont
  {{M{\'e}ndez}}}\ and\ \bibinfo {author} {\bibfnamefont {M.}~\bibnamefont
  {{van der Klis}}},\ }\bibfield  {title} {\enquote {\bibinfo {title} {{The
  harmonic and sideband structure of the kilohertz quasi-periodic oscillations
  in Sco X-1}},}\ }\href {\doibase 10.1046/j.1365-8711.2000.03788.x} {\bibfield
   {journal} {\bibinfo  {journal} {Mon.\ Not.\ R.\ Astron.\ Soc.}\ }\textbf
  {\bibinfo {volume} {318}},\ \bibinfo {pages} {938--942} (\bibinfo {year}
  {2000})},\ \Eprint {http://arxiv.org/abs/astro-ph/0006243} {astro-ph/0006243}
  \BibitemShut {NoStop}\bibitem [{\citenamefont {{Skrutskie}}\ \emph {et~al.}(2006)\citenamefont
  {{Skrutskie}} \emph {et~al.}}]{2mass06}  \BibitemOpen
  \bibfield  {author} {\bibinfo {author} {\bibfnamefont {M.~F.}\ \bibnamefont
  {{Skrutskie}}} \emph {et~al.},\ }\bibfield  {title} {\enquote {\bibinfo
  {title} {{The Two Micron All Sky Survey (2MASS)}},}\ }\href {\doibase
  10.1086/498708} {\bibfield  {journal} {\bibinfo  {journal} {The Astronomical
  Journal}\ }\textbf {\bibinfo {volume} {131}},\ \bibinfo {pages} {1163--1183}
  (\bibinfo {year} {2006})}\BibitemShut {NoStop}\bibitem [{\citenamefont {{Fomalont}}\ \emph {et~al.}(2001)\citenamefont
  {{Fomalont}}, \citenamefont {{Geldzahler}},\ and\ \citenamefont
  {{Bradshaw}}}]{2001ApJ...558..283F}  \BibitemOpen
  \bibfield  {author} {\bibinfo {author} {\bibfnamefont {E.~B.}\ \bibnamefont
  {{Fomalont}}}, \bibinfo {author} {\bibfnamefont {B.~J.}\ \bibnamefont
  {{Geldzahler}}}, \ and\ \bibinfo {author} {\bibfnamefont {C.~F.}\
  \bibnamefont {{Bradshaw}}},\ }\bibfield  {title} {\enquote {\bibinfo {title}
  {{Scorpius X-1: The Evolution and Nature of the Twin Compact Radio Lobes}},}\
  }\href {\doibase 10.1086/322479} {\bibfield  {journal} {\bibinfo  {journal}
  {\apj}\ }\textbf {\bibinfo {volume} {558}},\ \bibinfo {pages} {283--301}
  (\bibinfo {year} {2001})},\ \Eprint {http://arxiv.org/abs/astro-ph/0104372}
  {astro-ph/0104372} \BibitemShut {NoStop}\bibitem [{\citenamefont {{Bildsten}}\ \emph {et~al.}(1997)\citenamefont
  {{Bildsten}} \emph {et~al.}}]{1997ApJS..113..367B}  \BibitemOpen
  \bibfield  {author} {\bibinfo {author} {\bibfnamefont {L.}~\bibnamefont
  {{Bildsten}}} \emph {et~al.},\ }\bibfield  {title} {\enquote {\bibinfo
  {title} {{Observations of Accreting Pulsars}},}\ }\href {\doibase
  10.1086/313060} {\bibfield  {journal} {\bibinfo  {journal} {Astrophys.\ J.\
  Supp.}\ }\textbf {\bibinfo {volume} {113}},\ \bibinfo {pages} {367--408}
  (\bibinfo {year} {1997})},\ \Eprint {http://arxiv.org/abs/astro-ph/9707125}
  {astro-ph/9707125} \BibitemShut {NoStop}\bibitem [{\citenamefont {{Baykal}}\ and\ \citenamefont
  {{Oegelman}}(1993)}]{1993AandA...267..119B}  \BibitemOpen
  \bibfield  {author} {\bibinfo {author} {\bibfnamefont {A.}~\bibnamefont
  {{Baykal}}}\ and\ \bibinfo {author} {\bibfnamefont {H.}~\bibnamefont
  {{Oegelman}}},\ }\bibfield  {title} {\enquote {\bibinfo {title} {{An
  empirical torque noise and spin-up model for accretion-powered X-ray
  pulsars}},}\ }\href {http://adsabs.harvard.edu/abs/1993A&amp;A...267..119B}
  {\bibfield  {journal} {\bibinfo  {journal} {Astron.\ Astrophys.}\ }\textbf
  {\bibinfo {volume} {267}},\ \bibinfo {pages} {119--125} (\bibinfo {year}
  {1993})}\BibitemShut {NoStop}\bibitem [{\citenamefont {{King}}(2006)}]{2006csxs.book..507K}  \BibitemOpen
  \bibfield  {author} {\bibinfo {author} {\bibfnamefont {A.~R.}\ \bibnamefont
  {{King}}},\ }\enquote {\bibinfo {title} {{Accretion in compact binaries}},}\
  in\ \href@noop {} {\emph {\bibinfo {booktitle} {Compact stellar X-ray
  sources}}},\ \bibinfo {editor} {edited by\ \bibinfo {editor} {\bibfnamefont
  {W.~H.~G.}\ \bibnamefont {{Lewin}}}\ and\ \bibinfo {editor} {\bibfnamefont
  {M.}~\bibnamefont {{van der Klis}}}}\ (\bibinfo {year} {2006})\ pp.\ \bibinfo
  {pages} {507--546}\BibitemShut {NoStop}\bibitem [{\citenamefont {{van~der~Putten}}\ \emph {et~al.}(2010)\citenamefont
  {{van~der~Putten}}, \citenamefont {{Bulten}}, \citenamefont
  {{van~den~Brand}},\ and\ \citenamefont {{Holtrop}}}]{2010JPhCS.228a2005V}  \BibitemOpen
  \bibfield  {author} {\bibinfo {author} {\bibfnamefont {S.}~\bibnamefont
  {{van~der~Putten}}}, \bibinfo {author} {\bibfnamefont {H.~J.}\ \bibnamefont
  {{Bulten}}}, \bibinfo {author} {\bibfnamefont {J.~F.~J.}\ \bibnamefont
  {{van~den~Brand}}}, \ and\ \bibinfo {author} {\bibfnamefont {M.}~\bibnamefont
  {{Holtrop}}},\ }\bibfield  {title} {\enquote {\bibinfo {title} {{Searching
  for gravitational waves from pulsars in binary systems: An all-sky
  search}},}\ }\href {\doibase 10.1088/1742-6596/228/1/012005} {\bibfield
  {journal} {\bibinfo  {journal} {J.\ Phys.\ Conf.\ Ser.}\ }\textbf {\bibinfo
  {volume} {228}},\ \bibinfo {eid} {012005} (\bibinfo {year}
  {2010})}\BibitemShut {NoStop}\bibitem [{\citenamefont {Bonferroni}(1935)}]{Bonferroni35}  \BibitemOpen
  \bibfield  {author} {\bibinfo {author} {\bibfnamefont {C.~E.}\ \bibnamefont
  {Bonferroni}},\ }\bibfield  {title} {\enquote {\bibinfo {title} {Il calcolo
  delle assicurazioni su gruppi di teste},}\ }\bibfield  {booktitle} {\emph
  {\bibinfo {booktitle} {Studi in Onore del Professore Salvatore Ortu
  Carboni}},\ }\href@noop {} {\ ,\ \bibinfo {pages} {13--60} (\bibinfo {year}
  {1935})}\BibitemShut {NoStop}\bibitem [{\citenamefont {Bonferroni}(1936)}]{Bonferroni36}  \BibitemOpen
  \bibfield  {author} {\bibinfo {author} {\bibfnamefont {C.~E.}\ \bibnamefont
  {Bonferroni}},\ }\bibfield  {title} {\enquote {\bibinfo {title} {Teoria
  statistica delle classi e calcolo delle probabilit\`a},}\ }\href@noop {}
  {\bibfield  {journal} {\bibinfo  {journal} {Pubblicazioni del R Istituto
  Superiore di Scienze Economiche e Commerciali di Firenze}\ }\textbf {\bibinfo
  {volume} {8}},\ \bibinfo {pages} {3--62} (\bibinfo {year}
  {1936})}\BibitemShut {NoStop}\bibitem [{\citenamefont {Ballmer}(2006)}]{Ballmer2006CQG}  \BibitemOpen
  \bibfield  {author} {\bibinfo {author} {\bibfnamefont {Stefan~W.}\
  \bibnamefont {Ballmer}},\ }\bibfield  {title} {\enquote {\bibinfo {title} {{A
  radiometer for stochastic gravitational waves}},}\ }\href@noop {} {\bibfield
  {journal} {\bibinfo  {journal} {Class.\ Quant.\ Grav.}\ }\textbf {\bibinfo
  {volume} {23}},\ \bibinfo {pages} {S179--S186} (\bibinfo {year} {2006})},\
  \Eprint {http://arxiv.org/abs/gr-qc/0510096} {gr-qc/0510096} \BibitemShut
  {NoStop}\bibitem [{\citenamefont {Messenger}\ and\ \citenamefont
  {Woan}(2007)}]{Messenger2007CQG}  \BibitemOpen
  \bibfield  {author} {\bibinfo {author} {\bibfnamefont {C.}~\bibnamefont
  {Messenger}}\ and\ \bibinfo {author} {\bibfnamefont {G.}~\bibnamefont
  {Woan}},\ }\bibfield  {title} {\enquote {\bibinfo {title} {{A Fast search
  strategy for gravitational waves from low-mass X-ray binaries}},}\ }\href
  {\doibase 10.1088/0264-9381/24/19/S10} {\bibfield  {journal} {\bibinfo
  {journal} {Class.\ Quant.\ Grav.}\ }\textbf {\bibinfo {volume} {24}},\
  \bibinfo {pages} {S469--S480} (\bibinfo {year} {2007})},\ \Eprint
  {http://arxiv.org/abs/gr-qc/0703155} {arXiv:gr-qc/0703155} \BibitemShut
  {NoStop}\bibitem [{\citenamefont {{Sammut}}\ \emph {et~al.}(2014)\citenamefont
  {{Sammut}}, \citenamefont {{Messenger}}, \citenamefont {{Melatos}},\ and\
  \citenamefont {{Owen}}}]{2014PhRvD..89d3001S}  \BibitemOpen
  \bibfield  {author} {\bibinfo {author} {\bibfnamefont {L.}~\bibnamefont
  {{Sammut}}}, \bibinfo {author} {\bibfnamefont {C.}~\bibnamefont
  {{Messenger}}}, \bibinfo {author} {\bibfnamefont {A.}~\bibnamefont
  {{Melatos}}}, \ and\ \bibinfo {author} {\bibfnamefont {B.~J.}\ \bibnamefont
  {{Owen}}},\ }\bibfield  {title} {\enquote {\bibinfo {title} {{Implementation
  of the frequency-modulated sideband search method for gravitational waves
  from low mass x-ray binaries}},}\ }\href {\doibase
  10.1103/PhysRevD.89.043001} {\bibfield  {journal} {\bibinfo  {journal}
  {Phys.\ Rev.\ D}\ }\textbf {\bibinfo {volume} {89}},\ \bibinfo {eid} {043001}
  (\bibinfo {year} {2014})},\ \Eprint {http://arxiv.org/abs/1311.1379}
  {arXiv:1311.1379} \BibitemShut {NoStop}\bibitem [{\citenamefont {{Ransom}}\ \emph {et~al.}(2003)\citenamefont
  {{Ransom}}, \citenamefont {{Cordes}},\ and\ \citenamefont
  {{Eikenberry}}}]{2003ApJ...589..911R}  \BibitemOpen
  \bibfield  {author} {\bibinfo {author} {\bibfnamefont {S.~M.}\ \bibnamefont
  {{Ransom}}}, \bibinfo {author} {\bibfnamefont {J.~M.}\ \bibnamefont
  {{Cordes}}}, \ and\ \bibinfo {author} {\bibfnamefont {S.~S.}\ \bibnamefont
  {{Eikenberry}}},\ }\bibfield  {title} {\enquote {\bibinfo {title} {{A New
  Search Technique for Short Orbital Period Binary Pulsars}},}\ }\href
  {\doibase 10.1086/374806} {\bibfield  {journal} {\bibinfo  {journal}
  {Astrophys.\ J.}\ }\textbf {\bibinfo {volume} {589}},\ \bibinfo {pages}
  {911--920} (\bibinfo {year} {2003})},\ \Eprint
  {http://arxiv.org/abs/astro-ph/0210010} {astro-ph/0210010} \BibitemShut
  {NoStop}\bibitem [{\citenamefont {Jaranowski}\ \emph {et~al.}(1998)\citenamefont
  {Jaranowski}, \citenamefont {Krolak},\ and\ \citenamefont
  {Schutz}}]{1998PhRvD..58f3001J}  \BibitemOpen
  \bibfield  {author} {\bibinfo {author} {\bibfnamefont {Piotr}\ \bibnamefont
  {Jaranowski}}, \bibinfo {author} {\bibfnamefont {Andrzej}\ \bibnamefont
  {Krolak}}, \ and\ \bibinfo {author} {\bibfnamefont {Bernard~F.}\ \bibnamefont
  {Schutz}},\ }\bibfield  {title} {\enquote {\bibinfo {title} {{Data analysis
  of gravitational-wave signals from spinning neutron stars. I: The signal and
  its detection}},}\ }\href {\doibase 10.1103/PhysRevD.58.063001} {\bibfield
  {journal} {\bibinfo  {journal} {Phys.\ Rev.\ D}\ }\textbf {\bibinfo {volume}
  {58}},\ \bibinfo {pages} {063001} (\bibinfo {year} {1998})},\ \Eprint
  {http://arxiv.org/abs/gr-qc/9804014} {arXiv:gr-qc/9804014} \BibitemShut
  {NoStop}\bibitem [{\citenamefont {{Goetz}}\ and\ \citenamefont
  {{Riles}}(2011)}]{2011CQGra..28u5006G}  \BibitemOpen
  \bibfield  {author} {\bibinfo {author} {\bibfnamefont {E.}~\bibnamefont
  {{Goetz}}}\ and\ \bibinfo {author} {\bibfnamefont {K.}~\bibnamefont
  {{Riles}}},\ }\bibfield  {title} {\enquote {\bibinfo {title} {{An all-sky
  search algorithm for continuous gravitational waves from spinning neutron
  stars in binary systems}},}\ }\href {\doibase 10.1088/0264-9381/28/21/215006}
  {\bibfield  {journal} {\bibinfo  {journal} {Classical and Quantum Gravity}\
  }\textbf {\bibinfo {volume} {28}},\ \bibinfo {eid} {215006} (\bibinfo {year}
  {2011})},\ \Eprint {http://arxiv.org/abs/1103.1301} {arXiv:1103.1301}
  \BibitemShut {NoStop}\bibitem [{\citenamefont {Dhurandhar}\ \emph {et~al.}(2008)\citenamefont
  {Dhurandhar}, \citenamefont {Krishnan}, \citenamefont {Mukhopadhyay},\ and\
  \citenamefont {Whelan}}]{Dhurandhar:2007vb}  \BibitemOpen
  \bibfield  {author} {\bibinfo {author} {\bibfnamefont {Sanjeev}\ \bibnamefont
  {Dhurandhar}}, \bibinfo {author} {\bibfnamefont {Badri}\ \bibnamefont
  {Krishnan}}, \bibinfo {author} {\bibfnamefont {Himan}\ \bibnamefont
  {Mukhopadhyay}}, \ and\ \bibinfo {author} {\bibfnamefont {John~T.}\
  \bibnamefont {Whelan}},\ }\bibfield  {title} {\enquote {\bibinfo {title}
  {{The cross-correlation search for periodic gravitational waves}},}\ }\href
  {\doibase 10.1103/PhysRevD.77.082001} {\bibfield  {journal} {\bibinfo
  {journal} {Phys.\ Rev.\ D}\ }\textbf {\bibinfo {volume} {77}},\ \bibinfo
  {pages} {082001} (\bibinfo {year} {2008})},\ \Eprint
  {http://arxiv.org/abs/0712.1578} {arXiv:0712.1578} \BibitemShut {NoStop}\bibitem [{\citenamefont {Chung}\ \emph {et~al.}(2011)\citenamefont {Chung},
  \citenamefont {Melatos}, \citenamefont {Krishnan},\ and\ \citenamefont
  {Whelan}}]{Chung:2011da}  \BibitemOpen
  \bibfield  {author} {\bibinfo {author} {\bibfnamefont {Christine}\
  \bibnamefont {Chung}}, \bibinfo {author} {\bibfnamefont {Andrew}\
  \bibnamefont {Melatos}}, \bibinfo {author} {\bibfnamefont {Badri}\
  \bibnamefont {Krishnan}}, \ and\ \bibinfo {author} {\bibfnamefont {John~T.}\
  \bibnamefont {Whelan}},\ }\bibfield  {title} {\enquote {\bibinfo {title}
  {{Designing a cross-correlation search for continuous-wave gravitational
  radiation from a neutron star in the supernova remnant SNR 1987A}},}\ }\href
  {\doibase 10.1111/j.1365-2966.2011.18585.x} {\bibfield  {journal} {\bibinfo
  {journal} {Mon.\ Not.\ R.\ Astron.\ Soc.}\ }\textbf {\bibinfo {volume}
  {414}},\ \bibinfo {pages} {2650--2663} (\bibinfo {year} {2011})},\ \Eprint
  {http://arxiv.org/abs/1102.4654} {arXiv:1102.4654} \BibitemShut {NoStop}\bibitem [{\citenamefont {{Whelan}}\ \emph {et~al.}(2015)\citenamefont
  {{Whelan}}, \citenamefont {{Sundaresan}}, \citenamefont {{Zhang}},\ and\
  \citenamefont {{Peiris}}}]{LMXBCrossCorr}  \BibitemOpen
  \bibfield  {author} {\bibinfo {author} {\bibfnamefont {John~T.}\ \bibnamefont
  {{Whelan}}}, \bibinfo {author} {\bibfnamefont {Santosh}\ \bibnamefont
  {{Sundaresan}}}, \bibinfo {author} {\bibfnamefont {Yuanhao}\ \bibnamefont
  {{Zhang}}}, \ and\ \bibinfo {author} {\bibfnamefont {Prabath}\ \bibnamefont
  {{Peiris}}},\ }\bibfield  {title} {\enquote {\bibinfo {title} {{A Model-Based
  Cross-Correlation Search for Gravitational Waves from Scorpius X-1}},}\
  }\href {https://dcc.ligo.org/LIGO-P1200142/public} {\bibfield  {journal}
  {\bibinfo  {journal} {LIGO DCC}\ }\textbf {\bibinfo {volume} {P1200142}}
  (\bibinfo {year} {2015})}\BibitemShut {NoStop}\bibitem [{\citenamefont {website}()}]{eath}  \BibitemOpen
  \bibfield  {author} {\bibinfo {author} {\bibfnamefont {Einstein@Home}\
  \bibnamefont {website}},\ }\href@noop {} {}\bibinfo {howpublished}
  {http://einstein.phys.uwm.edu/}\BibitemShut {NoStop}\bibitem [{\citenamefont {{Aasi}}\ \emph {et~al.}(2013)\citenamefont {{Aasi}}
  \emph {et~al.}}]{2013PhRvD..87d2001A}  \BibitemOpen
  \bibfield  {author} {\bibinfo {author} {\bibfnamefont {J.}~\bibnamefont
  {{Aasi}}} \emph {et~al.} (\bibinfo {collaboration} {LIGO Scientific
  Collaboration and Virgo Collaboration}),\ }\bibfield  {title} {\enquote
  {\bibinfo {title} {{Einstein@Home all-sky search for periodic gravitational
  waves in LIGO S5 data}},}\ }\href {\doibase 10.1103/PhysRevD.87.042001}
  {\bibfield  {journal} {\bibinfo  {journal} {Phys.\ Rev.\ D}\ }\textbf
  {\bibinfo {volume} {87}},\ \bibinfo {eid} {042001} (\bibinfo {year}
  {2013})},\ \Eprint {http://arxiv.org/abs/1207.7176} {arXiv:1207.7176}
  \BibitemShut {NoStop}\bibitem [{\citenamefont {Knispel}\ \emph {et~al.}(2011)\citenamefont {Knispel}
  \emph {et~al.}}]{Knispel:2011ss}  \BibitemOpen
  \bibfield  {author} {\bibinfo {author} {\bibfnamefont {B.}~\bibnamefont
  {Knispel}} \emph {et~al.},\ }\bibfield  {title} {\enquote {\bibinfo {title}
  {{Arecibo PALFA Survey and Einstein@Home: Binary Pulsar Discovery by
  Volunteer Computing}},}\ }\href {\doibase 10.1088/2041-8205/732/1/L1}
  {\bibfield  {journal} {\bibinfo  {journal} {Astrophys.\ J.}\ }\textbf
  {\bibinfo {volume} {732}},\ \bibinfo {pages} {L1} (\bibinfo {year} {2011})},\
  \Eprint {http://arxiv.org/abs/1102.5340} {arXiv:1102.5340} \BibitemShut
  {NoStop}\bibitem [{\citenamefont {Knispel}\ \emph {et~al.}(2013)\citenamefont {Knispel}
  \emph {et~al.}}]{Knispel:2013da}  \BibitemOpen
  \bibfield  {author} {\bibinfo {author} {\bibfnamefont {B.}~\bibnamefont
  {Knispel}} \emph {et~al.},\ }\bibfield  {title} {\enquote {\bibinfo {title}
  {{Einstein@Home Discovery of 24 Pulsars in the Parkes Multi-beam Pulsar
  Survey}},}\ }\href {\doibase 10.1088/0004-637X/774/2/93} {\bibfield
  {journal} {\bibinfo  {journal} {Astrophys.\ J.}\ }\textbf {\bibinfo {volume}
  {774}},\ \bibinfo {pages} {93} (\bibinfo {year} {2013})},\ \Eprint
  {http://arxiv.org/abs/1302.0467} {arXiv:1302.0467} \BibitemShut {NoStop}\bibitem [{\citenamefont {Allen}\ \emph {et~al.}(2013)\citenamefont {Allen}
  \emph {et~al.}}]{Allen:2013sua}  \BibitemOpen
  \bibfield  {author} {\bibinfo {author} {\bibfnamefont {B.}~\bibnamefont
  {Allen}} \emph {et~al.},\ }\bibfield  {title} {\enquote {\bibinfo {title}
  {{The Einstein@Home search for radio pulsars and PSR J2007+2722
  discovery}},}\ }\href {\doibase 10.1088/0004-637X/773/2/91} {\bibfield
  {journal} {\bibinfo  {journal} {Astrophys.\ J.}\ }\textbf {\bibinfo {volume}
  {773}},\ \bibinfo {pages} {91} (\bibinfo {year} {2013})},\ \Eprint
  {http://arxiv.org/abs/1303.0028} {arXiv:1303.0028} \BibitemShut {NoStop}\bibitem [{\citenamefont {Pletsch}\ \emph {et~al.}(2013)\citenamefont {Pletsch}
  \emph {et~al.}}]{Pletsch:2013iva}  \BibitemOpen
  \bibfield  {author} {\bibinfo {author} {\bibfnamefont {H.~J.}\ \bibnamefont
  {Pletsch}} \emph {et~al.},\ }\bibfield  {title} {\enquote {\bibinfo {title}
  {{Einstein@Home discovery of four young gamma-ray pulsars in $Fermi$ LAT
  data}},}\ }\href {\doibase 10.1088/2041-8205/779/1/L11} {\bibfield  {journal}
  {\bibinfo  {journal} {Astrophys.\ J.}\ }\textbf {\bibinfo {volume} {779}},\
  \bibinfo {pages} {L11} (\bibinfo {year} {2013})},\ \Eprint
  {http://arxiv.org/abs/1311.6427} {arXiv:1311.6427} \BibitemShut {NoStop}\bibitem [{\citenamefont {Prix}(2007)}]{2007PhRvD..75b3004P}  \BibitemOpen
  \bibfield  {author} {\bibinfo {author} {\bibfnamefont {Reinhard}\
  \bibnamefont {Prix}},\ }\bibfield  {title} {\enquote {\bibinfo {title}
  {{Search for continuous gravitational waves: Metric of the multi-detector
  F-statistic}},}\ }\href {\doibase 10.1103/PhysRevD.75.023004} {\bibfield
  {journal} {\bibinfo  {journal} {Phys.\ Rev.\ D}\ }\textbf {\bibinfo {volume}
  {75}},\ \bibinfo {pages} {023004} (\bibinfo {year} {2007})},\ \Eprint
  {http://arxiv.org/abs/gr-qc/0606088} {arXiv:gr-qc/0606088} \BibitemShut
  {NoStop}\bibitem [{\citenamefont {Brady}\ and\ \citenamefont
  {Creighton}(2000)}]{2000PhRvD..61h2001B}  \BibitemOpen
  \bibfield  {author} {\bibinfo {author} {\bibfnamefont {Patrick~R.}\
  \bibnamefont {Brady}}\ and\ \bibinfo {author} {\bibfnamefont {Teviet}\
  \bibnamefont {Creighton}},\ }\bibfield  {title} {\enquote {\bibinfo {title}
  {{Searching for periodic sources with LIGO. 2. Hierarchical searches}},}\
  }\href {\doibase 10.1103/PhysRevD.61.082001} {\bibfield  {journal} {\bibinfo
  {journal} {Phys.\ Rev.\ D}\ }\textbf {\bibinfo {volume} {61}},\ \bibinfo
  {pages} {082001} (\bibinfo {year} {2000})},\ \Eprint
  {http://arxiv.org/abs/gr-qc/9812014} {arXiv:gr-qc/9812014} \BibitemShut
  {NoStop}\bibitem [{\citenamefont {Wette}(2012)}]{Wette:2011jr}  \BibitemOpen
  \bibfield  {author} {\bibinfo {author} {\bibfnamefont {Karl}\ \bibnamefont
  {Wette}},\ }\bibfield  {title} {\enquote {\bibinfo {title} {{Estimating the
  sensitivity of wide-parameter-space searches for gravitational-wave
  pulsars}},}\ }\href {\doibase 10.1103/PhysRevD.85.042003} {\bibfield
  {journal} {\bibinfo  {journal} {Phys.\ Rev.\ D}\ }\textbf {\bibinfo {volume}
  {85}},\ \bibinfo {pages} {042003} (\bibinfo {year} {2012})},\ \Eprint
  {http://arxiv.org/abs/1111.5650} {arXiv:1111.5650} \BibitemShut {NoStop}\bibitem [{\citenamefont {Prix}\ and\ \citenamefont
  {Shaltev}(2012)}]{2012PhRvD..85h4010P}  \BibitemOpen
  \bibfield  {author} {\bibinfo {author} {\bibfnamefont {Reinhard}\
  \bibnamefont {Prix}}\ and\ \bibinfo {author} {\bibfnamefont {Miroslav}\
  \bibnamefont {Shaltev}},\ }\bibfield  {title} {\enquote {\bibinfo {title}
  {{Search for Continuous Gravitational Waves: Optimal StackSlide method at
  fixed computing cost}},}\ }\href {\doibase 10.1103/PhysRevD.85.084010}
  {\bibfield  {journal} {\bibinfo  {journal} {Phys.\ Rev.\ D}\ }\textbf
  {\bibinfo {volume} {85}},\ \bibinfo {pages} {084010} (\bibinfo {year}
  {2012})},\ \Eprint {http://arxiv.org/abs/1201.4321} {arXiv:1201.4321}
  \BibitemShut {NoStop}\bibitem [{\citenamefont {Meadors}\ \emph {et~al.}(2015)\citenamefont
  {Meadors}, \citenamefont {Goetz},\ and\ \citenamefont
  {Riles}}]{TwoSpectMDCMethods2015}  \BibitemOpen
  \bibfield  {author} {\bibinfo {author} {\bibfnamefont {G.D.}\ \bibnamefont
  {Meadors}}, \bibinfo {author} {\bibfnamefont {E.}~\bibnamefont {Goetz}}, \
  and\ \bibinfo {author} {\bibfnamefont {K.}~\bibnamefont {Riles}},\ }\bibfield
   {title} {\enquote {\bibinfo {title} {Twospect: search for a simulated
  scorpius x-1},}\ }\href {https://dcc.ligo.org/LIGO-P1500037/public}
  {\bibfield  {journal} {\bibinfo  {journal} {LIGO DCC}\ }\textbf {\bibinfo
  {volume} {P1500037}} (\bibinfo {year} {2015})}\BibitemShut {NoStop}\bibitem [{lal()}]{lalsuite}  \BibitemOpen
  \href {https://www.lsc-group.phys.uwm.edu/daswg/projects/lalsuite.html}
  {\enquote {\bibinfo {title} {{LSC Algorithm Library Suite}},}\ }\bibinfo
  {howpublished}
  {https://www.lsc-group.phys.uwm.edu/daswg/projects/lalsuite.html}\BibitemShut
  {NoStop}\bibitem [{\citenamefont {{D.~Buskulic}}\ \emph {et~al.}(2014)\citenamefont
  {{D.~Buskulic}}, \citenamefont {{I.~Fiori}}, \citenamefont {{F.~Marion}},\
  and\ \citenamefont {{B.~Mours}}}]{FrameLib}  \BibitemOpen
  \bibfield  {author} {\bibinfo {author} {\bibnamefont {{D.~Buskulic}}},
  \bibinfo {author} {\bibnamefont {{I.~Fiori}}}, \bibinfo {author}
  {\bibnamefont {{F.~Marion}}}, \ and\ \bibinfo {author} {\bibnamefont
  {{B.~Mours}}},\ }\href {http://lappweb.in2p3.fr/virgo/FrameL/} {\enquote
  {\bibinfo {title} {The frame vector library (frv)},}\ }\bibinfo
  {howpublished} {http://lappweb.in2p3.fr/virgo/FrameL/} (\bibinfo {year}
  {2014})\BibitemShut {NoStop}\bibitem [{\citenamefont {{Zhang}}\ \emph {et~al.}(2015)\citenamefont
  {{Zhang}}, \citenamefont {{Whelan}},\ and\ \citenamefont
  {{Krishnan}}}]{CrossCorrMDC}  \BibitemOpen
  \bibfield  {author} {\bibinfo {author} {\bibfnamefont {Yuanhao}\ \bibnamefont
  {{Zhang}}}, \bibinfo {author} {\bibfnamefont {John~T.}\ \bibnamefont
  {{Whelan}}}, \ and\ \bibinfo {author} {\bibfnamefont {Badri}\ \bibnamefont
  {{Krishnan}}},\ }\bibfield  {title} {\enquote {\bibinfo {title} {{Results of
  a Model-Based Cross-Correlation Search for Signals from Scorpius X-1 in Mock
  Gravitational-Wave Data}},}\ }\href
  {https://dcc.ligo.org/LIGO-P1400216/public} {\bibfield  {journal} {\bibinfo
  {journal} {LIGO DCC}\ }\textbf {\bibinfo {volume} {P1400216}} (\bibinfo
  {year} {2015})}\BibitemShut {NoStop}\bibitem [{\citenamefont {Prix}(2011)}]{T0900149-v5}  \BibitemOpen
  \bibfield  {author} {\bibinfo {author} {\bibfnamefont {Reinhard}\
  \bibnamefont {Prix}},\ }\href {https://dcc.ligo.org/LIGO-T0900149-v5/public}
  {\enquote {\bibinfo {title} {{The $\mathcal{F}$-statistic and its
  implementation in \texttt{ComputeFstatistic\_v2}}},}\ }\bibinfo
  {howpublished} {LIGO Technical Document LIGO-T0900149-v5} (\bibinfo {year}
  {2011})\BibitemShut {NoStop}\bibitem [{\citenamefont {Goodman}(1960)}]{Goodman}  \BibitemOpen
  \bibfield  {author} {\bibinfo {author} {\bibfnamefont {L.A.}\ \bibnamefont
  {Goodman}},\ }\bibfield  {title} {\enquote {\bibinfo {title} {On the exact
  variance of products},}\ }\href {\doibase 10.1080/01621459.1960.10483369}
  {\bibfield  {journal} {\bibinfo  {journal} {Journal of the American
  Statistical Association}\ }\textbf {\bibinfo {volume} {55}},\ \bibinfo
  {pages} {708--713} (\bibinfo {year} {1960})}\BibitemShut {NoStop}\end{thebibliography}
\end{document}